\documentclass[apj]{emulateapj}
\usepackage[section]{placeins}
\usepackage{subfigure}
\usepackage{rotating}
\usepackage{colortbl}
\usepackage{morefloats}
\usepackage{natbib}

\shorttitle{Kinematic Distance of Galactic Planetary Nebulae}
\shortauthors{Aiyuan Yang et al.}
\usepackage{rotating,color,array,threeparttable,booktabs,multirow,upgreek}
\usepackage{subfigure}
\usepackage{multirow}
\def\HII{{H \sc ii}}
\def\HI{{H \sc  i}}
\def\CO{{\sc CO}}
\def\Msol{M_\odot}

\begin{document}


\title{Kinematic Distances of Galactic Planetary Nebulae}


\author{
A. Y. Yang\altaffilmark{1,2},
W. W. Tian\altaffilmark{1,2,3},
H. Zhu\altaffilmark{1,2},
D. A. Leahy\altaffilmark{3},
D. Wu\altaffilmark{1,4,5}
}
\affil{$^1$Key Laboratory of Optical Astronomy, National Astronomical Observatories, \\ Chinese Academy of Sciences, Beijing~100012, China\\
$^2$University of Chinese Academy of Science, 19A Yuquan Road, Beijing 100049, China\\
$^3$Department of Physics $\&$ Astronomy, University of Calgary, Calgary, Alberta T2N 1N4, Canada\\
$^4$College of Information Science and Technology, Beijing Normal University, Beijing 100875, China\\
$^5$Beijing Key Laboratory of Digital Preservation and Virtual Reality for Cultural Heritage, Beijing 100875, China\\}
\email{Email: tww@bao.ac.cn}


\begin{abstract}

We construct \HI~absorption spectra for 18 planetary nebulae (PNe) and their background sources 
using the data from the International Galactic Plane Survey.
We estimate the kinematic distances of these PNe,
among which 15 objects' kinematic distances are obtained for the first time.
The distance uncertainties of 13 PNe range from 10$\%$ to 50$\%$, 
which is a significant improvement with uncertainties of a factor two or three smaller 
than most of previous distance measurements.
We confirm that PN G030.2$-$00.1 
is not a PN because of its large distance found here. 

\end{abstract}





\section{INTRODUCTION}

Distances, as a basic physical parameter of planetary
nebulae~(PNe), are crucial to study their size, luminosity, ionized mass, formation rate,
space density and Galactic distribution.
However, distances are still not well determined for the majority of total $\sim$3500 PNe \citep{Kwitter2014}.
For an individual PN, different methods may lead to different distances.
So far, there are only about thirty PNe having their distance measurements with uncertainties less than $~20\%$.

Nine popular methods have been used to measure the distances of PNe,
including trigonometric parallax \cite[e.g.,][]{Harris2007},
cluster member \cite[e.g.,][]{Jacoby1997},
expansion parallax \cite[e.g.,][]{Terzian1997},
spectroscopic parallax \cite[e.g.,][]{Ciardullo1999},
reddening \cite[e.g.,][]{Gathier1986},
Na~D absorption \cite[e.g.,][]{Napiwotzki1995},
determinations of central star gravities~\cite[e.g.,][]{Mendez1988},
statistical method \cite[revised Shklovsky method, e.g.,][]{Shklovskii1956}
and kinematics method \cite[e.g.,][]{Gathier1986a}.

Hydrogen is the most abundant element in
the universe and \HI~atom clouds
are broadly distributed in the Milky Way \citep{Dickey1990}.
The 21 cm \HI~absorption line has been widely used to measure kinematic
distances of \HI~clouds and radio strong sources associated.
When a \HI~cloud is located in front of or behind a strong radio source,
we are usually able to detect a \HI~absorption feature or only an emission
line from the cloud. The velocity of the emission/absorption
feature can be converted into a distance
based on the axisymmetric rotation curve model for the Galaxy.
The distance or distance limit of the source can be estimated from the distance of the \HI~cloud.
However, this method faces two main challenges.
One is the kinematic distance ambiguity (KDA) for sources located inside the solar circle, 
as each radial velocity along given line of sight
corresponds to two distances equally spaced on
either side of the tangent point. The KDA usually can be solved by integrated consideration
of \HI~absorption/self-absorption,
\CO~emission and \HI~absorption of background sources.
Another challenge is to construct a reliable \HI~absorption spectrum to a radio source
due to the uneven \HI~background in the Galaxy.
In order to minimize the second effect, \cite{Tian2007a} and \cite{Leahy2008} developed
revised methods to construct \HI~absorption spectra.
The methods have been applied to several types of Galactic objects successfully,
e.g., PNe \citep{Zhu2013}, supernova remnants and \HII~regions \cite[e.g.,][]{Leahy2008,Tian2008c}.

In this paper, we systematically construct the \HI~absorption spectra of
PNe which are located in the sky region of the International Galactic Plane Survey (IGPS).
~\HI~absorption features in the spectra are used to determine
the PNe's distances. This paper is organized as follows: the data and the revised methods are introduced in Section 2.
In Section 3, we apply the methods to estimate individual PN distance.
Summary is given in Section 4.

\section{DATA ANALYSIS}

\subsection{Data}

\begin{table*}[!hbt]
\scriptsize
 \centering
 \begin{minipage}{130mm}
 \caption {Information of PNe and background sources}
\footnotesize
\setlength{\extrarowheight}{0.5pt} 
  \begin{tabular}{llrlcc|cr}
  \hline
   \hline
 PN G           &~~~$l$          &$b$~~~~       &~~~~~$S_{\rm 1.4GHz}$       &Survey      &Ref.     & Background &Sources   \\
 Name          &~~~$\circ$     &$\circ$~~~~  &~~~~~mJy                      &            &        &$l$  & $b$~~~         \\
 \hline
\multirow{2}*{G020.9$-$01.1}&\multirow{2}*{20.999}&\multirow{2}*{-1.125}
&~~~~~\multirow{2}*{249.0$\pm$7.5}&\multirow{2}*{V}&\multirow{2}*{CK98}& 21.345~&~$-$0.630  \\
&       &             &         &                  &        &   21.500~&~$-$0.885      \\
\hline
\multirow{3}*{G029.0$+$00.4} &\multirow{3}*{29.079} &\multirow{3}*{0.454} &~~~~~\multirow{3}*{159.0$\pm$15} &\multirow{3}*{V}&\multirow{3}*{CK98}& 28.800~~&~~0.175\\
&       &         &         &                 &        &    28.825~~&~~$-$0.230 \\
&       &         &         &                  &        &    29.098~~&~~0.545 \\
\hline
\multirow{3}*{G030.2$-$00.1}& \multirow{3}*{30.234}&\multirow{3}*{-0.139}
&~~~~~\multirow{3}*{$\cdots^{a}$} &\multirow{3}*{V} &\multirow{3}*{A11}  &29.930~~&~~$-$0.055     \\
 &       &         &         &                  &          & 30.535~~&~~0.020  \\
 &       &         &         &                   &          &   30.685~~&~~$-$0.260   \\
 &       &         &         &                   &          & 29.960~~&~~$-$0.020  \\
\hline
\multirow{3}*{G051.5$+$00.2} &\multirow{3}*{51.509}&\multirow{3}*{0.167}
&~~~~~\multirow{3}*{84.0}&\multirow{3}*{V} &\multirow{3}*{LCY05} &51.775~~&~~0.800 \\
&       &         &         &                 &          & 50.625~~&~~$-$0.030   \\
&       &         &         &                &         &50.950~~&~~0.850 \\
\hline
\multirow{2}*{G052.1$+$01.0}&\multirow{2}*{52.099}&\multirow{2}*{1.043}
&~~~~~\multirow{2}*{455.1$\pm$13.7}&\multirow{2}*{V}&\multirow{2}*{IPHAS09}&52.235~~&~~0.745 \\
&       &        &         &                  &        &  52.750~~&~~0.335     \\
\hline
\multirow{3}*{G055.5$-$00.5} &\multirow{3}*{55.507} &\multirow{3}*{ -0.558}
&~~~~~\multirow{3}*{84.1$\pm$2.6} &\multirow{3}*{V} &\multirow{3}*{CK98} &55.525~~&~~$-$1.150\\
&       &        &         &                  &        &55.995~~&~~$-$1.195   \\
&       &        &         &                  &        &55.775~~&~~$-$0.250 \\
\hline
\multirow{2}*{G069.7$-$00.0}&\multirow{2}*{69.800}&\multirow{2}*{0.004}
&~~~~~\multirow{2}*{87.1$\pm$2.6}&\multirow{2}*{V}&\multirow{2}*{CK98}&68.755~~&~~0.275\\
&       &        &         &                  &        &70.155~~&~~0.090   \\
\hline
\multirow{3}*{G070.7$+$01.2}&\multirow{3}*{70.674} &\multirow{3}*{1.192}
&~~~~~\multirow{3}*{949.6$\pm$33.4 } &\multirow{3}*{C} &\multirow{3}*{IPHAS09}&71.225~~&~~1.445      \\
&       &       &         &                  &        &  70.690~~&~~0.630   \\
&       &       &         &                  &        &  70.600~~&~~1.380  \\
\hline
\multirow{2}*{G084.9$-$03.4}&\multirow{2}*{84.930 }&\multirow{2}*{-3.496} &~~~~~\multirow{2}*{1373$\pm$41}  & \multirow{2}*{C}  & \multirow{2}*{CK98}&85.120~~&~~$-$3.100           \\
&       &       &         &                  &        &  84.445~~&~~$-$2.915  \\
\hline
\multirow{3}*{G089.0$+$00.3}&\multirow{3}*{89.002} &\multirow{3}*{0.376}
&~~~~~\multirow{3}*{260.5$\pm$7.8}&\multirow{3}*{C}&\multirow{3}*{IPHAS09}& 88.825~~&~~0.925 \\
&       &       &         &                  &        &  88.465~~&~~0.004  \\
&       &       &         &                  &        &  89.650~~&~~0.925   \\
\hline
G107.8$+$02.3&107.845&2.314 &~~~~~581$\pm$22.8&C&IPHAS09&108.755~~&~~2.575     \\  [5pt]
\hline
G138.8$+$02.8 &38.816 & 2.805&~~~~~153.1$\pm$ 5.8&C&CK98&138.795~~&~~2.140 \\[5pt]
\multirow{2}*{G147.4$-$02.3} &\multirow{2}*{147.401}&\multirow{2}*{-2.307}
&~~~~~\multirow{2}*{77.8$\pm$ 2.4}&\multirow{2}*{C} &\multirow{2}*{IPHAS09}&146.625~~&~~$-$2.690 \\
&       &     &         &                 &        &  147.950~~&~~$-$2.645  \\
\hline
\multirow{2}*{G169.7$-$00.1}&\multirow{2}*{169.653}&\multirow{2}*{-0.077}
&~~~~~\multirow{2}*{135.3$\pm$5.1}&\multirow{2}*{C} &\multirow{2}*{IPHAS09} & 169.080~~&~~$-$0.245  \\
&       &     &         &                  &        &   170.330~~&~~$-$0.225     \\
\hline
\multirow{2}*{G259.1$+$00.9}&\multirow{2}*{259.149}&\multirow{2}*{0.940}
&~~~~~\multirow{2}*{345$\pm$12}&\multirow{2}*{S}&\multirow{2}*{CK98}&257.902~~&~~0.844 \\
&       &        &         &                  &        &257.913~~&~~0.655   \\
\hline
\multirow{2}*{G333.9$+$00.6} &\multirow{2}*{333.930}&\multirow{2}*{0.686}
&~~~~~\multirow{2}*{294.8$\pm$9.7}&\multirow{2}*{S} &\multirow{2}*{BPF11}&333.723~~&~~0.377  \\
&       &          &         &                  &        &  332.967~~&~~0.777  \\
\hline
\multirow{3}*{G352.6$+$00.1} &\multirow{3}*{352.675}&\multirow{3}*{0.144}
&~~~~~\multirow{3}*{378$\pm$38}&\multirow{3}*{S}&\multirow{3}*{CK98}&352.588~~&~~$-$0.167   \\
&       &         &         &                 &        & 351.616~~&~~0.177  \\
&       &         &         &                  &        &  353.408~~&~~$-$0.355 \\
\hline
\multirow{3}*{G352.8$-$00.2} &\multirow{3}*{352.828}&\multirow{3}*{-0.259}
&~~~~~\multirow{3}*{474$\pm$47} &\multirow{3}*{S} &\multirow{3}*{CK98} &352.588~~&~~$-$0.167\\
&       &        &         &                  &        &  352.854~~&~~$-$0.189   \\
&       &        &         &                  &        &  353.408~~&~~$-$0.355 \\
\hline
\end{tabular}
\tablecomments{
The first three columns give the name and Galactic coordinates. 
Column (4) gives the flux density at 1.4\,GHz.
The survey names and references are given in the next two columns.
The final two columns give the Galactic coordinates of background sources.
References related in table are as follow: IPHAS09-\cite{Viironen2009},
CK98-\cite{Condon1998}, LCY05-\cite{Luo2005}, BPF11-\cite{Boji2011}, A11-\cite{Anderson2011}.
a--the flux at 8.7\,{\rm GHz} is 752mJy for PN G030.2$-$00.1.
}
\end{minipage}
\label{table1}
\end{table*}

The 1420\,MHz radio continuum and \HI-line emission data come from
IGPS~(the Very Large Array Galactic Plane Survey (VGPS)\citep{Stil2006},
the Southern Galactic Plane Survey (SGPS) \citep{McClure-Griffiths2005},
and the Canadian Galactic Plane Survey (CGPS) \citep{Taylor2003}).
The project surveys the Galactic disk from longitudes 18$^{\circ}$ to 67$^{\circ}$,
255$^{\circ}$ to 357$^{\circ}$ and 65$^{\circ}$ to 175$^{\circ}$, respectively.
For CGPS, the continuum image at 1420MHz has a spacial resolution of 1$^{'}$
and \HI~spectra line images have a resolution of 1$^{'}\times$ 1$^{'}\times$ 1.56\,km $s^{-1} $.
At declination $\delta$, the synthesized beam is 49$^{''}\times$ 49$^{''}csc\delta$ for 1420\,MHz 
and 58$^{''}\times$ 58$^{''}csc\delta \times$ 1.32\,km $s^{-1}$for \HI~in the survey of CGPS. SGPS has a
resolution 100$^{''}$ for continuum and $\sim$ 2$^{'}\times$ 2$^{'}\times$ 1\,km $s^{-1} $ for \HI~data. The $^{13}\CO$~
spectral line data from the Galactic Ring Survey of the Five College Radio Astronomical Observatory~(FCRAO)
14\,m telescope \citep{Jackson2006} has an angular and spectral resolution of 46$^{''}$ and 0.21\,km $s^{-1} $
at longitudes from 18$^{\circ}$ to 52$^{\circ}$ and latitudes between $-$1$^{\circ}$ and 1$^{\circ}$. 
The $^{12}\CO$~ (J=1$-$0) spectral line data from the FCRAO~\CO~Survey of the Outer 
Galaxy has an angular and spectral resolution of 45$^{''}$ and 0.98\,km $s^{-1} $ between
Galactic longitudes 102.49$^{\circ}$~---~141.54$^{\circ}$ and
latitudes $-$3.03$^{\circ}$~---~5.41$^{\circ}$ \citep{Heyer1998}.

In this work, we construct the \HI~absorption spectra of the PNe 
with flux density larger than 50\,mJy at 1420\,MHz in the IGPS. 
By checking their \HI~spectra one by one visually, 18 of them show reliable absorption features. 
We analyze the 18 PNe in this paper. 
For each PN, at least one bright nearby background source has been chosen
as a comparison with the PN in order to understand the PN's \HI~ absorption spectrum.
The parameters of 18 PNe and their background sources are shown
in Table~\ref{table1}.

\subsection{To obtain a reliable~\HI~absorption spectrum}

Based on the knowledge of radiation transfer, 
the brightness temperature of source ($T_{\rm on}$) and background ($T_{\rm off}$)
that have continuum emission subtracted can be determined by the equations:
\begin{equation}
T_{\rm on}(\nu)=T_{\rm B} (\nu)(1-e^{\rm -\tau(\nu)})+T_{\rm S}^{\rm C}(e^{\rm -\tau_{c}(\nu)}-1)
\end{equation}
\begin{equation}
T_{\rm off}(\nu)=T_{\rm B} (\nu)(1-e^{\rm -\tau(\nu)})+T_{\rm bg}^{\rm C}(e^{\rm -\tau_{c}(\nu)}-1)
\end{equation}
So, we can obtain the absorption spectrum of \HI,
\begin{equation}
e^{\rm -\tau_{c}(\nu) }=1-{\frac{T_{\rm off}(\nu)-T_{\rm on}(\nu)}{T_{\rm S}^{\rm C}-T_{\rm bg}^{\rm C}}},
\end{equation}
Where, $T_{\rm B} (\nu)$ is the spin temperature of \HI~cloud, $T_{\rm S}^{\rm C}$ and $T_{\rm bg}^{\rm C}$ are the
continuum brightness temperatures of source and background.
The \HI~absorption spectrum is usually represented by $e^{-\tau_{\rm c}(\nu)}$ or sometimes by $\tau_{\rm c}$.

To construct a \HI~absorption spectrum, traditionally one usually
chooses the source and background regions separately.
This could increase the possibility of false absorption spectrum caused by
the different distributions of \HI~clouds along the two lines of sight.
Nevertheless, \cite{Tian2007a} and \cite{Leahy2008} proposed revised methods
by selecting the background region directly surrounding the source region to
minimize the possibility of a false \HI~absorption spectrum.
In addition, they extracted \CO~emission spectrum in the source direction
and constructed \HI~absorption spectra of nearby strong background sources with their angular
separation not exceeding 1$^{\circ}$ from target source to
understand the target source's absorption spectrum better.
What's more, when it is possible,~\HI~self absorption
resulting from cold \HI~cloud absorbing emission from background
warm \HI~cloud at the same velocity has been used to
reduce the KDA problem in the methods \cite[e.g.,][]{Leahy2010,Tian2010}.

\section{KINEMATIC DISTANCE MEASUREMENT TO PLANETARY NEBULAE}

\subsection{The model}
\begin{figure*}
 \centering
 \begin{tabular}{cc}
    \includegraphics[width = 0.31\textwidth]{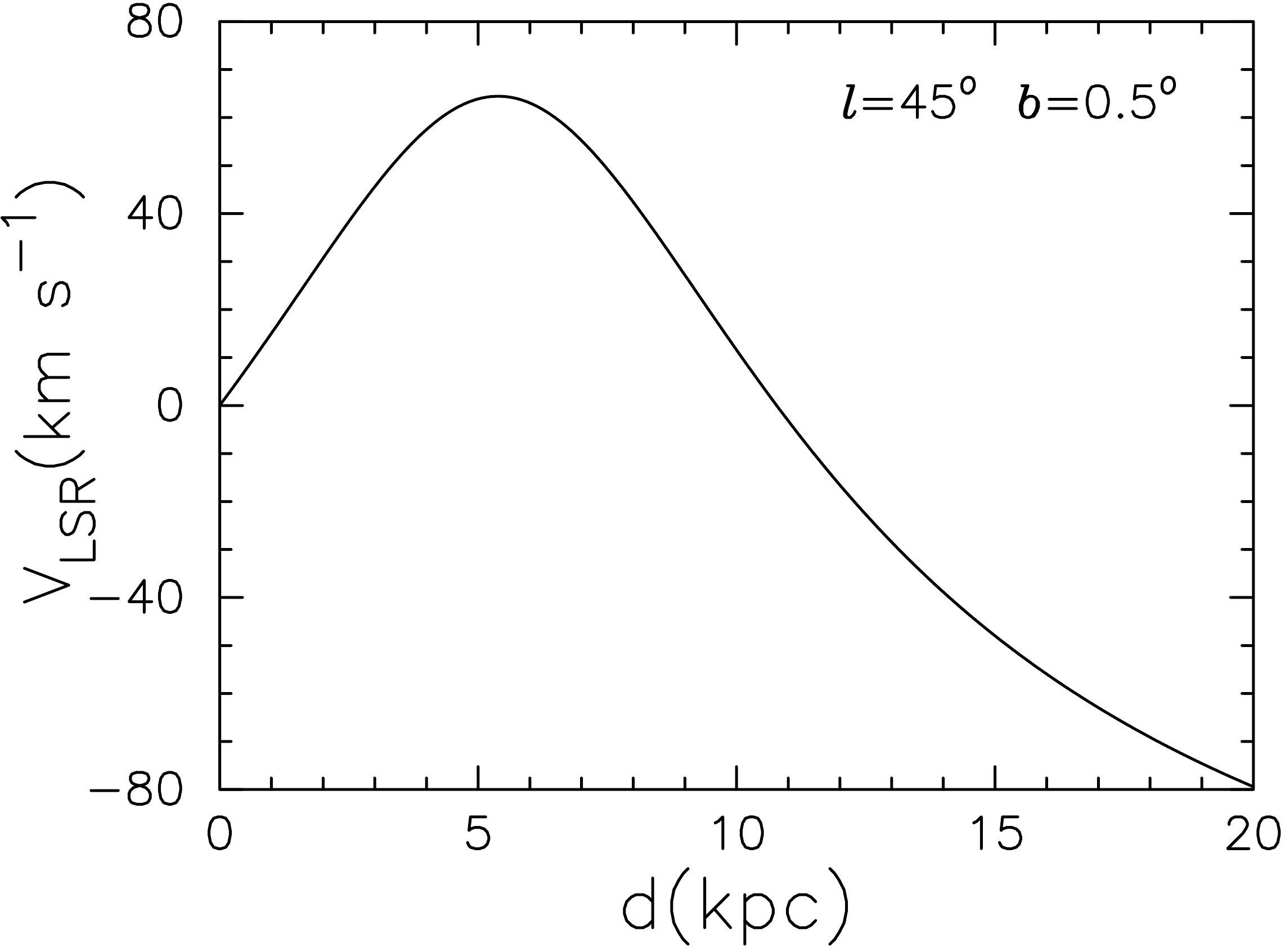}&
    \includegraphics[width = 0.315\textwidth]{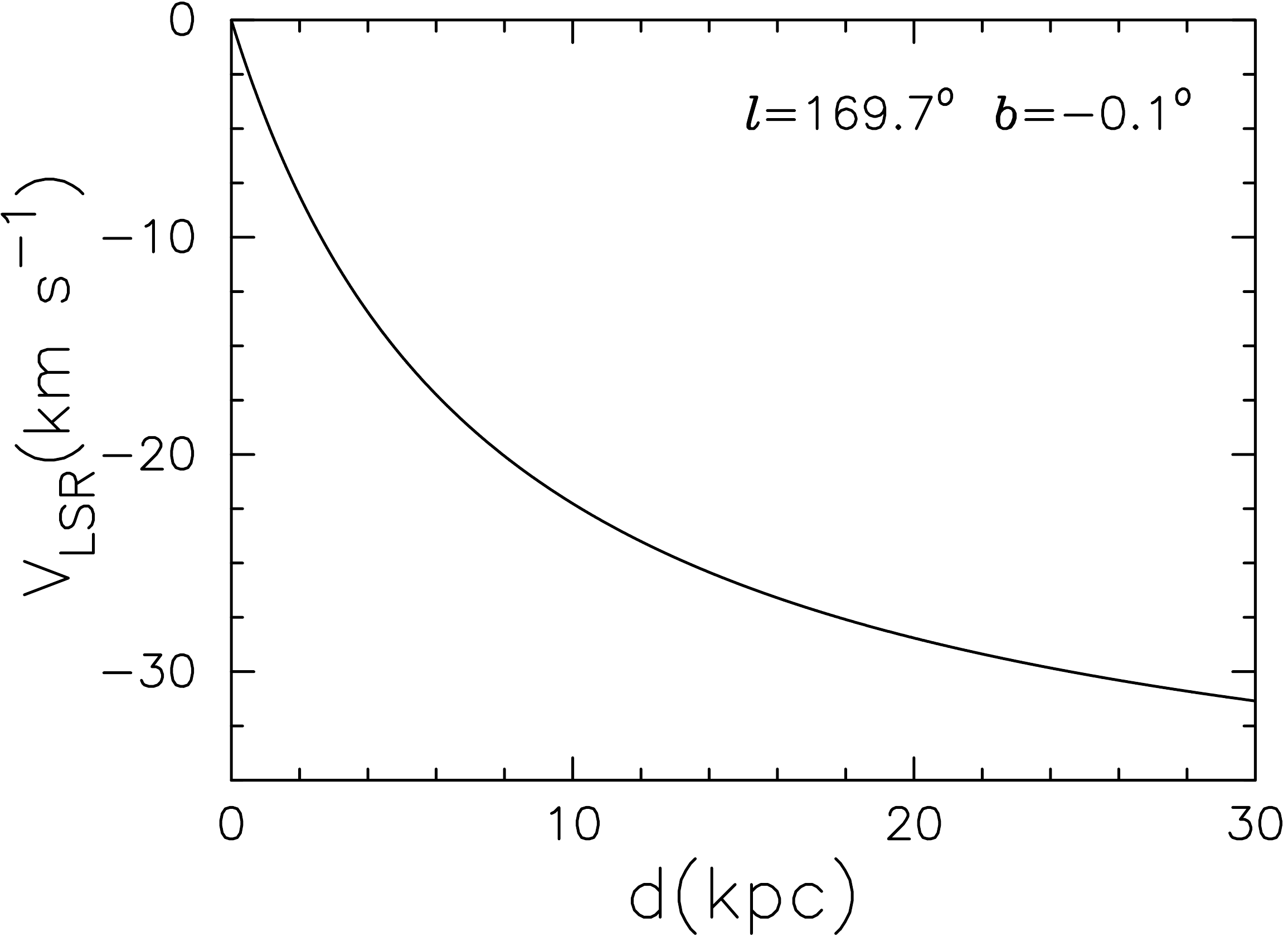}\\
    \includegraphics[width = 0.31\textwidth]{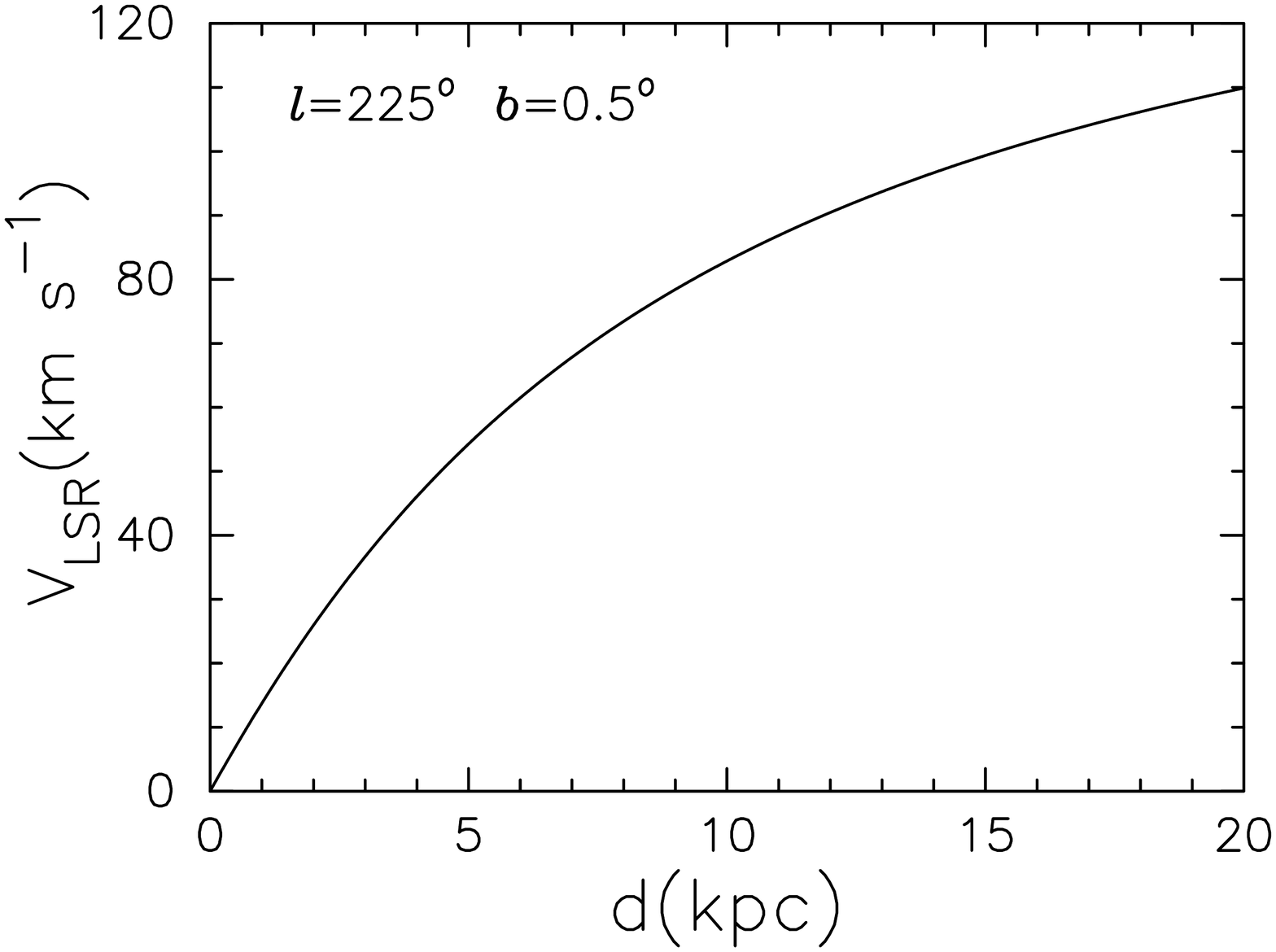}&
    \includegraphics[width = 0.315\textwidth]{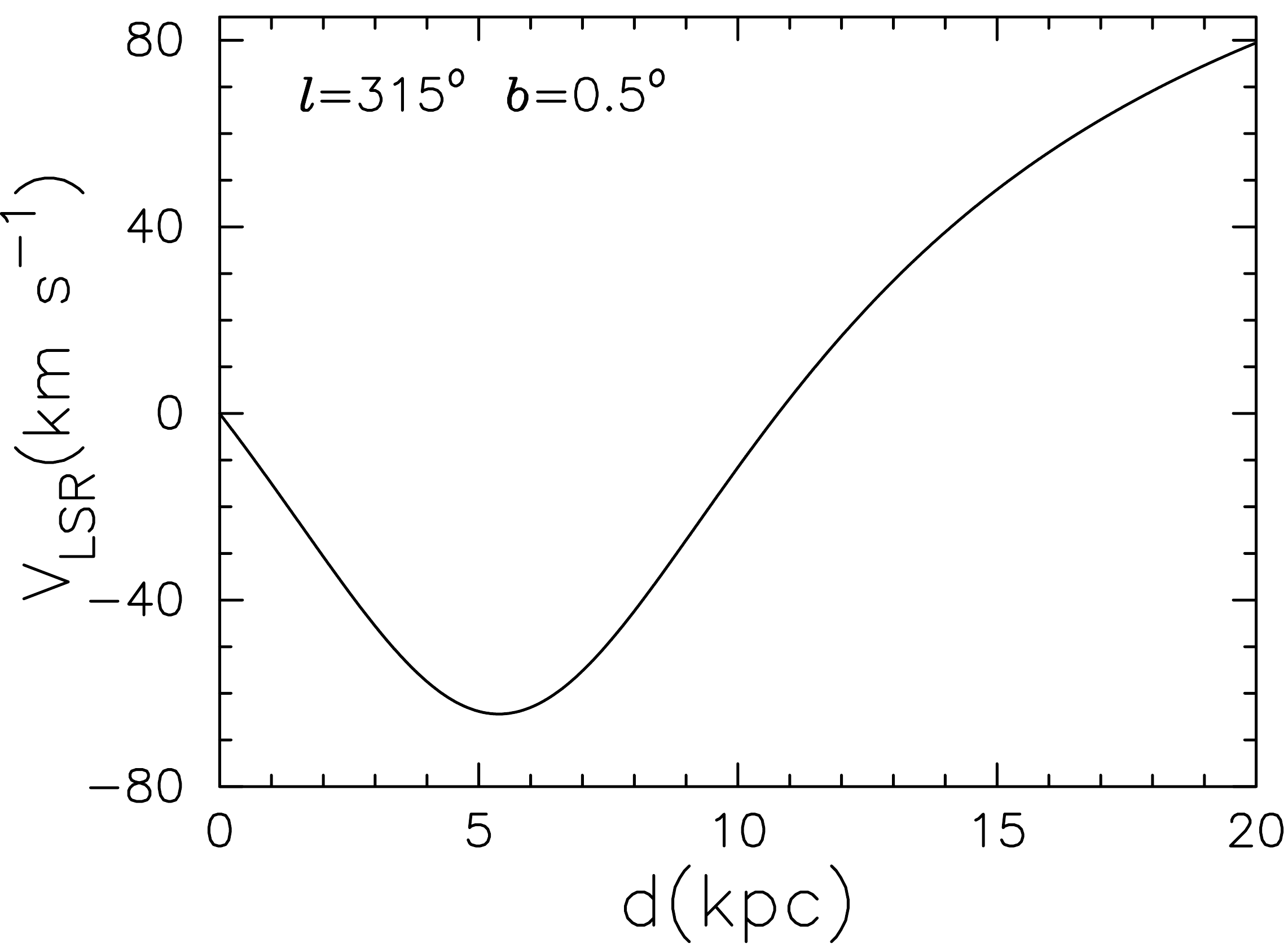}\\
   \end{tabular}
 \caption{The relation between heliocentric distance $d$ and 
 radial velocity $V_{r}$ in the four quadrants in the direction ($l$, $b$) in Galactic
coordinates, using a Galactic circular rotation curve model and adopting
a galactocentric distance of $R_{0}$=7.62\,{\rm kpc}
and a rotation velocity of $V_{0}$=220 \,km s$^{-1}$.}
 \label{fig1}
\end{figure*}

The methods of determining distances are based on the flat Galactic circular rotation curve model.
For a given PN at a distance $d$ from the Sun in the direction of ($l$, $b$) in Galactic
coordinates, the relation between the heliocentric distance $d$ and
the galactocentric distance $R$ can be written as
\begin{equation}\label{1}
R^{2}=R_{0}^{2}+d^2\cos^2b-2R_{0}d\cos{b}\cos{l} \, ,
\end{equation}
where $R_{0}$=7.62$\pm$0.32\,{\rm kpc}~\citep{Eisenhauer2005}, the distance
to the Galactic center from the Sun. However, $R_{0}$ is still uncertain \cite[e.g.][]{Bovy2012}.
Assuming circular orbits, the rotation velocity $V_{R}$ at
galactocentric distance $R$ is given by
\begin{equation}\label{2}
V_{R}={\frac{R}{R_{0}}}({\frac{V_{r}}{\cos{b}\sin{l}}}+V_{0}) \, ,
\end{equation}
where $V_{r}$ is the radial velocity corresponding to the Local Standard of Rest (LSR),
and $V_{0}$=220 \,km s$^{-1}$ is the IAU adopted velocity at the LSR. In this work,
we focus on the relation between the heliocentric distance $d$ and
the radial velocity $V_{r}$. Since $d$ can be expressed by

\begin{equation}\label{3}
d=\frac{R_{0}\cos{l}\pm \sqrt{R^{2}-{(R_{0}\sin l)}^2 } } {\cos{b}} \,  ,
\end{equation}

where R is related to  $V_{r}$ in equation (5), the heliocentric distance $d$ can be written as a function of the radial velocity $V_{r}$, 
which has different forms of expression in the four quadrants of the Galactic coordinates, 
see Fig.~\ref{fig1}. In general, we analyze the spectra of PNe and their background
sources to determine the kinematic distances of PNe assuming ${V}_{\rm R}={V}_{0}$. In some special cases, when the maximum observed
radial velocity (tangent point velocity)
in the PN spectrum is much larger than the expected value, i.e., ${V}_{\rm R}>{V}_{0}$,
we consider $V_{\rm R}$ linearly increasing from $V_{0}$=220\,km s$^{-1}$ to ${V}_{\rm R}$ as
R reduces to the value at the tangent point. The same situation has been discussed in \cite{Leahy2008a}.
We note that \cite{Reid2007} had updated the value $V_{0}/R_{0}$ 
with $V_{0}$=224\,km s$^{-1}$ and $V_{\rm R}$=242\,km s$^{-1}$.
\cite{Reid2009} also found a higher rotation velocity of $V_{0}$=254$\pm$16\,km s$^{-1}$
by measuring the trigonometric parallaxes and proper motions of masers with the Very Long Baseline Array data.
Likewise, \cite{Leahy2008a} and \cite{Levine2008} suggested a higher rotation velocity at longitude near 53$^{\circ}$.

\subsection{Application to Planetary Nebula}

We estimate distances of 18 PNe by taking 3$\,\sigma$ as the minimum
level of significance for the detection of a \HI~absorption feature,
where $\sigma$ is the standard deviation
calculated from the no emission baseline of PN \HI~spectrum.
The distance uncertainty includes that caused by an average random velocity of \HI~ clouds,
i.e., $\sim$ 6\,km s$^{-1}$ \citep{Crovisier1978,Shaver1982,Anantharamaiah1984}.
For each PN, at least one bright background source within an angular
separation of 1$^{\circ}$ from the PNe has been chosen.
The spectra are used to compare with the PN's in order to understand the PN's absorption spectrum.
The 1420\,MHz continuum images of both PNe and background sources are 
displayed in Fig.~\ref{fig2} to Fig.~\ref{fig19},
together with their \HI~absorption spectra. The distance for 
each individual PN is discussed below, taking $R_{0}$=7.62\,{\rm kpc}.

\vskip 3mm
PN G020.9$-$01.1~(Fig.~\ref{fig2}):

\vskip 1mm
Fig.~\ref{fig2}~shows the 1420\,MHz continuum image and \HI~spectra of PN G020.9$-$01.1
and its nearby background sources (G021.3$-$0.63, G021.5$-$0.89).
The PN is fainter in radio than both G021.3$-$0.63 and G021.5$-$0.89
so that the PN \HI~absorption spectrum shows more noise than the others.
Absence of absorption at the tangent point velocity~($\sim$100\,km s$^{-1}$) for  both PN G020.9$-$01.1
and G021.5$-$0.89 implies they are likely located in front of the tangent point~(7.1$\pm$0.6\,{\rm kpc}).
The spectrum of G021.3$-$0.63 shows absorption features at 105\,km s$^{-1}$ and negative velocities, which supports that this source is further than both PN G020.9$-$01.1 and G021.5$-$0.89.
The absorption feature at 62\,km s$^{-1}$ is probably not real
since the absorption is very close to 3$\,\sigma$.
The reliable absorption feature at 45\,km s$^{-1}$~(e$^{-\tau}$=0.44) 
indicates  a lower limit distance of 3.1$\pm$0.3\,{\rm kpc}
for the PN, i.e., the nearside distance for this velocity.

For PN G020.9$-$01.1, \cite{Cazetta2001} found a distance of 2.4\,{\rm kpc}
based on the relation between distance and the surface gravity of central star of
PN (~$d^2\propto \Msol F_{*}g^{-1}10^{0.4V_0}$~).
This kinematic distance of 3.1$\pm$0.3\,{\rm kpc} is reasonably
consistent with the surface gravity distance, and larger than previous statistical
distances of 1.75\,{\rm kpc} by \cite{Cahn1992},
1.66\,{\rm kpc} by \cite{Steene1995}, 1.74\,{\rm kpc} by \cite{Zhang1995},
2.29\,{\rm kpc} by \cite{Stanghellini2008} and 1.59\,{\rm kpc} by \cite{Phillips2004}.

\vskip 3mm
PN G029.0$+$00.4~(Fig.~\ref{fig3}):

\vskip 1mm

Although the background sources near PN G029.0$+$00.4 show clear absorption spectra, the PN spectrum is complex.
The prominent emission at the tangent point ($\sim$100\,km s$^{-1}$) and the absence of an absorption feature at the velocity in the PN spectrum indicate an upper limit distance of 6.6$\pm$1.0\,{\rm kpc} for the PN. One probable absorption feature at 60\,km s$^{-1}$ implies a lower limit distance of 3.5$\pm$0.3\,{\rm kpc}.

For this PN, a distance of 1.2\,{\rm kpc} has been derived by \cite{Maciel1984}
assuming a relationship between the nebular ionized mass and radius,
which is smaller than our result.

\vskip 3mm
PN G030.2$-$00.1~(Fig.~\ref{fig4}):

\vskip 1mm
PN G030.2$-$00.1 is a PN candidate suggested by \cite{Anderson2011}.
The absorption features of the PN candidate and four~\HII~regions appear
up to the tangent point velocity~($\sim$110\,km s$^{-1}$), which
indicates all five objects are beyond the tangent point, i.e., 6.6$\pm$0.9\,{\rm kpc}.
The absence of absorption at negative velocity in the PN spectrum implies the PN is 
inside the solar circle (13.2$\pm$0.5\,{\rm kpc}). The obvious absorption feature in the PN spectrum and absence of absorption in spectra of all background sources at 40\,km s$^{-1}$ imply that the PN is likely beyond the far-side distance of 40\,km s$^{-1}$, i.e., 10.7$\pm$0.3\,{\rm kpc}. Based on these information, the PN sits between 10.7$\pm$0.3\,{\rm kpc} and 13.2$\pm$0.5\,{\rm kpc}.

\vskip 3mm
PN G051.5$+$00.2~(Fig.~\ref{fig5}):

\vskip 1mm
Based on the Galactic circular rotation curve model and the
IAU adopted parameters of $V_{0}$=220\,km s$^{-1}$, the tangent
velocity $V_{\perp }$ is expected to be 48\,km $s^{-1}$.
This is much smaller than the observed value of $V_{\perp }\simeq 70$\,km s$^{-1}$
obtained from the \HI~emission spectrum in Fig.~\ref{fig5}. 
This higher $V_{\perp }$ could be due to spiral arm velocity perturbation 
near the tangent point in the direction of $l$=51.5$^{\circ}$ \citep{Dobbs2006}. 
If $V_{\perp }=70$\,km s$^{-1}$, the rotation velocity $V_{\rm R}$ would be as high as 242\,km s$^{-1}$. 
In fact, \cite{Levine2008} found a high rotation
velocity of $\sim$ 236\,km s$^{-1}$ at longitude near 53$^{\circ}$.
In addition, the high rotation velocity is also obtained in the \HI~spectra of
PN G052.1$+$01.0 and PN G055.5$-$00.5.
Altogether, we calculate the kinematic distance to the PN 
using $R_{0}$=7.62\,{\rm kpc},
$V_{0}$=220\,km s$^{-1} $ and $V_{\rm R}$=243\,km s$^{-1}$.

The PN spectrum reveals
absorptions appear up to the tangent point velocity,
giving a lower limit distance of 4.7$\pm$1.4\,{\rm kpc}.
The absence of any absorption feature at negative velocities in the PN spectrum means that the PN
is within the solar circle, giving an upper limit distance of 9.5$\pm$0.4\,{\rm kpc}.

\vskip 3mm
PN G052.1$+$01.0~(Fig.~\ref{fig6}):

\vskip 1mm
Similar to PN G051.5$+$00.2~, the observed tangent point velocity
of 68\,km s$^{-1}$ from the PN \HI~spectrum is larger than
the expected value of 46\,km s$^{-1}$ when taking commonly used parameters of $V_{0}$=220\,km s$^{-1}$. 
This leads to a rotation velocity up to $V_{R}$=242\,km s$^{-1}$.
Fig.~\ref{fig6} shows absorption features in the spectra of the PN and two \HII~regions
(G052.2$+$0.75, G052.7$+$0.3) up to the tangent point velocity~(68\,km $s^{-1}$),
revealing all three objects are beyond the tangent point.
So the lower limit distance for this PN is 4.7$\pm$1.4\,{\rm kpc}.
Bright \HI~emission is detected at 48\,km $s^{-1}$ in the three spectra,
whereas the absorption feature at this velocity is detected only in the spectra of two \HII~regions.
This implies that the PN is in front of \HI~at its far-side distance of 5.6$\pm$0.8\,{\rm kpc}.

\vskip 3mm
PN G055.5$-$00.5~(Fig.~\ref{fig7}):

\vskip 1mm
The observed tangent point velocity (55\,km s$^{-1}$ ) 
in the PN spectrum is larger than the expected value (38\,km s$^{-1}$ ), 
which suggests that the rotation velocity  is $V_{\rm R}$= 236\,km s$^{-1}$ at the tangent point. 
Absorption features at the tangent
point velocity in the spectra of the PN as well as three background sources~
(G055.5$-$01.1, G055.9$-$01.2 and G055.7$-$00.2 )
indicate that all four objects are beyond the tangent point. Therefore, the lower
limit distance is 4.3$\pm$1.5\,{\rm kpc} for the PN. Absence of absorption in the PN spectrum at negative velocity implies that the PN is within the solar cycle.
So we suggest that the distance of the PN is between 4.3$\pm$1.5\,{\rm kpc} and 8.6$\pm$0.4\,{\rm kpc}.

\cite{Giammanco2011} suggested a distance of 2.9$\pm$0.4 based on
distance-extinction relationship in the direction toward the PN.
\cite{Zhang1995} derived a distance of 3.17\,{\rm kpc} by using the relation between
the radio continuum surface brightness and the nebular radius.
\cite{Stanghellini2008} obtained a distance of 3.68\,{\rm kpc}
by the revised relation of ionized mass and optical thickness.
So our lower limit distance of 4.3$\pm$1.5\,{\rm kpc}
for the PN G055.5$-$00.5 is reasonable.

\vskip 3mm
PN G069.7$-$00.0~(Fig.~\ref{fig8}):

\vskip 1mm
The PN spectrum has low S/N due to its low brightness. 
Significant absorption features are detected at the tangent
point velocity and at negative velocities in the spectra of both the PN and its background sources, 
hinting that they are all beyond the solar circle ($d$= 5.3$\pm$0.7\,{\rm kpc}).
According to the noise level~(3\,$\sigma$=0.23), the \HI~emission and absorption features
at $-$64\,km s$^{-1}$ reveal a lower limit distance of 11.1$\pm$0.6\,{\rm kpc} for the PN.
The presence of \HI~ emission at -73\,km\,s$^{-1}$ and the absence of an absorption feature at
the same velocity indicate an upper limit distance of 12.0$\pm$0.7\,{\rm kpc}.
In summary, PN G069.7$-$00.0 has distance between 11.1$\pm$0.6\,{\rm kpc} and 12.0$\pm$0.7\,{\rm kpc}, which is much larger than previous distances, e.g., 3.31\,{\rm kpc} by \citep{Cahn1992},
3.26\,{\rm kpc} by \cite{Zhang1995}, 4.23\,{\rm kpc} by \cite{Stanghellini2008},
3.09\,{\rm kpc} by \cite{Steene1995} and 2.96\,{\rm kpc} by \cite{Phillips2004}. 
So we prefer to keep an open question for its distance.

\vskip 3mm
PN G070.7$+$01.2~(Fig.~\ref{fig9}):

\vskip 1mm
Based on the observed tangent point velocity of 22\,km s$^{-1}$ in the PN spectrum,
we take a rotational velocity $V_{\rm R}$=230\,km s$^{-1}$
at the tangent point in the direction $l$=70.7$^{\circ}$.
The fact that the \HI~absorption features of the PN and three background sources
appear at the tangent point velocity implies  all sources are beyond the tangent point. 
So we obtain a lower limit distance of 2.5$\pm$1.6\,{\rm kpc} for the PN .
There are clear absorption features at negative velocities in nearby sources spectra,
while no absorption feature is detected at these velocities in the PN spectrum.
This implies the PN is within solar circle (5.0$\pm$0.7\,{\rm kpc}).
So PN G070.7$+$01.2 is between 2.5$\pm$1.6\,{\rm kpc} and 5.0$\pm$0.7\,{\rm kpc}.

\cite{Bally1989} have given a distance of  4.5$\pm$1.0\,{\rm kpc}
using the line width of \CO~emission and angular radius of \CO~cloud.
Therefore, the upper distance of 5.0$\pm$0.7\,{\rm kpc} is suitable for the PN.

\vskip 3mm
PN G084.9$-$03.4~(Fig.~\ref{fig10}):

\vskip 1mm
Fig.~\ref{fig10} shows the tangent point velocity (18\,km s$^{-1}$) towards the PN, 
which is larger than the expected value (0.8\,km s$^{-1}$). 
We obtain the rotation velocity $V_{\rm R}$=237\,km s$^{-1}$
at the tangent point in the direction $l$=84.9$^{\circ}$. 
The \HI~spectra of the PN and two background sources show clear \HI~
absorption features from $\sim$0\,km s$^{-1}$ up to the tangent point~(18\,km s$^{-1}$),
hinting the PN and two background sources are beyond the tangent point.
So we obtain the lower limit distance of $\sim$0.7\,{\rm kpc} for this PN.
The prominent \HI~emission at $-$20\,km s$^{-1}$ appears in the spectra of the PN as well as two nearby background sources, whereas its respective absorption feature is detected only in two background sources. This gives an upper limit distance of 4.3$\pm$0.6\,{\rm kpc} for the PN. 
Hence, the distance of the PN is between $\sim$0.7\,{\rm kpc} and 4.3$\pm$0.6\,{\rm kpc}.

This PN, named NGC\,7027 as the most luminous Galactic PN, has a distance measured by various methods, such as statistical distances~(0.7\,{\rm kpc}, \cite{Maciel1984}; 0.63\,{\rm kpc}, \cite{Steene1995}; 0.64\,{\rm kpc}, \cite{Zhang1995}), reddening \cite[$<$1.15\,{\rm kpc},][]{Navarro2012}, expansion parallax (0.703$\pm$0.095\,{\rm kpc}, \cite{Hajian1993}; 0.98$\pm$1.0\,{\rm kpc}, \cite{Zijlstra2008}; 0.88$\pm$0.15\,{\rm kpc}, \cite{Masson1989a}; 0.68$\pm$0.17\,{\rm kpc}, \cite{Mellema2004}). 
In fact, \cite{Pottasch1982} gave a upper limit distance 4.5\,{\rm kpc} also measured by comparing the \HI~absorption feature at $-$20\,km s$^{-1}$ between the PN and a background source.
So the lower limit distance 0.7\,{\rm kpc} is reliable for PN G084.9$-$03.4.

\vskip 3mm
PN G089.0$+$00.3~(Fig.~\ref{fig11}):

\vskip 1mm
The disagreement between the observed tangent point velocity (10\,km s$^{-1}$) shown in Fig.~\ref{fig11} and the expected value (0.03\,km $s^{-1}$) may be due to random motions of \HI~clouds at the tangent point in the direction $l$=89$^{\circ}$.
Fig.~\ref{fig11} shows \HI~spectra of the PN and background sources.
The absorption features at $-$70\,km s$^{-1}$ and $-$40\,km s$^{-1}$ are likely not real, and two significant absorption features at $-$20\,km s$^{-1}$ and $\sim$0\,km s$^{-1}$ indicate that a reliable lower limit distance for this PN is 3.5$\pm$0.6\,{\rm kpc}. No reliable upper limit distance can be determined for this PN.

The distance of this PN~(also named NGC\,7026)~has been 
investigated previously by using \HI~absorption \cite[2.5$\pm$1.0\,{\rm kpc},][]{Gathier1986a},
statistical method (e.g., 2.35\,{\rm kpc}, \cite{Stanghellini2008}; 2.03\,{\rm kpc}, \cite{Zhang1995}), reddening (e.g., 1.57$\pm$0.65\,{\rm kpc}, \cite{Kaler1985}; 2.3\,{\rm kpc}, \cite{Pottasch1983a}), surface gravity method (e.g., 3.5\,{\rm kpc}, \cite{Zhang1993}; 4.2\,{\rm kpc}, \cite{Cazetta2000}), 
spectroscopic method \cite[1.9\,{\rm kpc},][]{Gruendl2004}. Our result is reasonably consistent with surface gravity distance and larger than others. Actually, \cite{Gathier1986a} found a weak absorption at $\sim$$-$20\,km s$^{-1}$ in the PN spectrum, but they did not take this absorption as clear evidence to constrain its distance. The $-$7\,km s$^{-1}$ absorption feature they chose for the lower limit distance of 2.5$\pm$1.0\,{\rm kpc} is also detected in our data. Since our data has higher resolution and more sensitive than before, the absorption at $-$20\,km s$^{-1}$ detected in our work can be used to determine a more reliable lower limit distance of 3.5$\pm$0.6\,kpc for this PN.

\vskip 3mm
PN G107.8$+$02.3~(Fig.~\ref{fig12}):

\vskip 1mm
The absorption features at positive velocity in both the PN and G108.7$+$02.57 are partly due to a cloud with anomalous motion. Clear absorption features are present at $\sim$$-$30\,km s$^{-1}$, $\sim$$-$60\,km s$^{-1}$ and $\sim$$-$105\,km s$^{-1}$ in the spectrum of background source G108.7$+$02.57, while no absorption features are detected at the velocities in the PN spectrum. This implies the PN is in front of the \HI~cloud at $\sim$$-$30\,km s$^{-1}$, i.e, 2.8$\pm$0.5\,{\rm kpc}. The \CO~emission, \HI~emission and absorption features
at $-$12\,km s$^{-1}$ in the PN spectrum indicate a lower limit distance of 1.2$\pm$0.6\,{\rm kpc}. Therefore, the distance of PN G107.8$+$02.3 is between 1.2$\pm$0.6\,{\rm kpc} and 2.8$\pm$0.5\,{\rm kpc}.

For this PN, also named NGC 7354, 
the distance has been measured by various methods. \cite{Gathier1986a} suggested a \HI~absorption~distance 1.5$\pm$0.5 determined 
by the nearby \HII~regions of the PN. \cite{Giammanco2011} obtained a distance of 1.0$\pm$0.15\,{\rm kpc} based on the distance-extinction relationship in the direction toward the PN. \cite{Zhang1993} gave a distance of 2.1\,{\rm kpc} by the surface gravity method. The revised Shklovshy method suggested distances of 1.27\,{\rm kpc} by \cite{Cahn1992}, 1.3\,{\rm kpc} by \cite{Zhang1995}, 1.23\,{\rm kpc} by \cite{Steene1995}, 1.19\,{\rm kpc} by \cite{Phillips2004}, and 1.70\,{\rm kpc} by \cite{Stanghellini2008}. 
So the lower distance of 1.2$\pm$0.6\,{\rm kpc} seems reasonable for the PN.

\vskip 3mm
PN G138.8$+$02.8~(Fig.~\ref{fig13}):

\vskip 1mm

There are clear \HI~absorption and \CO~emission features at $-$42\,km s$^{-1}$ in the direction of G138.8$+$02.14, while similar absorption is not detected in the PN spectrum. This indicates that the PN is in front of the \HI~cloud at$-$42\,km s$^{-1}$. The PN spectrum reveals one reliable \HI~absorption feature at $-$20\,km s$^{-1}$. Therefore, we obtain a lower limit distance of 1.6$\pm$0.5\,{\rm kpc} and an upper limit distance of 3.8$\pm$0.8\,{\rm kpc.

PN G138.8$+$02.8~(also named IC 289) has statistical distances of 1.43\,{\rm kpc} by \cite{Cahn1992}, 1.68\,{\rm kpc} by \cite{Zhang1995}), 1.45\,{\rm kpc} by \cite{Stanghellini2008}, 1.18\,{\rm kpc} by \cite{Phillips2004}, and 1.48\,{\rm kpc} by \cite{Steene1995}; as well as a reddening distance of 2.71$\pm$0.195\,{\rm kpc} \citep{Kaler1985}. So we suggest a distance of 1.6$\pm$0.5\,{\rm kpc} for the PN.

\vskip 3mm
PN G147.4$-$02.3~(Fig.~\ref{fig14}):

\vskip 1mm

Both \HI~emission and absorption features at $-$35\,km s$^{-1}$  appear in
the spectra of the PN as well as its nearby background sources.
This implies that the PN is behind the \HI~cloud at 3.6$\pm$0.9\,{\rm kpc}.
In addition, the presence of clear \HI~emission and the absence of
absorption at $-$60\,km s$^{-1}$ in the PN spectrum indicates that the PN
is in front of the \HI~cloud at  8.7$\pm$1.7\,{\rm kpc}.
Overall, the distance of the PN is between 3.6$\pm$0.9\,{\rm kpc} and 8.7$\pm$1.7\,{\rm kpc}.

The distance of the PN was measured by several methods previously, i.e., surface gravity distance of 2.2$-$3.1\,{\rm kpc} by \cite{Cazetta2001}, the revised Shklovsky distances of 3.39\,{\rm kpc} by \cite{Phillips2004} and 3.53\,{\rm kpc} by \cite{Zhang1995}, as well as a reddening distance of 3.3$\pm$0.35\,{\rm kpc} by \cite{Giammanco2011}. These are consistent with our lower limit distance 3.6$\pm$0.9\,{\rm kpc}.
\vskip 3mm

PN G169.7$-$00.1~(Fig.~\ref{fig15}):
\vskip 1mm
This PN is most likely a \HII~region as suggested by \cite{Zijlstra1990}, which is close to the Galactic plane (see Fig.~\ref{fig15}). The \HI~emission and absorption features at $-$32\,km s$^{-1}$ in spectra of the PN and the background sources imply they are beyond the \HI~cloud at $\sim$14.7\,{\rm kpc}, see Fig.~\ref{fig1} (upper-right panel). This lower limit distance is derived by considering the  average random velocity 6\,km s$^{-1}$ of \HI~clouds. No upper limit distance can be derived. Our lower limit is much larger than previous measurements, i.e., 1.37\,{\rm kpc} by \cite{Cahn1992}, 1.86\,{\rm kpc} by \cite{Zhang1995}, 1.26\,{\rm kpc} by \cite{Phillips2004}, 1.39\,{\rm kpc} by \cite{Stanghellini2008}. So we prefer to keep an open question for its distance. 

\vskip 3mm
PN G259.1$+$00.9~(Fig.~\ref{fig16}):

\vskip 1mm
There is strong continuous \HI~emission and absorption between 0 to 12\,km s$^{-1}$ in the PN spectrum (Fig.~\ref{fig16}), but no absorption features appear up to the tangent point velocity after 12\,km s$^{-1}$ (3$\,\sigma$=0.27).
This means that the PN is beyond the near side for 12\,km s$^{-1}$, i.e., 1.6$\pm$0.6\,{\rm kpc}. Unlike the PN spectrum, the two background sources show absorption features at velocities of 20\,km s$^{-1}$ and 40\,km s$^{-1}$.
This implies that the PN is in front of the \HI~at 20\,km s$^{-1}$,i.e., 2.4$\pm$0.6\,{\rm kpc}. So the distance of the PN is between 1.6$\pm$0.6\,{\rm kpc} and 2.4$\pm$0.6\,{\rm kpc}. In comparison with previous work obtained by statistical method (0.9\,{\rm kpc}, \cite{Cahn1992}; 1.07\,{\rm kpc}, \cite{Zhang1995}) and Spectroscopic distance \cite[0.7\,{\rm kpc},][]{Jones2014}. We suggest a distance of 1.6$\pm$0.6\,{\rm kpc} for this PN.

\vskip 3mm

PN G333.9$+$00.6~(Fig.~\ref{fig17}):

\vskip 1mm
Fig.~\ref{fig17} shows this PN has a poor absorption spectrum.
One possible absorption feature at $-$26\,km s$^{-1}$
can be used to provide a lower limit distance 1.9$\pm$0.4\,{\rm kpc}.
No upper limit distance can be derived for the PN.
In fact, the previous spectroscopic distance (1.0$-$1.5\,{\rm kpc})
obtained by \cite{Morgan2003} agrees with our result.
This PN likely has a distance of 1.9$\pm$0.4\,{\rm kpc}.

\vskip 3mm
PN G352.6$+$00.1~(Fig.~\ref{fig18}):
\vskip 1mm

Two background sources (G352.5$-$0.17, G351.6$+$0.17) show reliable absorption features at the tangent point velocity, while this does not appear in the PN spectrum. This implies that the PN is in front of the tangent point.
There appear \HI~emission and absorption at $-$20\,km s$^{-1}$ in the spectra of the PN and \HII~region G353.4$-$00.3. In fact the PN is behind the \HII~region G353.4$-$00.3 (3.2$\pm$0.8\,{\rm kpc}, \cite{TianT2008}).
In addition, an absorption feature at $-$50\,km s$^{-1}$ is seen in the spectrum of G351.6$+$0.17, while \HI~emission is seen but no absorption at the same velocity in the PN spectrum. This implies the PN is in front of near distance of $-$50\,km s$^{-1}$, i.e., 5.0$\pm$0.3\,{\rm kpc}. So PN G352.6$+$00.1 is between 3.2$\pm$0.8\,{\rm kpc} and 5.0$\pm$0.3\,{\rm kpc}.

\cite{Maciel1984} gave a lower limit 0.9\,{\rm kpc} for this PN by
assuming a relationship between the nebular ionized mass and radius. \cite{Zhang1995} and \cite{Steene1995} suggested distances of 1.40\,{\rm kpc} and 1.37\,{\rm kpc} based on the revised relation between radio surface brightness temperature and nebula radius. Therefore, we adopt 3.2$\pm$0.8\,{\rm kpc} for the PN.

\vskip 3mm
PN G352.8$-$00.2~(Fig.~\ref{fig19}):

\vskip 1mm
The weak absorption features at the tangent velocity,  $-$36\,km s$^{-1}$, and $-$90\,km s$^{-1}$ in the \HI~spectrum of PN G352.8$-$00.2 have a S/N of about  3\,$\sigma$ (0.24), so we do not regard it as real absorption.  The absorption feature at $-$20\,km s$^{-1}$ is detected in the spectra of three background sources, but  does not appear in the PN spectrum, implying the PN is in front of \HI~\citep[3.2$\pm$0.8\,kpc,][]{TianT2008}. In addition, the presence of clear \HI~emission and absence of absorption feature at $-$10\,km s$^{-1}$ in the PN spectrum reveals an lower limit distance of 2.1$\pm$0.9\,{\rm kpc} for the PN. We conclude that the distance of the PN is between 2.1$\pm$0.9\,{\rm kpc} and 3.2$\pm$0.8\,{\rm kpc}. 

The distance of the PN was also suggested
by statistical methods (e.g. 0.8\,{\rm kpc}, \cite{Maciel1984};
1.36\,{\rm kpc}, \cite{Phillips2004}; 1.53\,{\rm kpc}, \cite{Cahn1992};
1.39\,{\rm kpc}, \cite{Zhang1995}; 1.34\,{\rm kpc}, \cite{Steene1995}). 
Hence, we prefer the distance of 2.1$\pm$0.9\,{\rm kpc} for the PN.

\section{SUMMARY}

\begin{table*}[!htp]
\begin{minipage}{185mm}
\centering
\setlength{\extrarowheight}{0.05pt}
\small
 \caption {\label{table2} Final Kinematic Distances of Planetary Nebulae}
 \footnotesize
  \begin{tabular}{llllll}
  \hline
   \hline
 PN   G         &PN       &Distance limits  &Final distance             & Distance by others(Ref.)  &Method        \\
 Name           &Name   &{\rm kpc}        &{\rm kpc}                          &{\rm kpc}            &               \\
 \hline
\multirow{4}*{G020.9$-$01.1} &\multirow{4}*{M 1$-$51}&\multirow{4}*{3.1$\pm$0.3$-$7.1$\pm$0.6}  &\multirow{4}*{3.1$\pm$0.3}   &2.4(C01)        &surface gravity \\
                  &                         &          &                 &1.75(C92),1.66(S95)      &statistical \\
                  &                          &         &                 &1.74(Z95),2.29(S08)     &statistical \\
                  &                           &        &                 &1.59(Ph04)              & statistical \\
                &                           &        &                 & 2.31$\pm$0.75(F16)              &  statistical   \\

\hline
G029.0$+$00.4   &A66 48   &3.5$\pm$0.3$-$6.6$\pm$1.0      &3.5$\pm$0.3 &1.2(M84) &statistical \\  [2pt]
                  &                           &        &                 & 1.41$\pm$0.43(F16)            &  statistical  \\

\hline
G030.2$-$00.1    &-   &10.9$\pm$0.3$-$13.2$\pm$0.5 &10.9$\pm$0.3$-$13.2$\pm$0.5    &- &- \\  [2pt]
\hline
G051.5$+$00.2   & KLW 1 &4.7$\pm$1.4$-$9.5$\pm$0.4    &4.7$\pm$1.4$-$9.5$\pm$0.4     &- &- \\  [2pt]
\hline
G052.1$+$01.0    & -  &4.7$\pm$1.4$-$5.4$\pm$0.7    &4.7$\pm$1.4$-$5.6$\pm$0.8 &- &- \\  [2pt]
\hline
\multirow{3}*{G055.5$-$00.5 } & \multirow{3}*{M 1$-$71}  &\multirow{3}*{4.3$\pm$1.5$-$8.6$\pm$0.4} &\multirow{3}*{4.3$\pm$1.5}   &1.66(C92),3.68(S08)     &statistical       \\
                  &               &             &                     &3.17(Z95),3.17(S95)     & statistical         \\
                                    &                &            &                    & 2.88$\pm$0.91(F16)             &  statistical \\

                  &               &             &                     &2.9$\pm$0.4(G11)          & reddening       \\
\hline
\multirow{3}*{G069.7$-$00.0} &\multirow{3}*{K 3$-$55}    &\multirow{3}*{11.07$\pm$0.61$-$12.02$\pm$0.67}  &\multirow{3}*{?}   &3.31(C92),4.23(S08)     &statistical   \\
                 &                &             &                    &3.26(Z95),3.09(S95)           &statistical      \\
                  &                 &           &                     &2.96(Ph04)                 & statistical     \\
                  &                           &        &                 & 3.54$\pm$1.32(F16)             &  statistical  \\

 \hline
G070.7$+$01.2 &M 3$-$60    &2.5$\pm$1.6$-$5.0$\pm$0.7   &5.0$\pm$0.7        &4.5$\pm$1.0(B89) &\HI~absorption \\
[2pt]  \hline
\multirow{10}*{G084.9$-$03.4}&\multirow{10}*{NGC 7027}     &\multirow{10}*{0.7$-$4.3$\pm$0.6}  &\multirow{10}*{0.7 }     &0.7(M84),0.64(Z95)        &statistical      \\
                  &           &                &                      &0.27(C92),0.63(S95)        & statistical      \\
                  &                           &        &                 & 0.94$\pm$0.27(F16)              & statistical \\
                  &           &                &                      &2.3$-$2.9(C01)            &surface gravity          \\
                  &           &                 &                     &0.703$\pm$0.095(H93)      & expansion parallax      \\
                  &           &                 &                     &0.98$\pm$1.0(Z08)         & expansion parallax     \\
                  &           &                 &                     &0.94$\pm$0.2(M86)         & expansion parallax     \\
                  &           &                 &                     &0.88$\pm$0.15(M89)         & expansion parallax    \\
                  &           &                 &                     &0.68$\pm$0.17(M04)         & expansion parallax \\
                  &           &                 &                     &$<$1.15(N12)              & reddening      \\
                   &          &                 &                     &$<$4.5(P82)              & \HI~absorption      \\
 \hline
\multirow{8}*{G089.0$+$00.3} &\multirow{8}*{NGC 7026}  &\multirow{8}*{$\geq$3.5$\pm$0.6}  &\multirow{8}*{3.5$\pm$0.6}  &2.5$\pm$1.0(G86)            &\HI~absorption  \\
                  &           &             &                         &1.9(C92),2.35(S08)         & statistical     \\
                  &           &             &                         &2.03(Z95),1.91(Ph04)         & statistical     \\
                  &           &             &                         &1.94(S95)                 & statistical    \\
                  &                           &        &               &1.67$\pm$0.48(F16)             & statistical  \\
                  &            &            &                         &1.57$\pm$0.65(K85)        & reddening    \\
                  &           &             &                         &2.3(P83),$<$1.5(G11)        & reddening     \\
                  &           &             &                         &3.5(Z93),4.2(C00)           &surface gravity       \\
                  &           &             &                         &1.9(G04)                  & Spectroscopic parallax\\
\multirow{7}*{G107.8$+$02.3}&  \multirow{7}*{NGC 7354 }   &\multirow{7}*{1.2$\pm$0.6$-$2.8$\pm$0.5}  &\multirow{7}*{1.2$\pm$0.6}  &1.5$\pm$0.5(G86)    &\HI~absorption  \\
                  &          &              &                         &1.27(C92),1.19(Ph04)         & statistical    \\
                  &          &              &                         &1.3(Z95),1.70(S08)         & statistical   \\
                  &          &              &                         &1.23(S95)                 & statistical   \\
                  &                           &        &                 &1.26$\pm$0.37(F16)            & statistical  \\
                  &          &              &                         &1.0$\pm$0.15(G11)          & reddening     \\
                  &          &              &                         &3.43$\pm$0.62(K85)        & reddening   \\
                  &          &              &                        &2.1(Z93),3.2(C00)          &surface gravity     \\
\hline
\multirow{4}*{G138.8$+$02.8}&\multirow{4}*{IC 289}     &\multirow{4}*{1.6$\pm$0.6$-$3.8$\pm$0.6}   &\multirow{4}*{1.6$\pm$0.6}  &2.71$\pm$0.195(K85)       & reddening     \\
                  &           &             &                         &1.43(C92),1.45(S08)        & statistical    \\
                  &           &             &                         &1.68(Z95),1.18(Ph04)         & statistical   \\
                  &           &             &                         &1.48(S95)                 & statistical   \\
                  &                           &        &                 &1.88$\pm$0.58(F16)             & statistical \\
\hline
\multirow{5}*{G147.4$-$02.3}   &\multirow{5}*{M 1$-$4}  &\multirow{5}*{3.6$\pm$0.9$-$8.7$\pm$1.7} &\multirow{5}*{3.6$\pm$0.9}  &2.2$-$3.1(C01)             &surface gravity  \\
                  &             &           &                         &2.99(C92),3.36(S95)         & statistical  \\
                  &            &            &                        &3.53(Z95),3.39(Ph04)         & statistical   \\
                  &            &            &                       &6.6(S08)                  & statistical    \\
                  &                           &        &             &5.18$\pm$1.55(F16)              & statistical \\
                  &             &           &                         &3.3$\pm$0.35(G11)          & reddening     \\
\hline
\multirow{2}*{G169.7$-$00.1}   &\multirow{2}*{IC 2120}  &\multirow{2}*{$\geq$14.7} &\multirow{2}*{? }                 &1.37(C92),1.388(S08)  &statistical     \\
                          &        &        &                                 &1.86(Z95),1.26(Ph04)         & statistical     \\
 \hline
\multirow{2}*{G259.1$+$00.9} &\multirow{2}*{Hen 2$-$11}   &\multirow{2}*{1.6$\pm$0.6$-$2.4$\pm$0.6}  &\multirow{2}*{1.6$\pm$0.6}  &0.9(C92),1.07(Z95)         &statistical     \\
                  &                           &        &                 &0.80$\pm$0.24(F16)             & statistical  \\
                  &                &        &                        &0.7(J14)                 & Spectroscopic    \\
                   \hline
G333.9$+$00.6  & PMR 5   &$\geq$1.9$\pm$0.4   &1.9$\pm$0.4   &1.0$-$1.5(M03)           & Spectroscopic   \\
[2pt] \hline
\multirow{2}*{G352.6$+$00.1}& \multirow{2}*{H 1$-$12}   &\multirow{2}*{3.2$\pm$0.8$-$5.0$\pm$0.3}      &\multirow{2}*{3.2$\pm$0.8}   &$>$0.9(M84)               &statistical    \\
                  &               &         &                         &1.40(Z95),1.37(S95)        & statistical    \\
                                    &                           &        &                 &2.42$\pm$0.84(F16)               & statistical \\
 \hline
\multirow{3}*{G352.8$-$00.2} &\multirow{3}*{H 1$-$13}  &\multirow{3}*{2.1$\pm$0.9$-$3.2$\pm$0.8}  &\multirow{3}*{2.1$\pm$0.9 } &0.8(M84),1.36(Ph04)         &statistical    \\
                  &               &         &                         &1.53(C92),1.39(Z95)         & statistical   \\
                  &                &         &                        &1.34(S95)                & statistical   \\
                 &                           &        &                 & 1.74$\pm$0.59(F16)           & statistical \\
\hline
\end{tabular}
\begin{tabular}{llll}
Ref.:~C01- \cite{Cazetta2001},~C92-\cite{Cahn1992},~S95-\cite{Steene1995},~Z95-\cite{Zhang1995}\\
~~~~~~~S08-\cite{Stanghellini2008},~B89-\cite{Bally1989},~N12-\cite{Navarro2012},~G11-\cite{Giammanco2011}~\\
~~~~~~~K85-\cite{Kaler1985},~G86-\cite{Gathier1986a},~C00-\cite{Cazetta2000},~G04-\cite{Gruendl2004}\\
~~~~~~~P83-\cite{Pottasch1983a},~Z08-\cite{Zijlstra2008},~H93-\cite{Hajian1993},~M84-\cite{Maciel1984},~Z93-\cite{Zhang1993}\\
~~~~~~~Ph04-\cite{Phillips2004},~M04-\cite{Mellema2004},~M86-\cite{Masson1986},~M03-\cite{Morgan2003},~M89-\cite{Masson1989a}\\
~~~~~~~J14-\cite{Jones2014}, ~P82-\cite{Pottasch1982},~F16-\cite{Frew2016}\\
~~~~~~~$?$--represent that we keep an open question for its distance.\\
\end{tabular}
\end{minipage}
\end{table*}

We analyze the \HI~absorption spectra of 18 Galactic plane PNe  and estimate their kinematic distances in this paper. 
The final results are shown in Table~\ref{table2}. 
We compare new kinematic distances  of 15 PNe with the previous results determined from other methods, such as ~surface gravity~\citep[5 PNe, e.g.][]{Zhang1993,Cazetta2000},~expansion parallax~\citep[1 PN, e.g.][]{Hajian1993,Zijlstra2008}, reddening~\citep[5 PNe, e.g.][]{Kaler1985,Giammanco2011}, statistical~\citep[13 PNe, e.g.][]{Cahn1992,Zhang1995}, \HI~absorption~\citep[4 PNe, e.g.][]{Bally1989,Gathier1986a}.
By considering the additional distance information, we determine distances for 13 PNe with uncertainties ranging from 10$\%$ to 50$\%$.  For 8 out of 13 PNe, the kinematic distances are determined with uncertainties less than $25\%$. For three objects~(PN G020.9$-$01.1, PN G029.0$+$00.4, PN G084.9$-$03.4)~ the kinematic distances are derived with uncertainties
less than 10$\%$, and for the other ten cases the kinematic distances are estimated with uncertainties less than 50$\%$. 
This is a significant improvement compared against most of the previous measurements with uncertainties of two or three factor smaller.
For three cases the kinematic distance are derived with lower and upper distance limits (see Table~\ref{table2}).  We do not suggest  distances for PN G069.7$-$00.0 and PN G169.7$-$00.1 based on our current \HI~measurements only.  
For PN candidate PN G030.2$-$00.1, which was discussed by \cite{Anderson2011}, its luminosity is $\sim$ 225 times stronger than the most luminous Galactic PN NGC\,7027 (506\,mJy at $\sim$8.6\,GHz, see \citealt{Zijlstra2008}), so G030.2$-$00.1 might not be a PN.

In addition, our spectra have revealed that five of the PNe show larger tangent point velocities than expected from the rotation curve model when adopting IAU value of $V_{0}$=220\,km s$^{-1}$. Three of them are located near 53$^{\circ}$.
This is consistent with the previous studies in \cite{Levine2008} and \cite{Leahy2008a}.

We compare our distances with the previous measurements based on the data in table 2 (also see Fig.~\ref{fig20}). We find that all other work (except statistical results, see Fig.~\ref{fig20}, left panel) show obvious dispersion between each other, and that our distance measurements are well consistent with the most reliable method, i.e. expansion parallax.  Our distance measurements are consistent with \cite{Frew2016} when considering the uncertainties, but larger than the other statistical results (see Fig.~\ref{fig20}, middle and right panels). 
In order to investigate the possible effects of some PNe parameters on our measurements, we have tried to find the correlation between these parameters (e.g., radius, reddening) and the residuals which obtained by subtracting our distances from those obtained by other methods. No obvious correlation is found.  A total number of 22 PNe have \HI~absorption measurements in the literatures.
Our work significantly increases the number of Galactic PNe with \HI~absorption measurements and known kinematic distances.

\begin{acknowledgements}

\vskip 3mm
We acknowledge support from the NSFC~(11473038, 11273025, 1126114046) and also support from China's Ministry of Science and Technology under the State Key Development Program for Basic Research~(2012CB821800, 2013CB837901). D. A. Leahy is supported by a grant from the Natural Sciences and Engineering Research Council of Canada. The work has made used of data from the International Galactic Plane Survey (http://www.ras.ucalgary.ca/IGPS/). 

\end{acknowledgements}

\vskip 3mm
\bibliographystyle{apj}
\bibliography{ref}
\newpage
\begin{figure*}
 \centering
 \begin{tabular}{cc}
    \includegraphics[width = 0.27\textwidth,height = 0.215\textwidth]{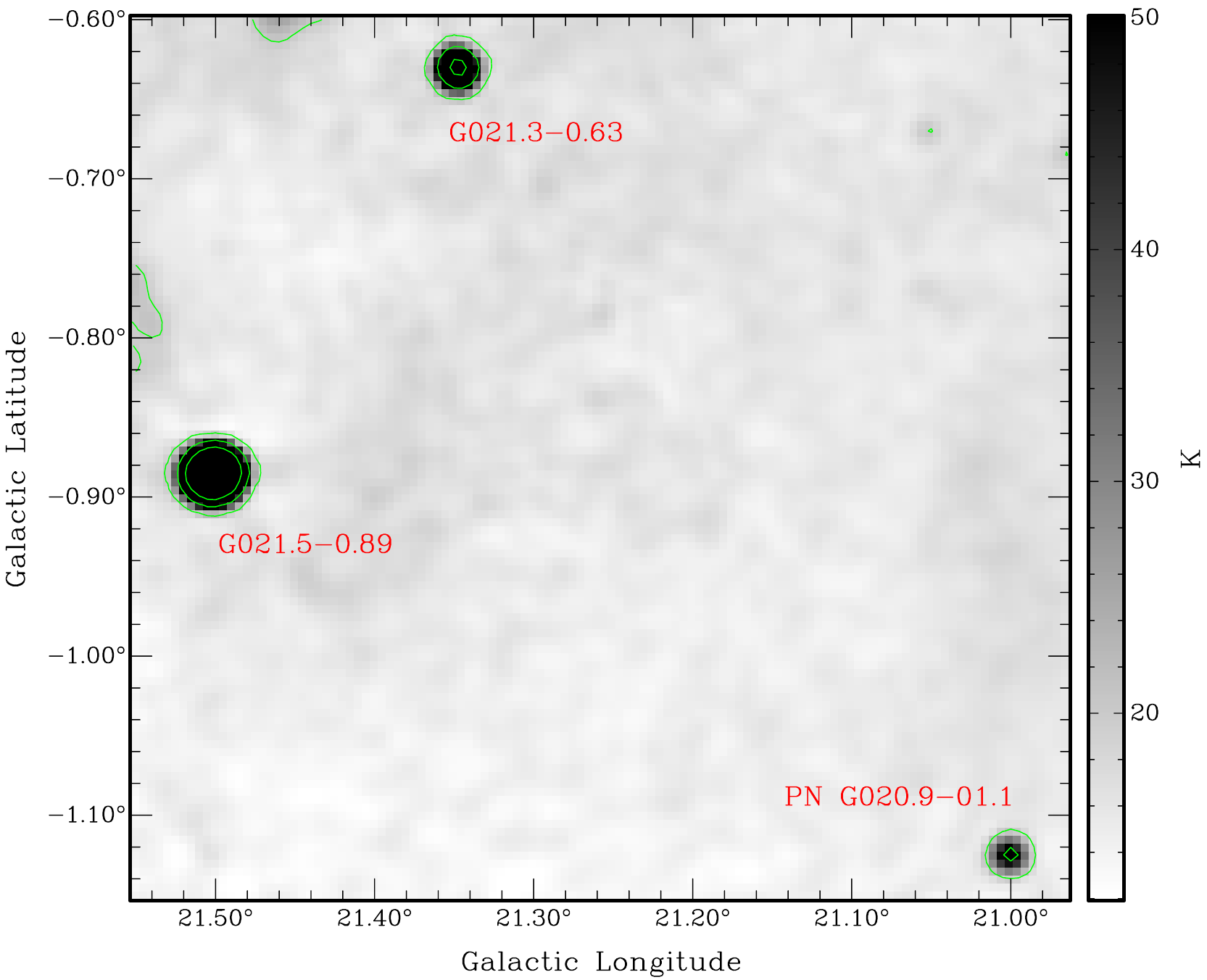}&
    \includegraphics[width = 0.3\textwidth]{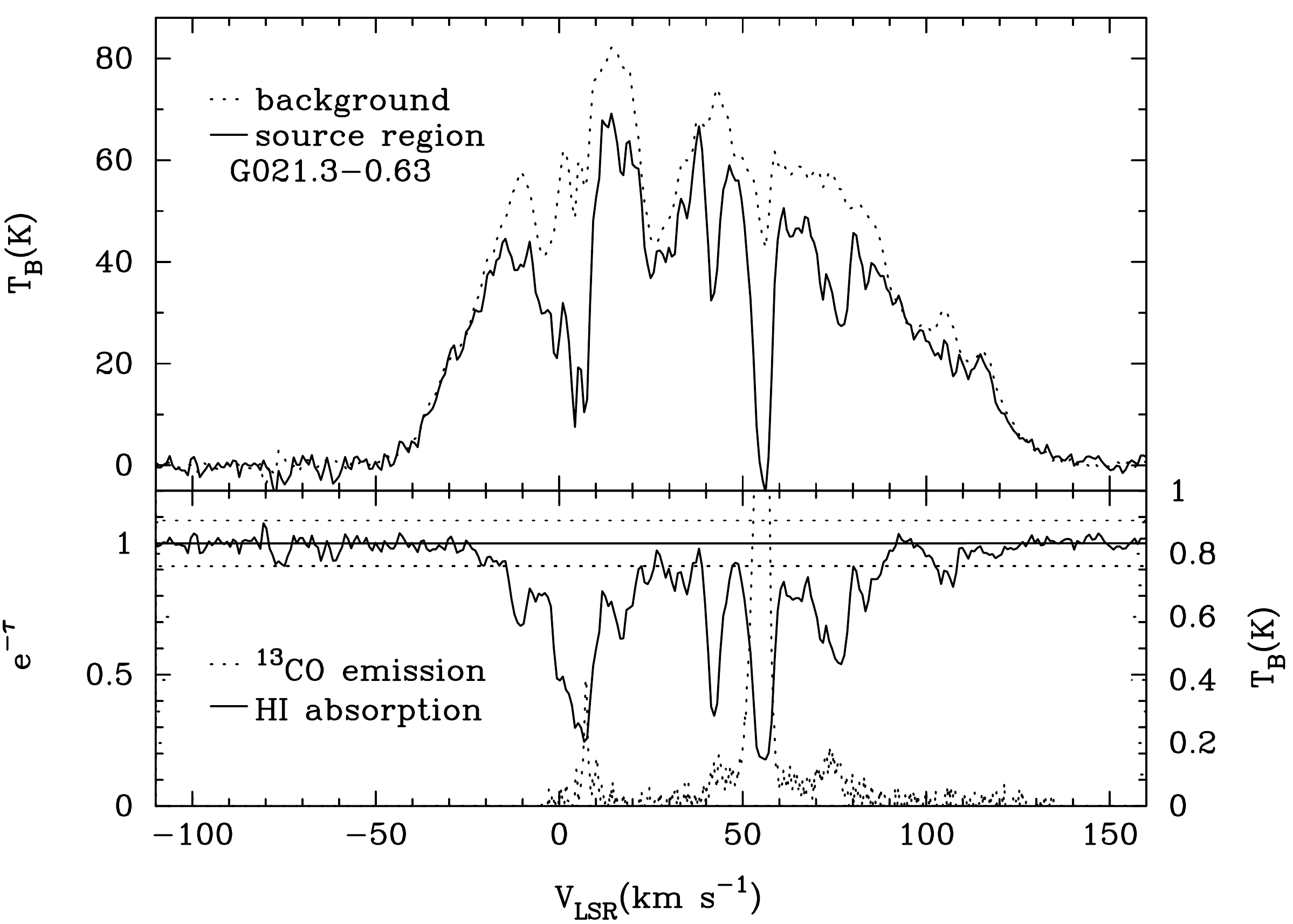}\\
    \includegraphics[width = 0.27\textwidth]{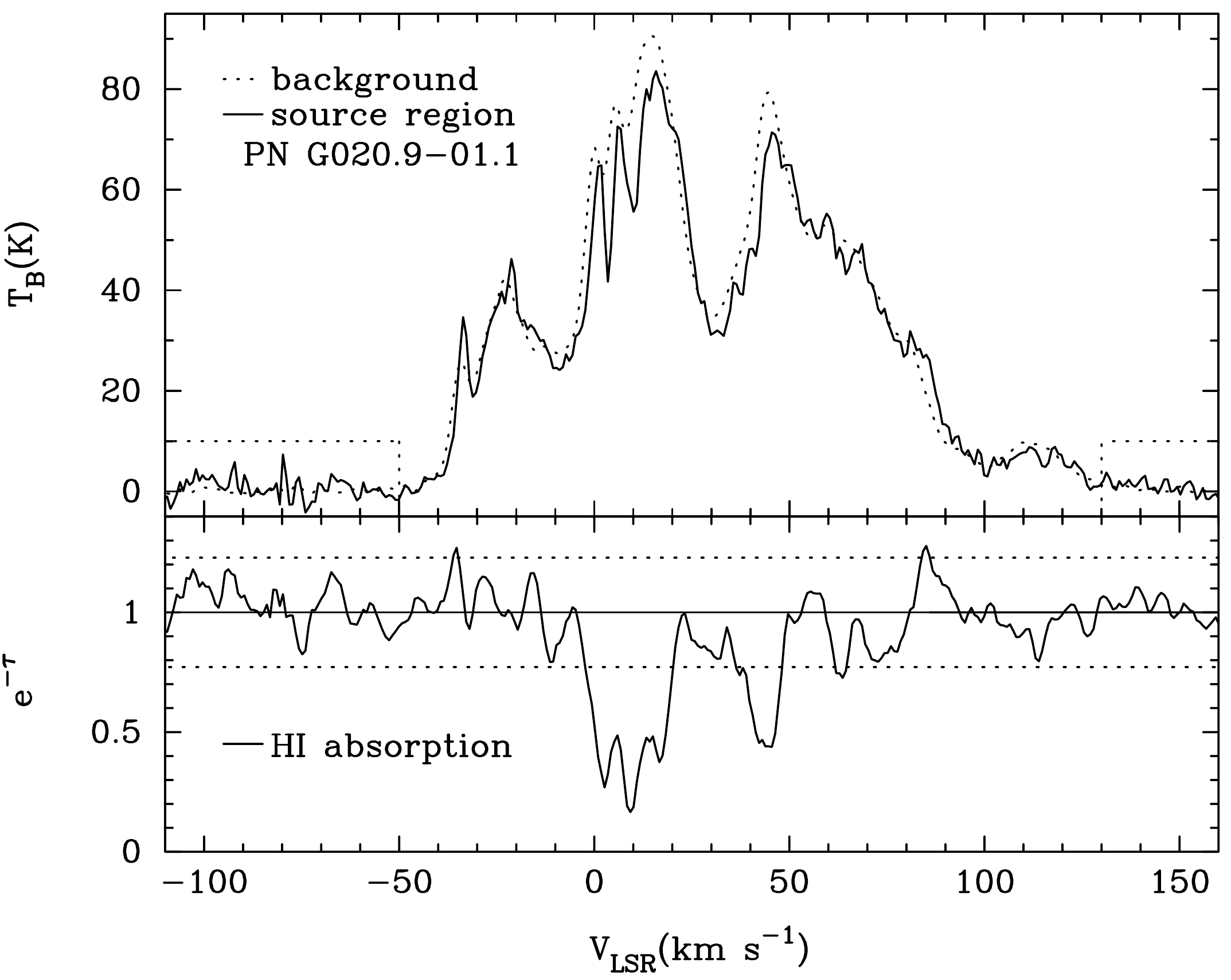}&
    \includegraphics[width = 0.3\textwidth]{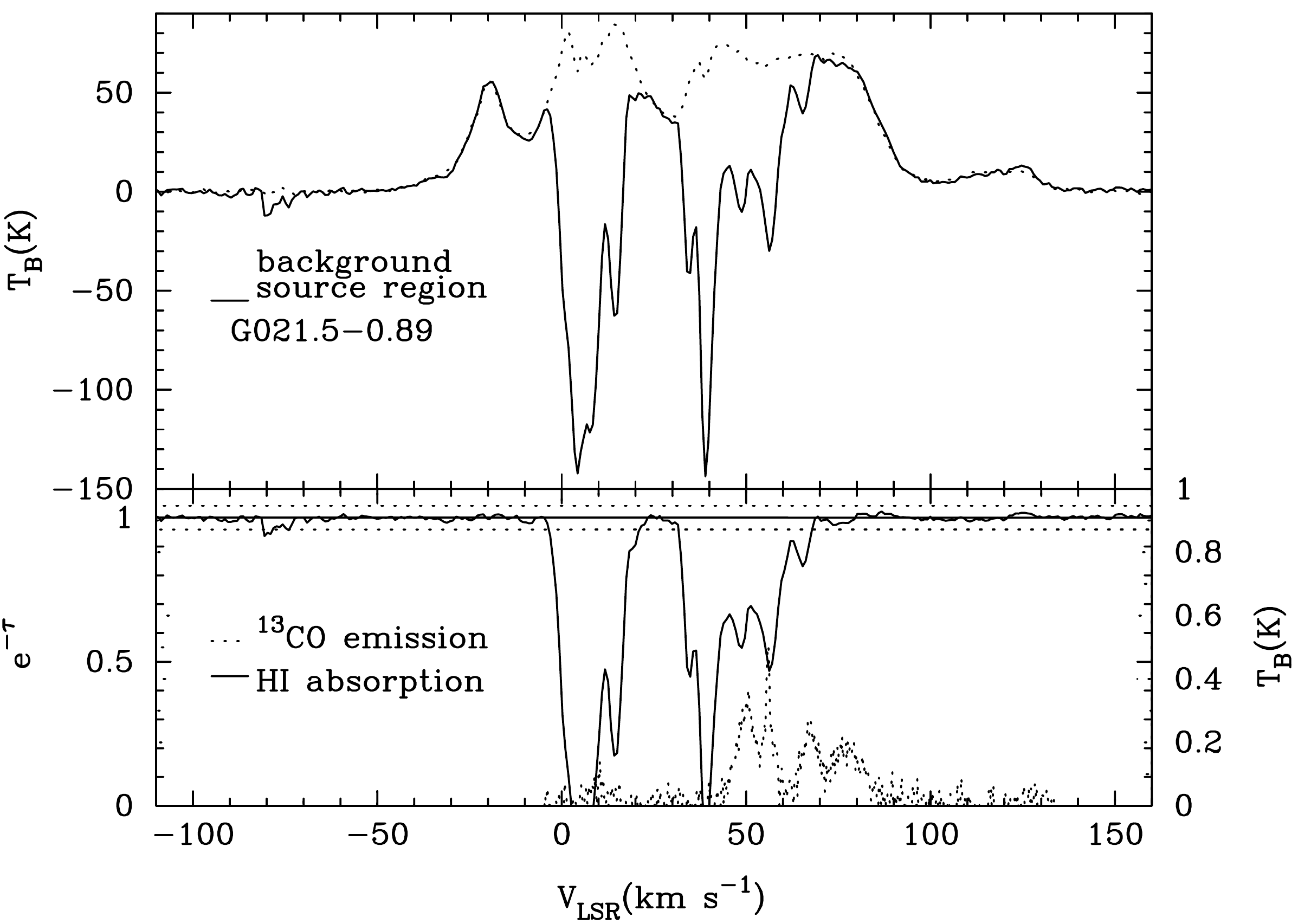}\\
   \end{tabular}
 \caption{1420\,MHz continuum image of PN G020.9$-$01.1 and its background sources (top left),
 and the \HI~spectra of PN G020.9$-$01.1 (bottom left), G021.3$-$0.63 (top right) and G021.5$-$0.89 (bottom right).
 The map has superimposed contours (20,50,130\,K) of 1420\,MHz continuum emission. 
  The dotted horizontal lines in the lower panel of the PN spectrum show the 3$\sigma$ noise level. 
 This description applies for all the spectra of PNe and background sources from Fig.~\ref{fig2} to Fig.~\ref{fig19}.}
 \label{fig2}
\end{figure*}

\begin{figure*}
 \centering
 \begin{tabular}{cc}
    \includegraphics[width = 0.29\textwidth,height = 0.215\textwidth]{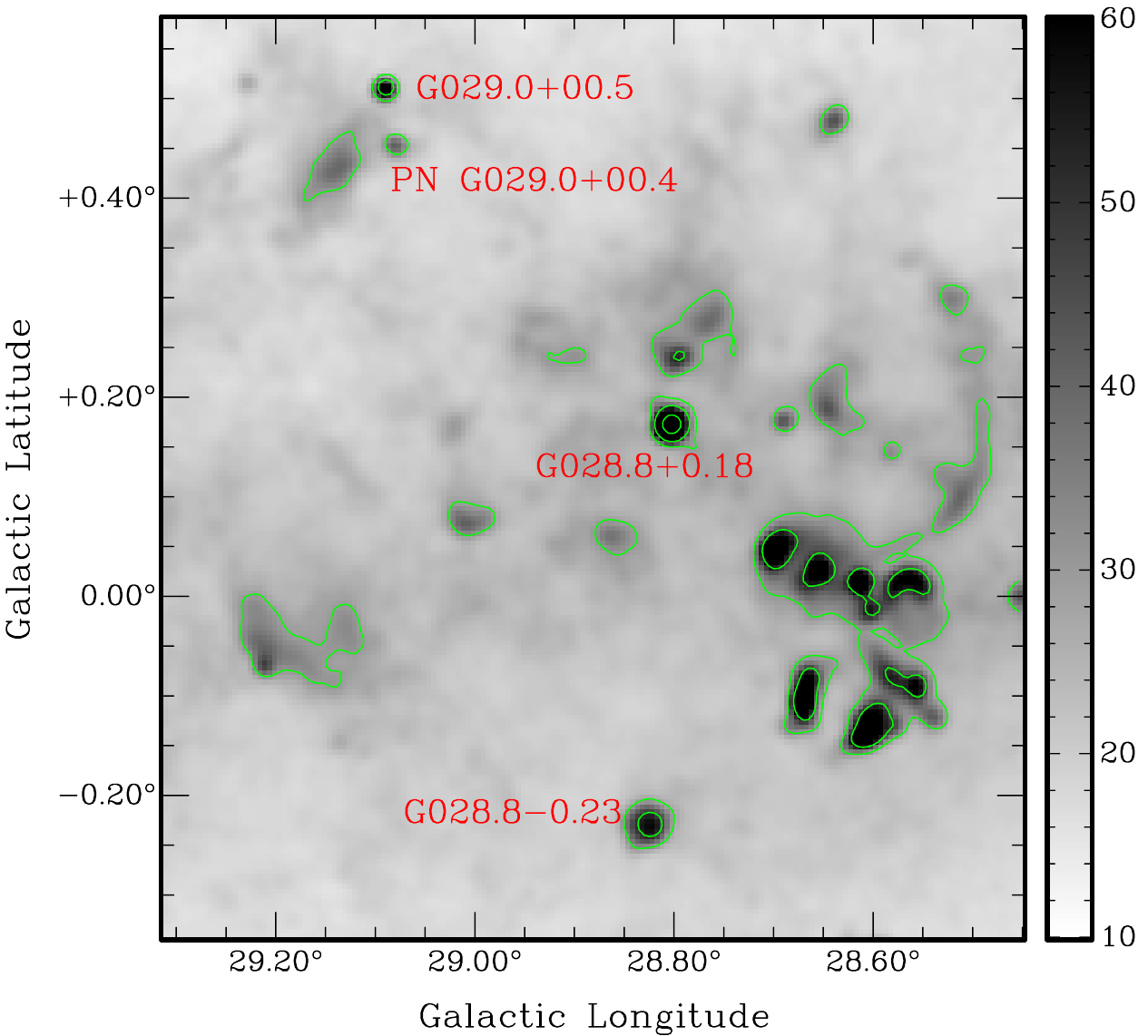}&
    \includegraphics[width = 0.3\textwidth]{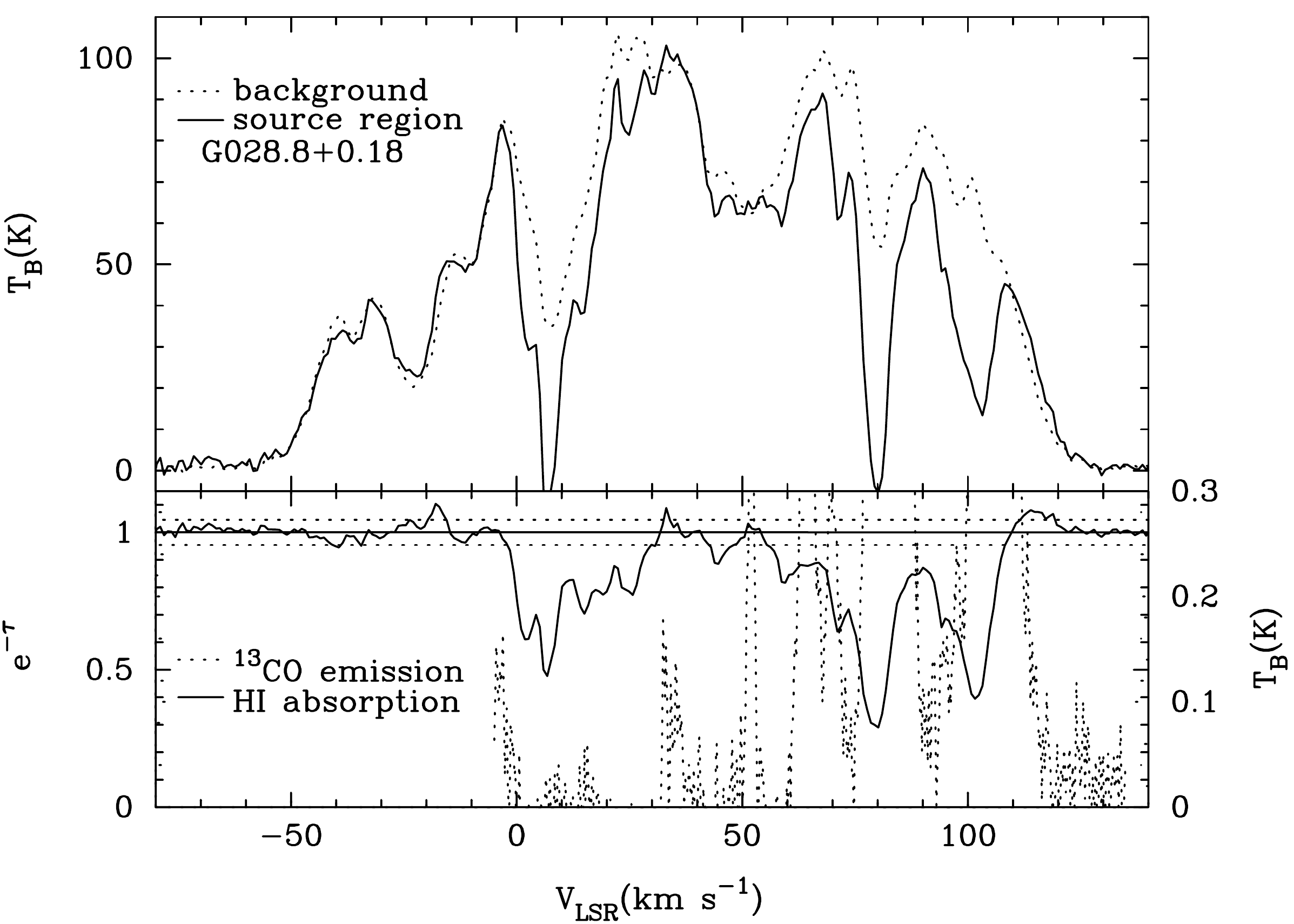}\\
    \includegraphics[width = 0.3\textwidth]{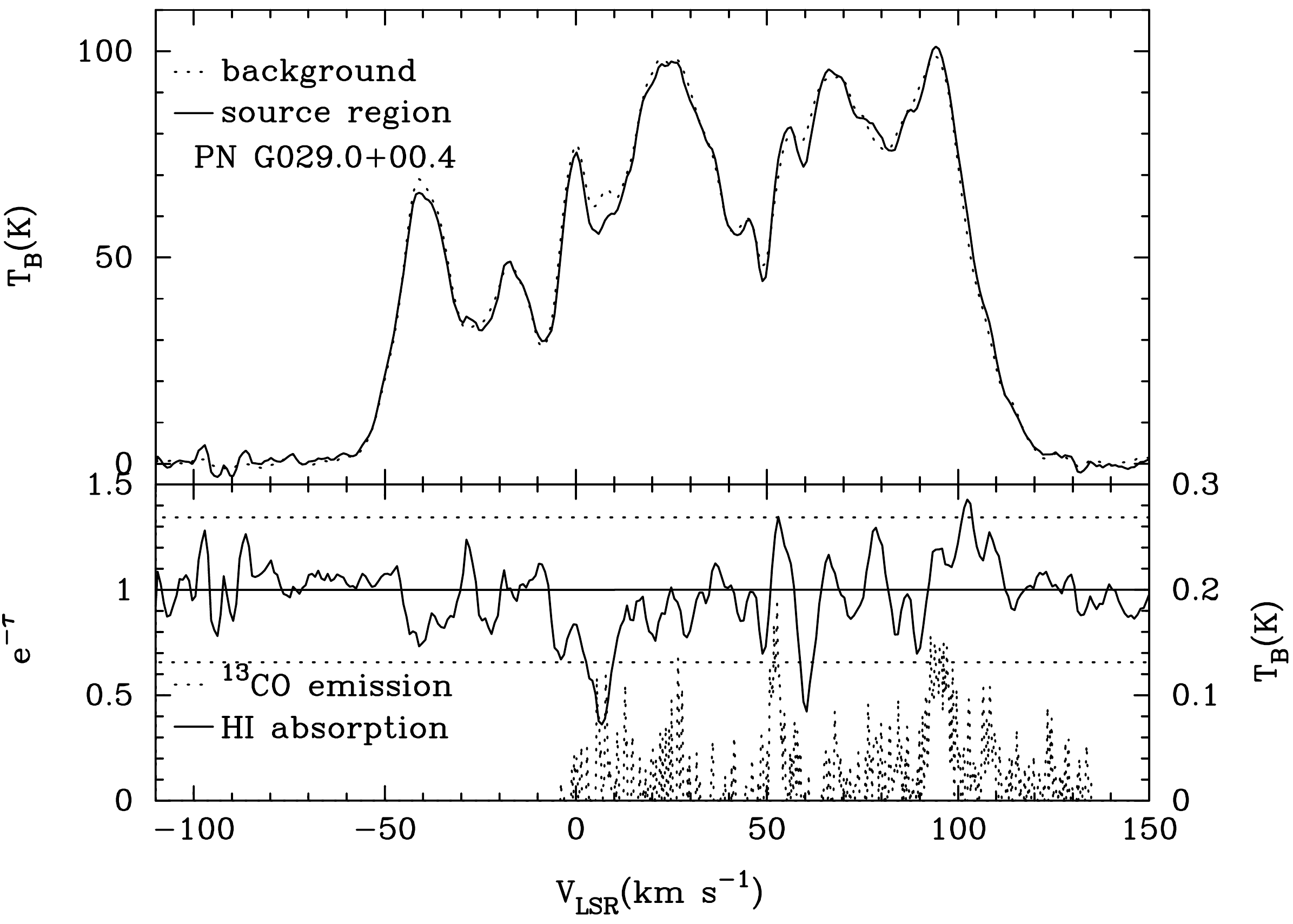}&
    \includegraphics[width = 0.3\textwidth]{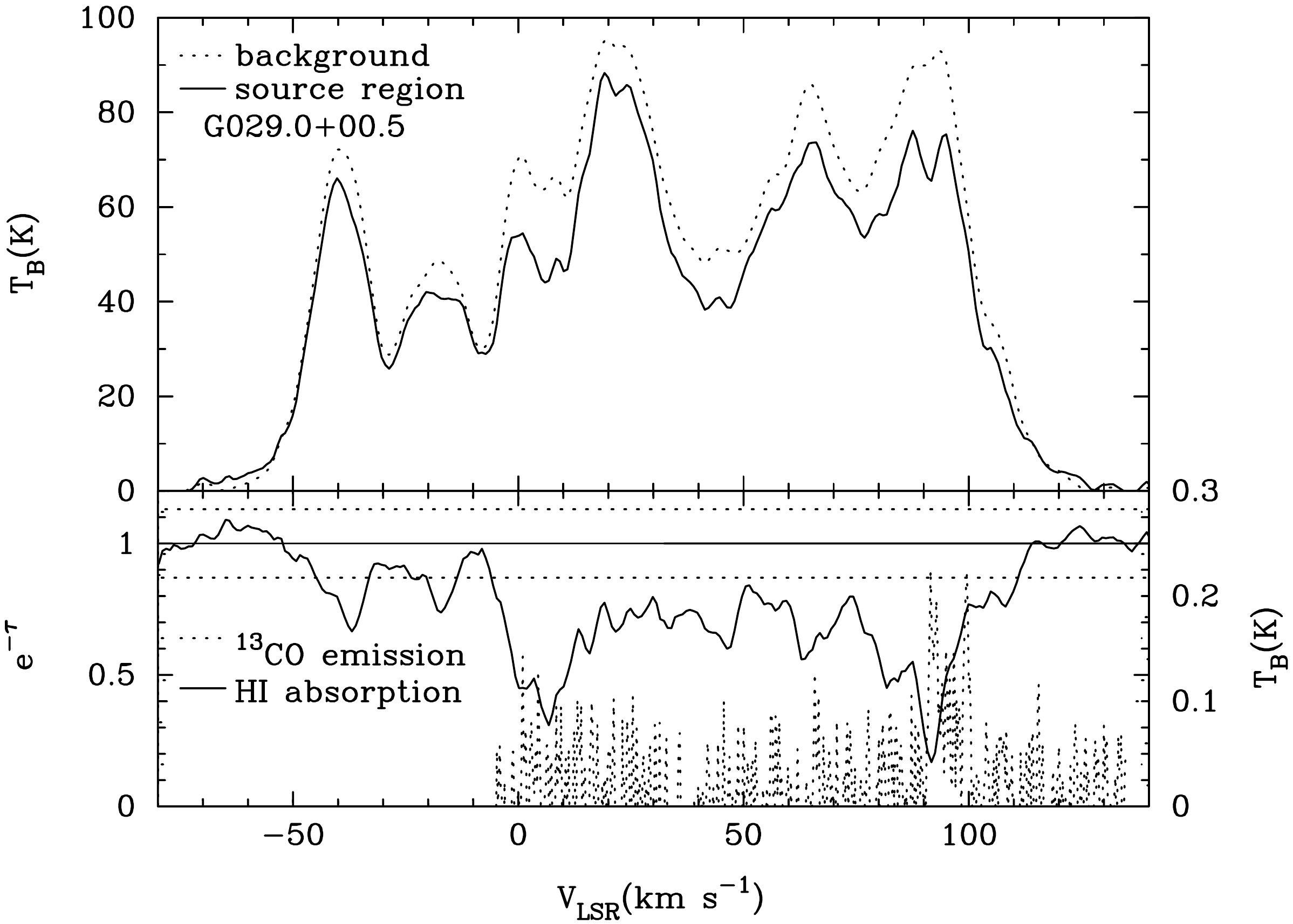}\\
   \end{tabular}
 \caption{1420\,MHz continuum image of PN G029.0$+$00.4, and its background sources (top left),
 and the \HI~spectra of PN G029.0$+$00.4 (bottom left), G028.8$+$0.18 (top right), and G029.0$+$00.5 (bottom right). 
  The \HI~spectra of G028.8$-$0.23 is listed in appendix Fig.~\ref{fig21}.
 The map has superimposed contours (30,50,120\,K) of 1420\,MHz continuum emission.}
 \label{fig3}
\end{figure*}

\begin{figure*}
 \centering
 \begin{tabular}{cc}
    \includegraphics[width = 0.29\textwidth]{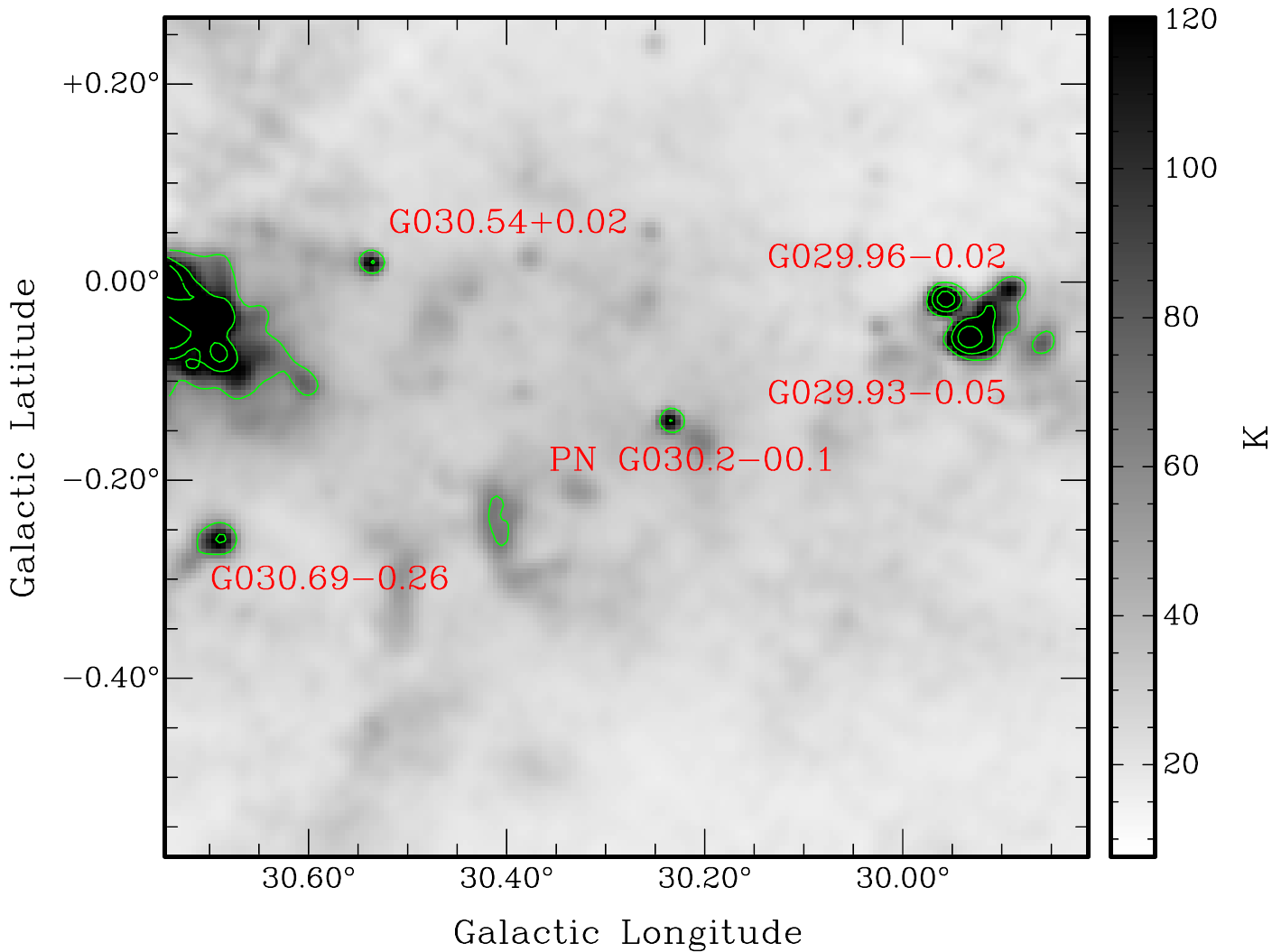}&
    \includegraphics[width = 0.3\textwidth]{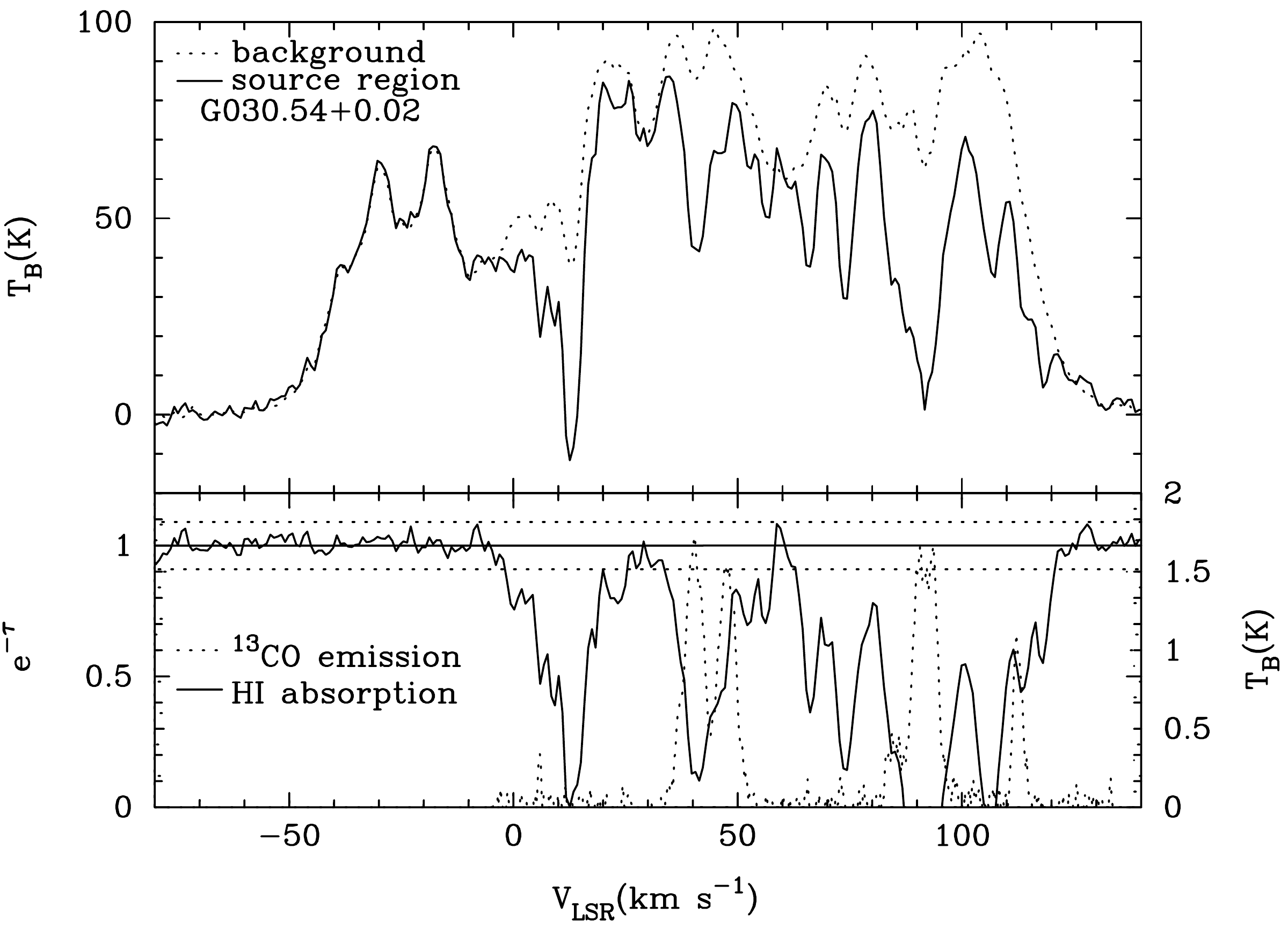}\\
    \includegraphics[width = 0.3\textwidth]{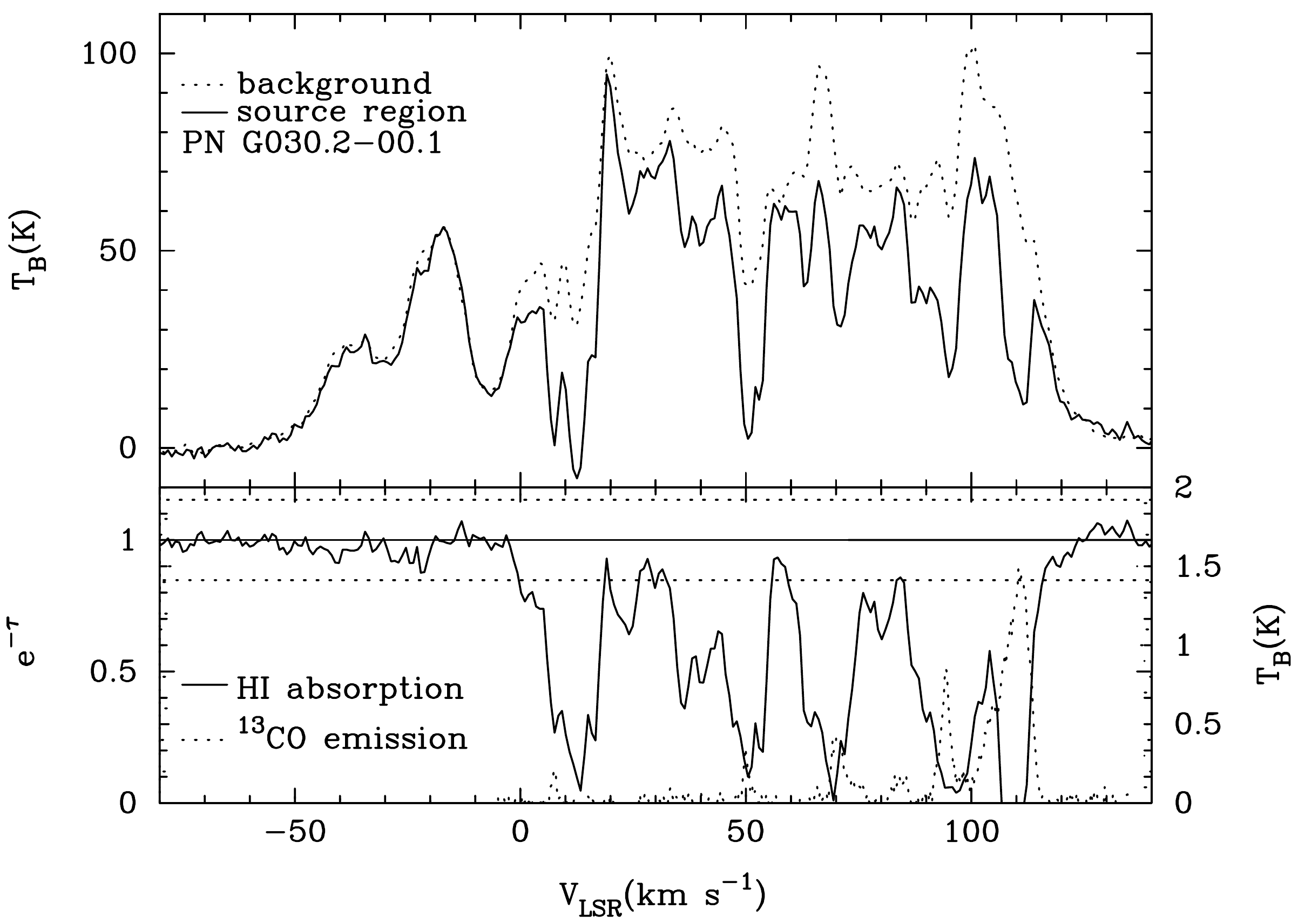}&
    \includegraphics[width = 0.3\textwidth]{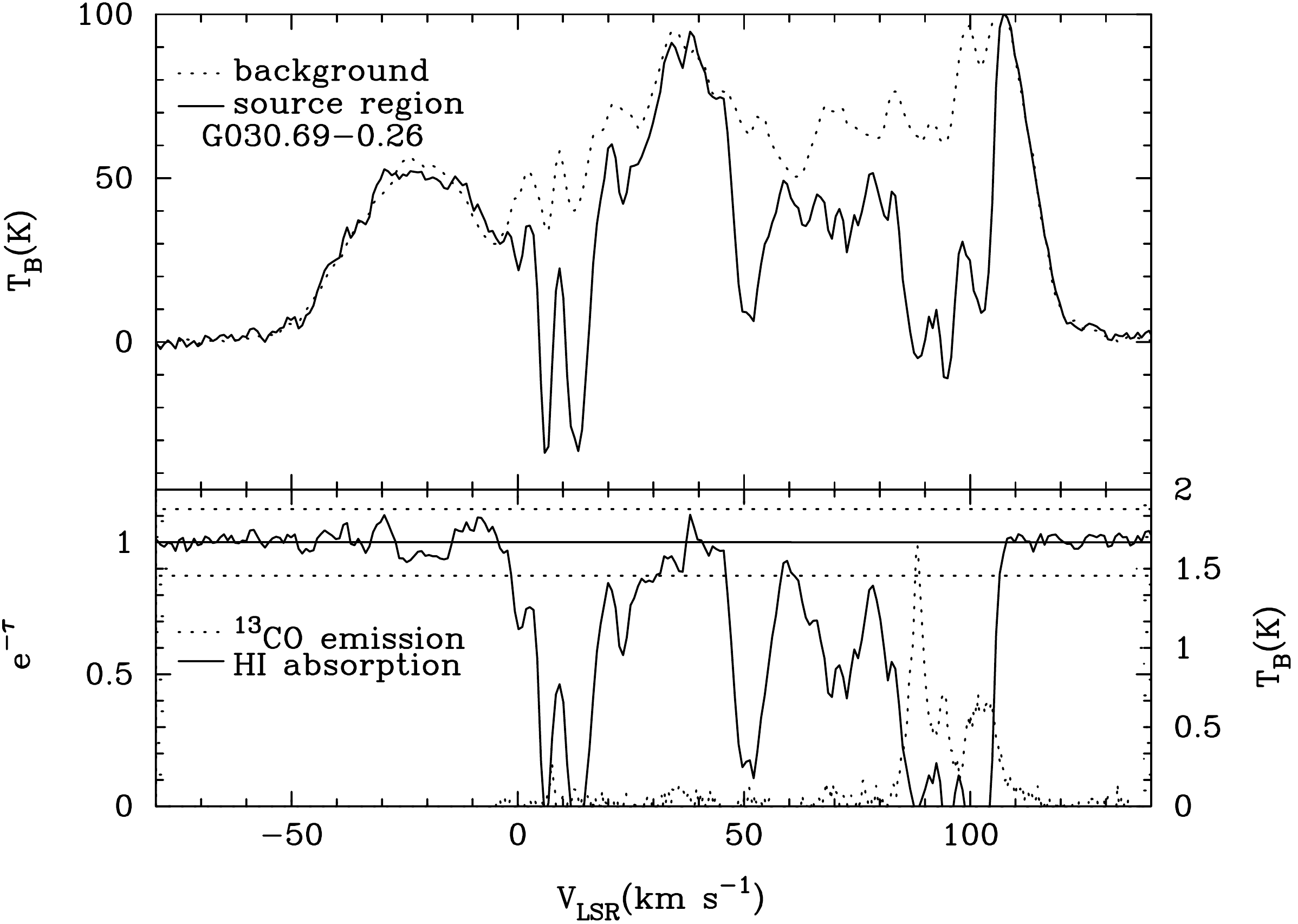}\\
       \end{tabular}
 \caption{1420\,MHz continuum image of PN G030.2$-$00.1 and its nearby background source (top left),
 and the \HI~spectra of PN G030.2$-$00.1 (bottom left),  G030.54$+$0.02 (top right), and G030.69$-$0.26 (bottom right), The \HI~spectra of G029.93$-$0.05 and G029.96$-$0.02 are listed in appendix Fig.~\ref{fig21}.
 The map has superimposed contours (65,125,230\,K) of 1420\,MHz continuum emission.}
 \label{fig4}
\end{figure*}
\begin{figure*}
 \centering
 \begin{tabular}{cc}
    \includegraphics[width = 0.29\textwidth,height = 0.215\textwidth]{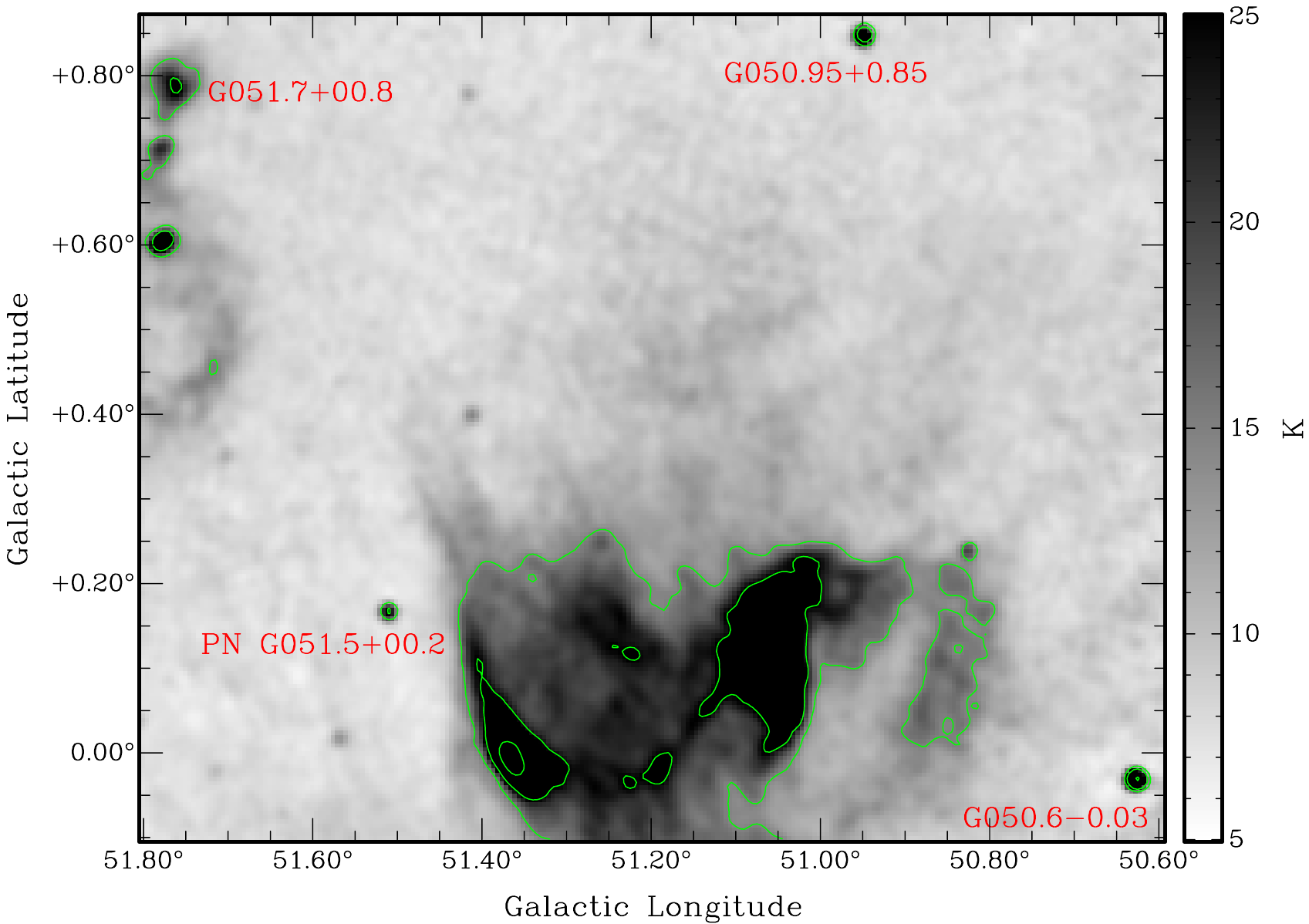}&
    \includegraphics[width = 0.3\textwidth]{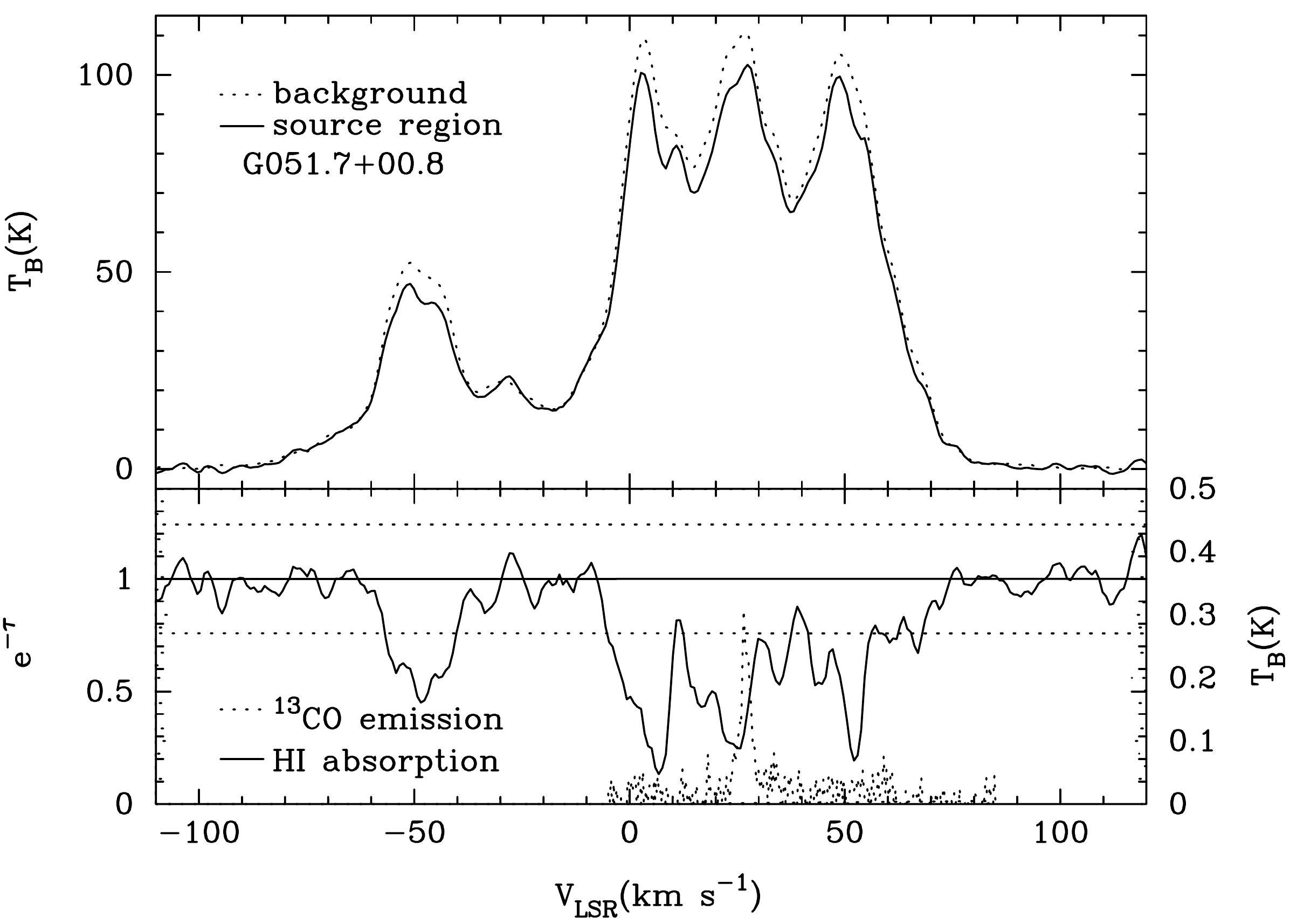}\\
    \includegraphics[width = 0.3\textwidth]{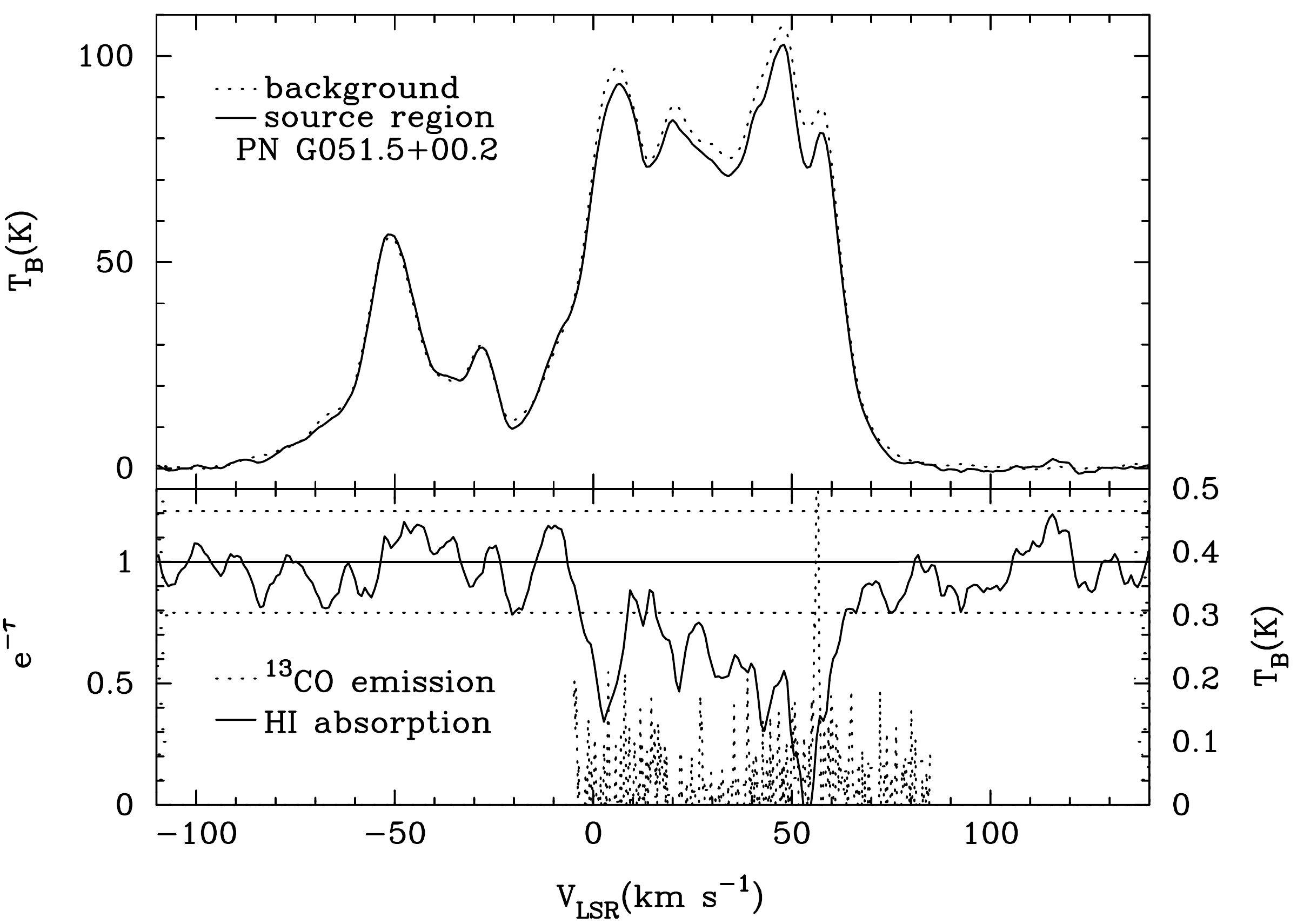}&
    \includegraphics[width = 0.3\textwidth]{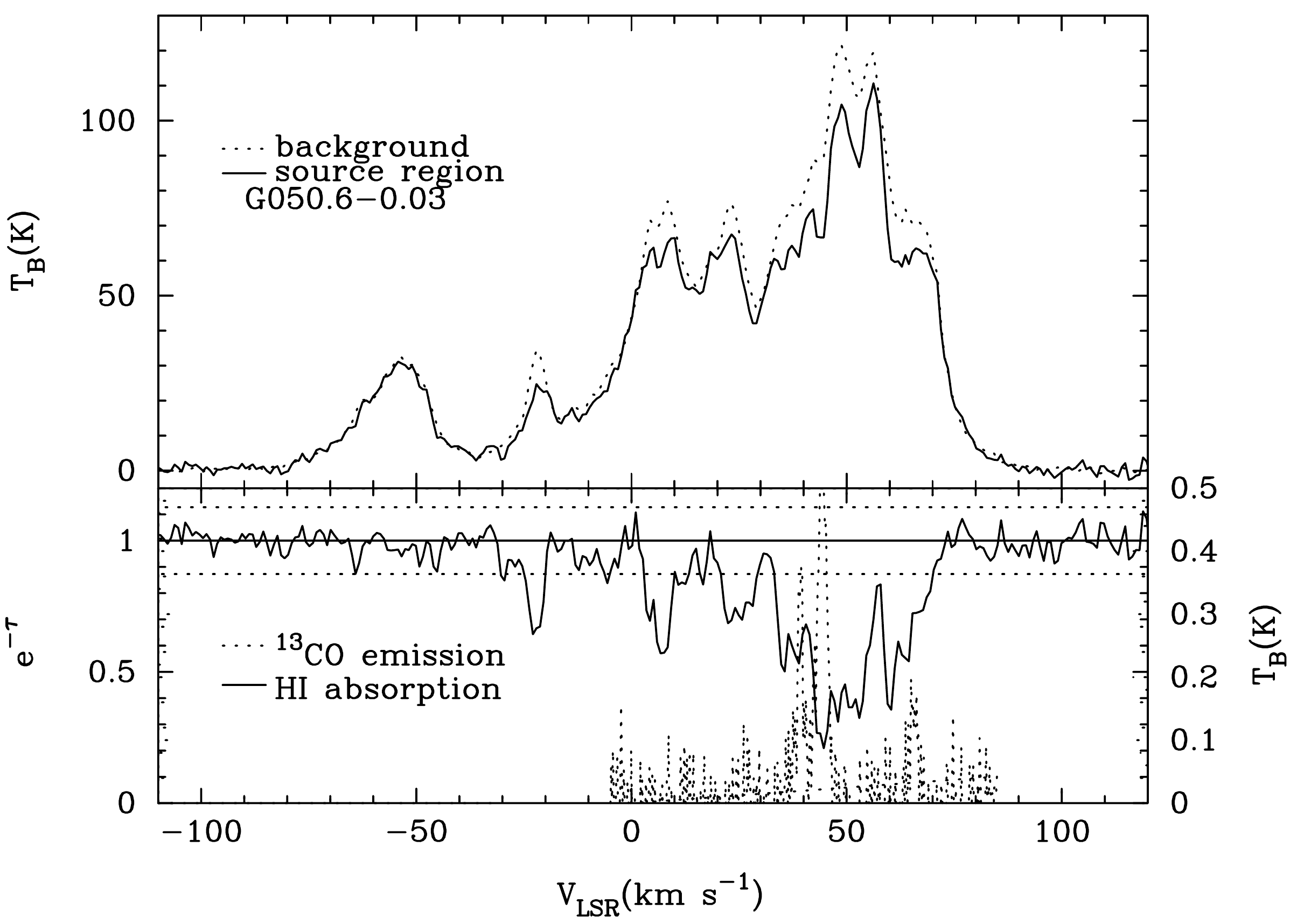}\\
   \end{tabular}
 \caption{1420\,MHz continuum image of PN G051.5$+$00.2 and its nearby background sources (top left),
 and the \HI~spectra of PN G051.5$+$00.2 (bottom left),
 G051.7$+$00.8 (top right), and G050.6$-$00.3 (bottom right). 
 The \HI~spectra of G050.95$+$0.85 is shown in appendix Fig.~\ref{fig21}.
 The map has superimposed contours (15,25,60\,K) of 1420\,MHz continuum emission.}
 \label{fig5}
\end{figure*}
\newpage
\begin{figure*}
 \centering
 \begin{tabular}{cc}
    \includegraphics[width = 0.29\textwidth,height = 0.215\textwidth]{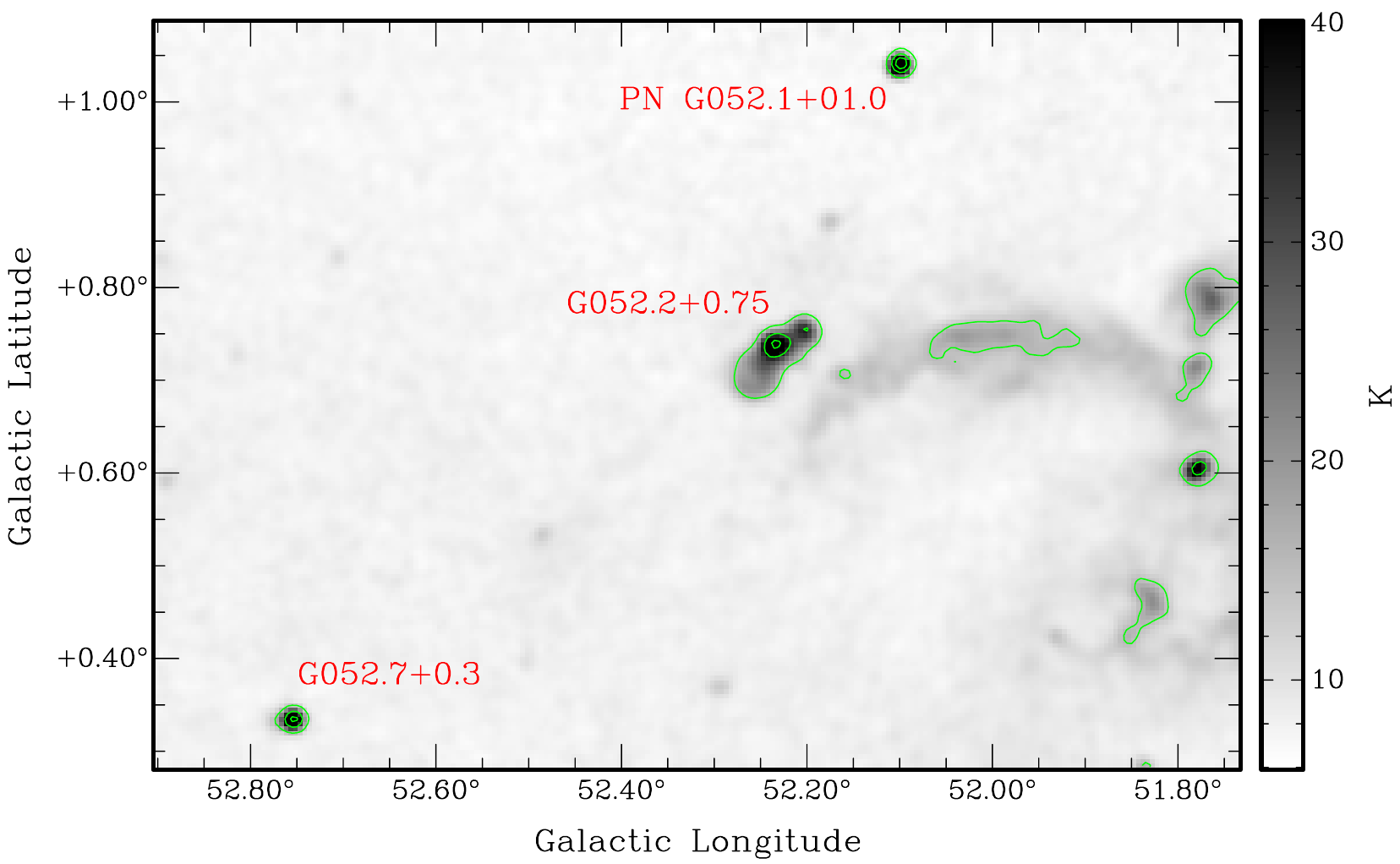}&
    \includegraphics[width = 0.3\textwidth]{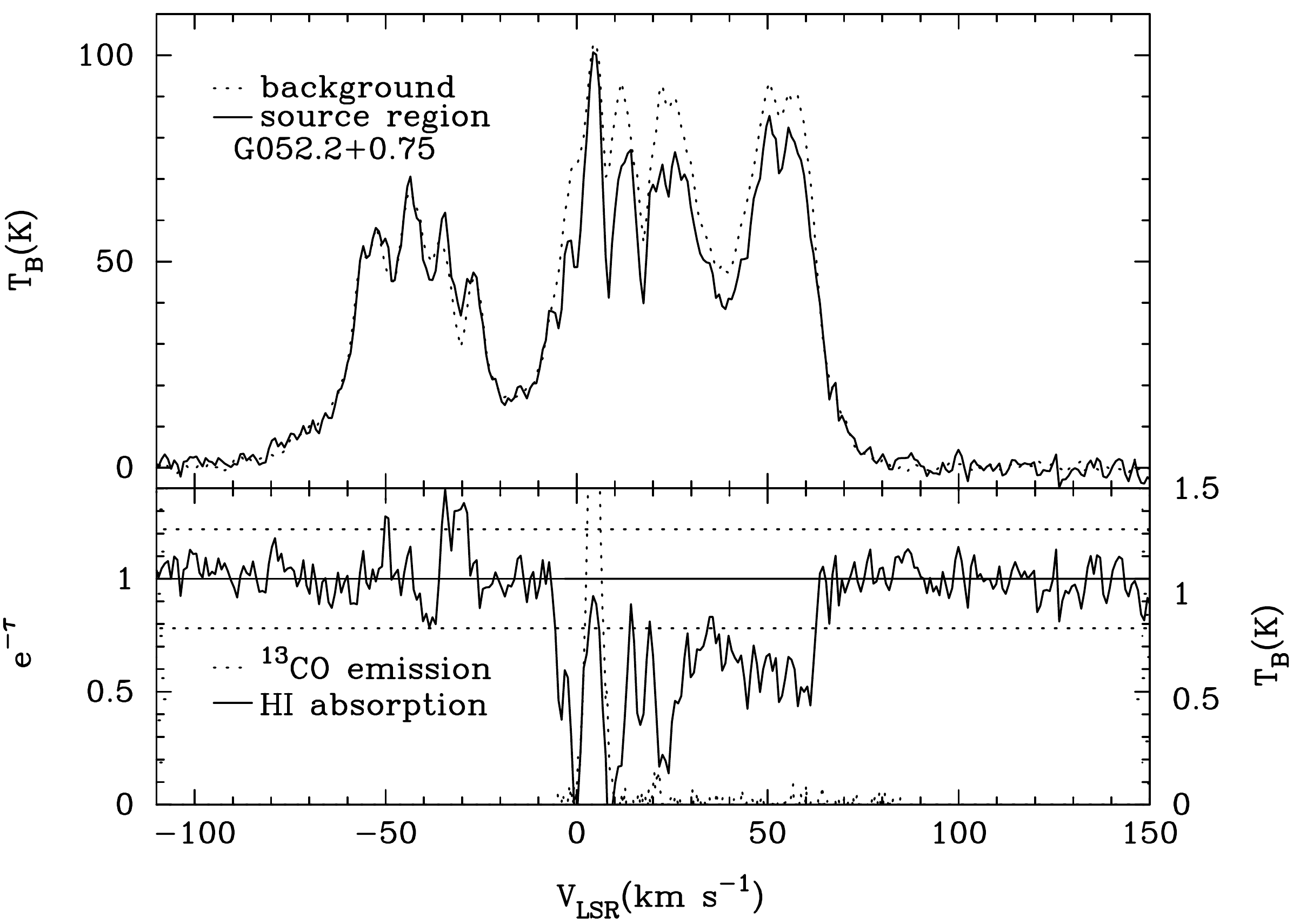}\\
    \includegraphics[width = 0.3\textwidth]{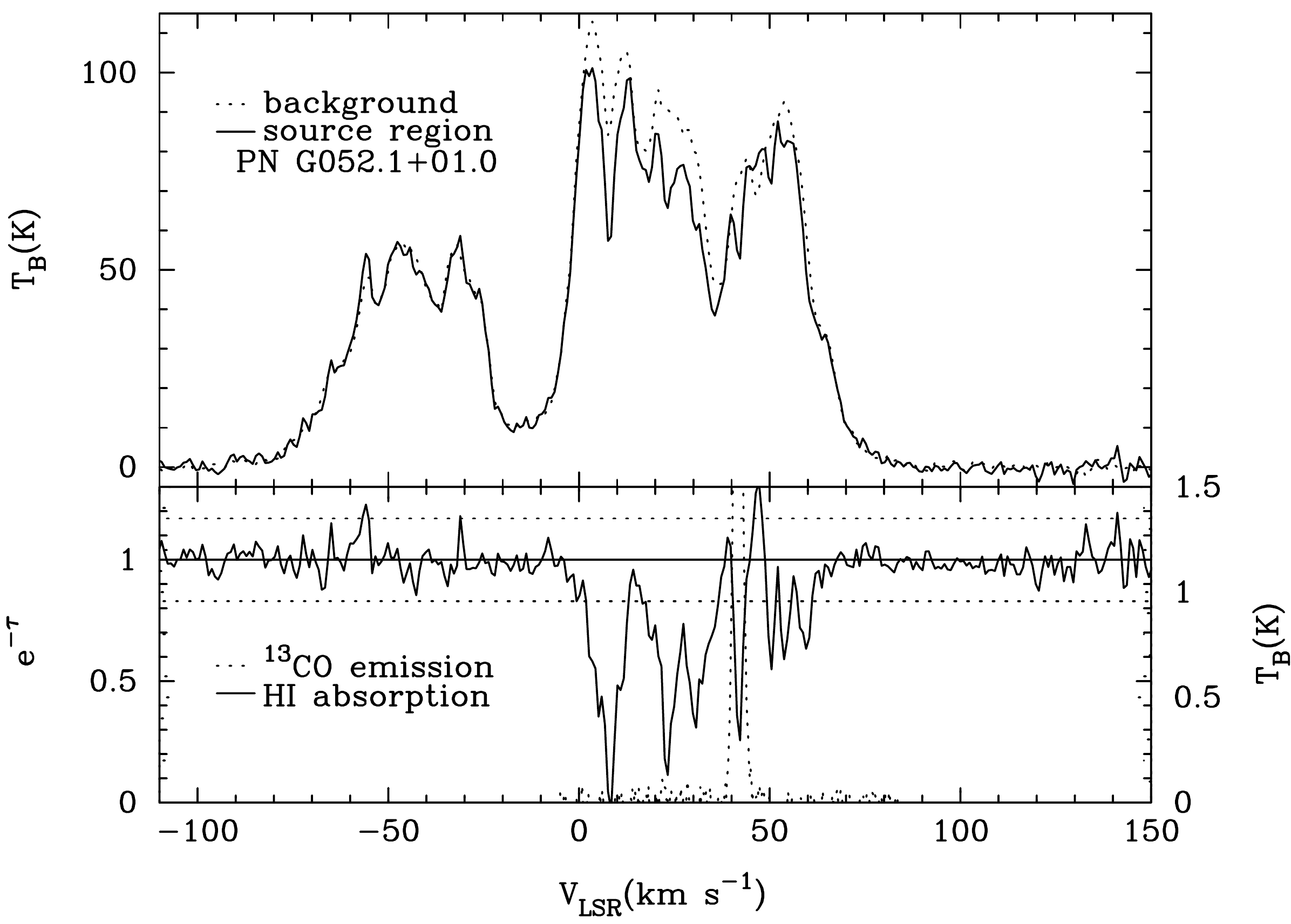}&
    \includegraphics[width = 0.3\textwidth]{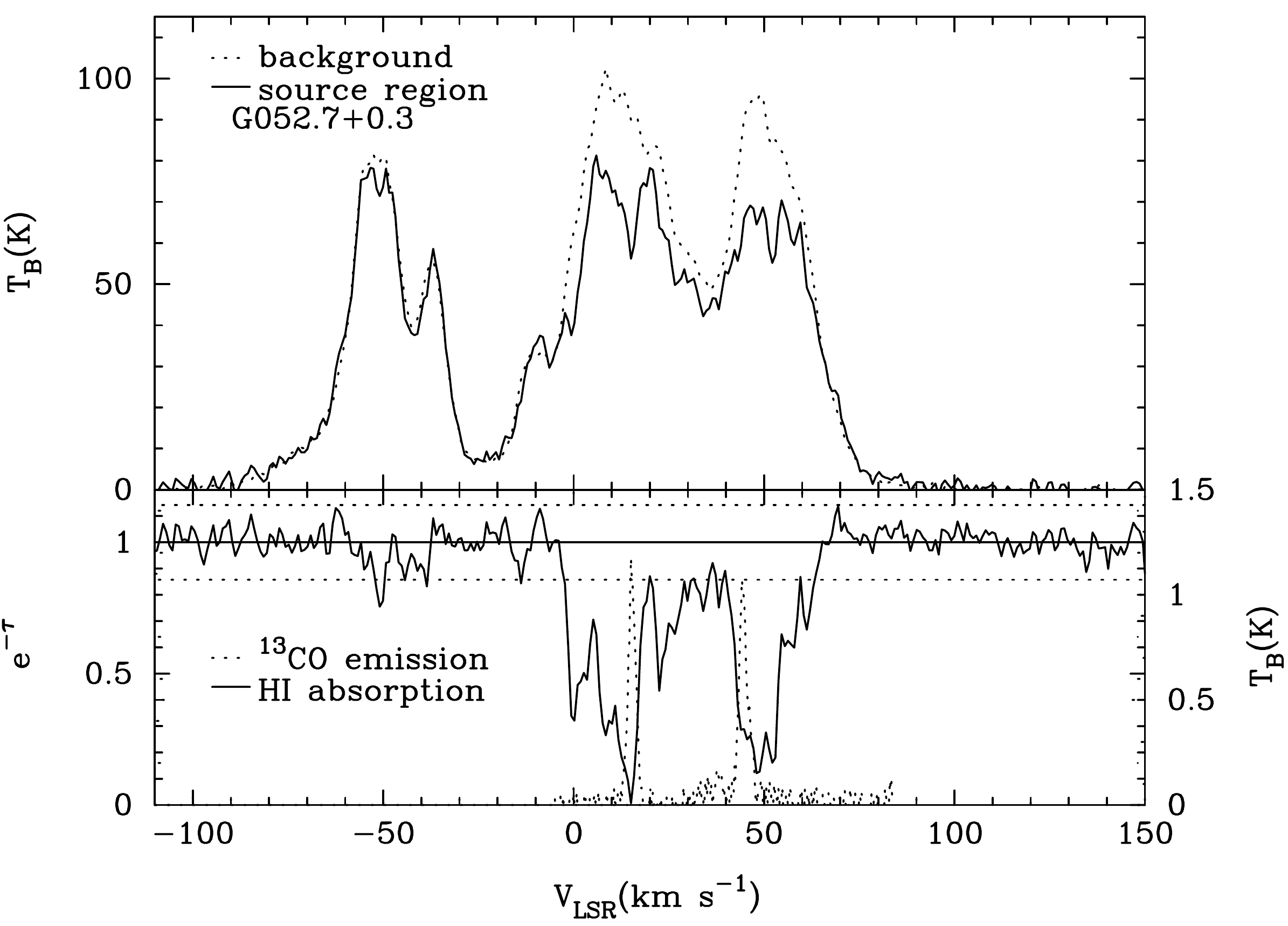}\\
   \end{tabular}
 \caption{1420\,MHz continuum image of PN G052.1$+$01.0, and its background sources (top left),
 and the \HI~spectra of PN G052.1$+$01.0 (bottom left), G052.2$+$0.75 (top right) and G052.7$+$0.3 (bottom right).
 The map has superimposed contours (15,35,55\,K) of 1420\,MHz continuum emission.}
 \label{fig6}
\end{figure*}
\begin{figure*}
 \centering
 \begin{tabular}{cc}
    \includegraphics[width = 0.29\textwidth,height = 0.23\textwidth]{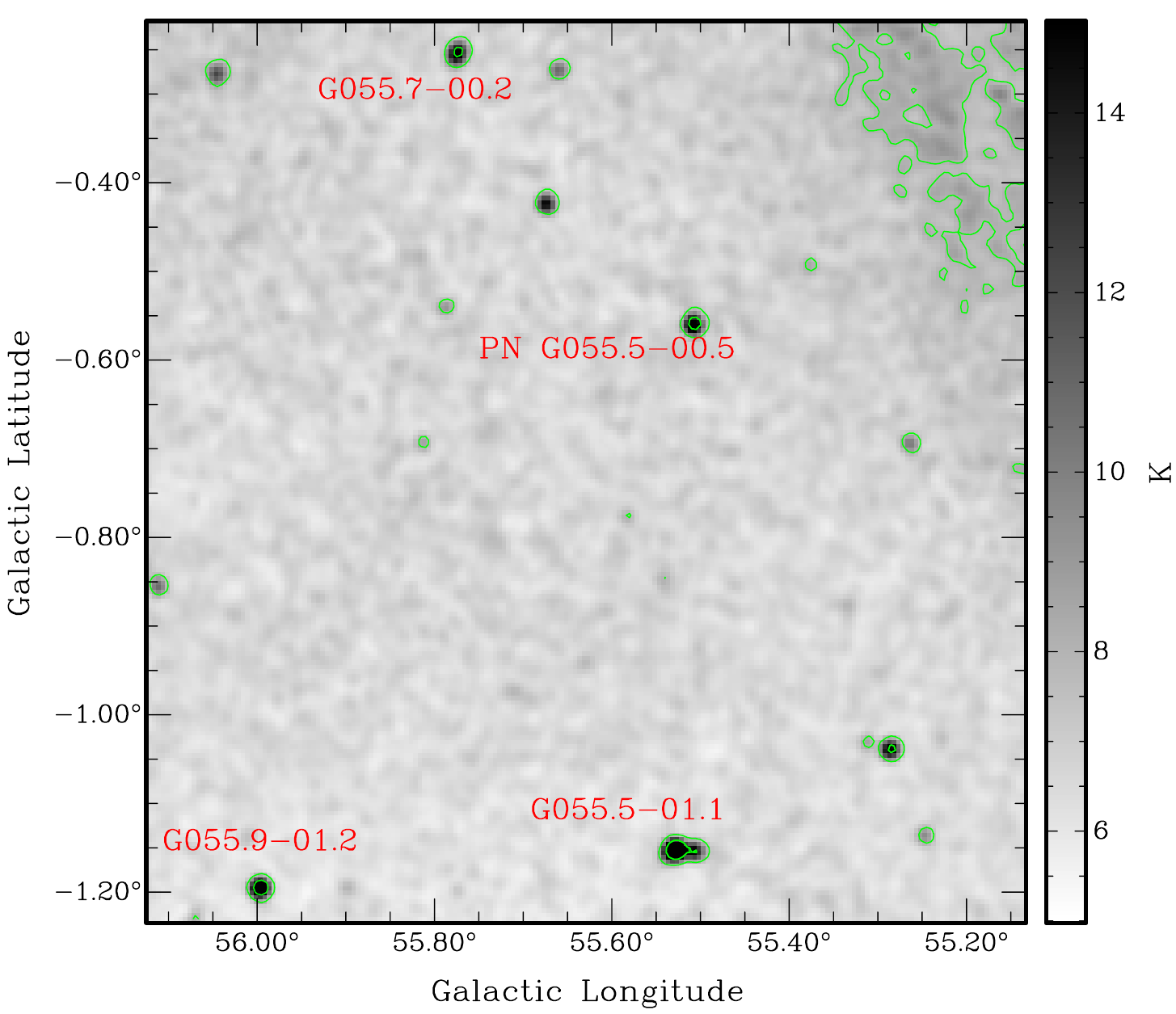}&
    \includegraphics[width = 0.29\textwidth]{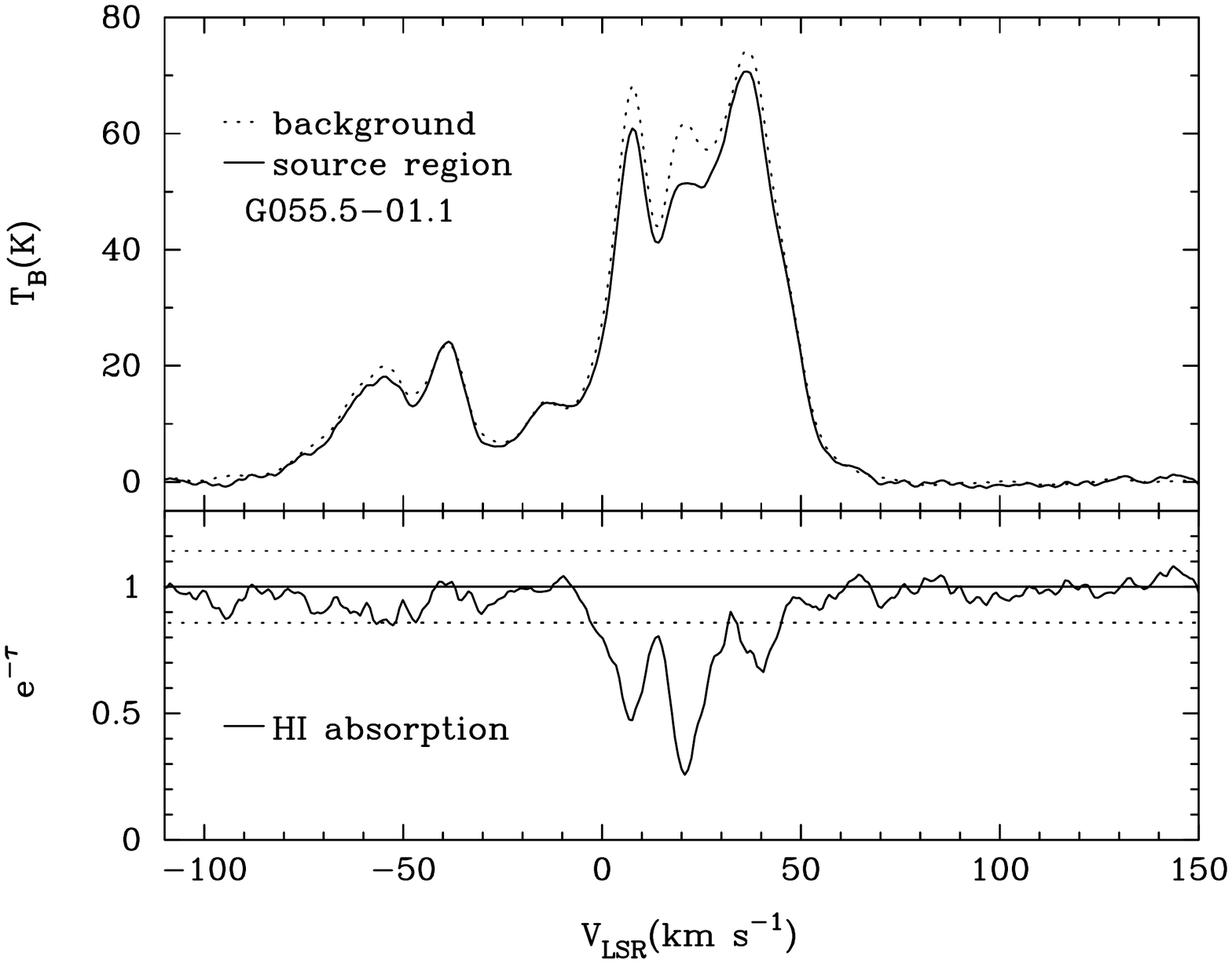}\\
    \includegraphics[width = 0.33\textwidth]{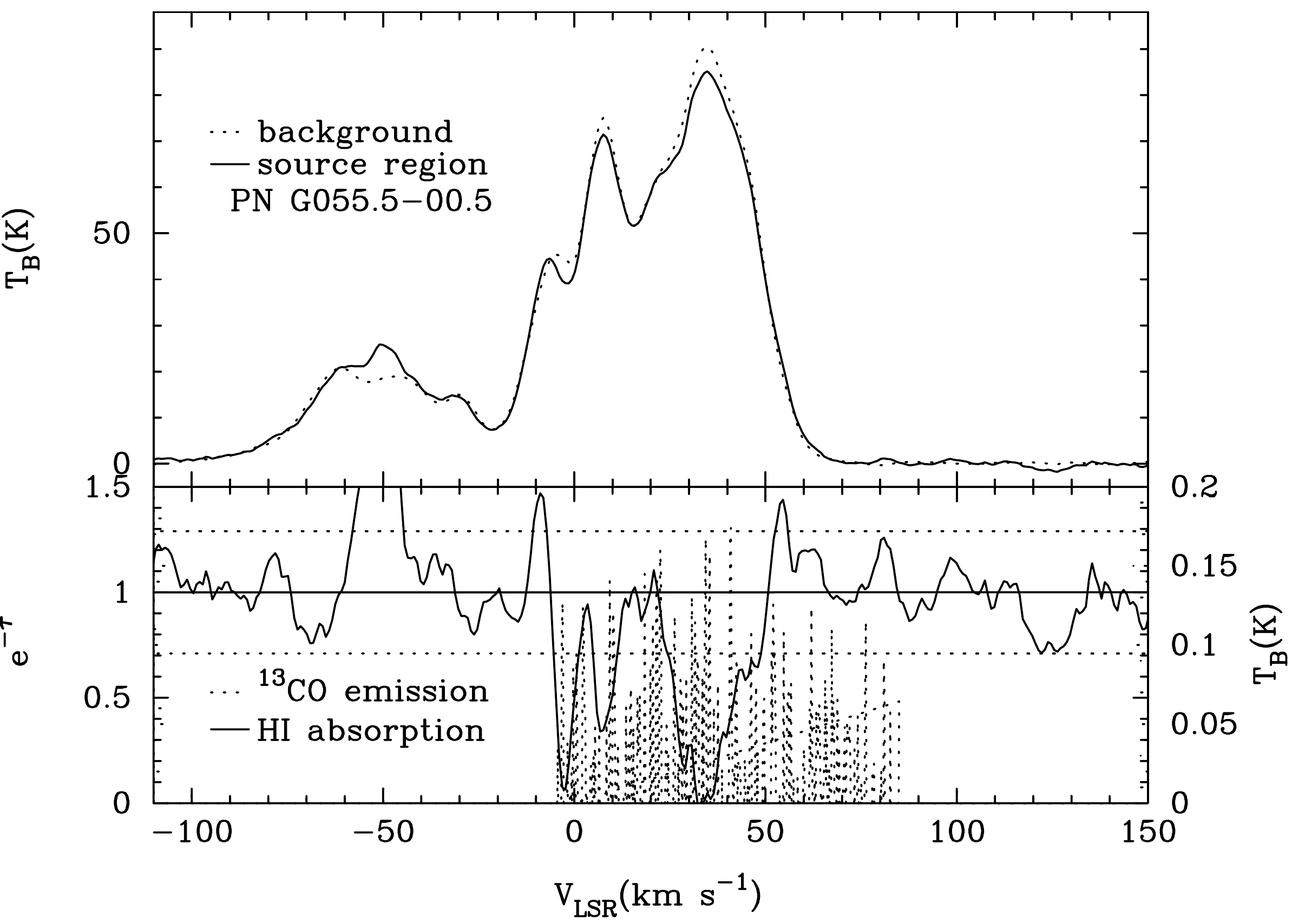}&
    \includegraphics[width = 0.29\textwidth]{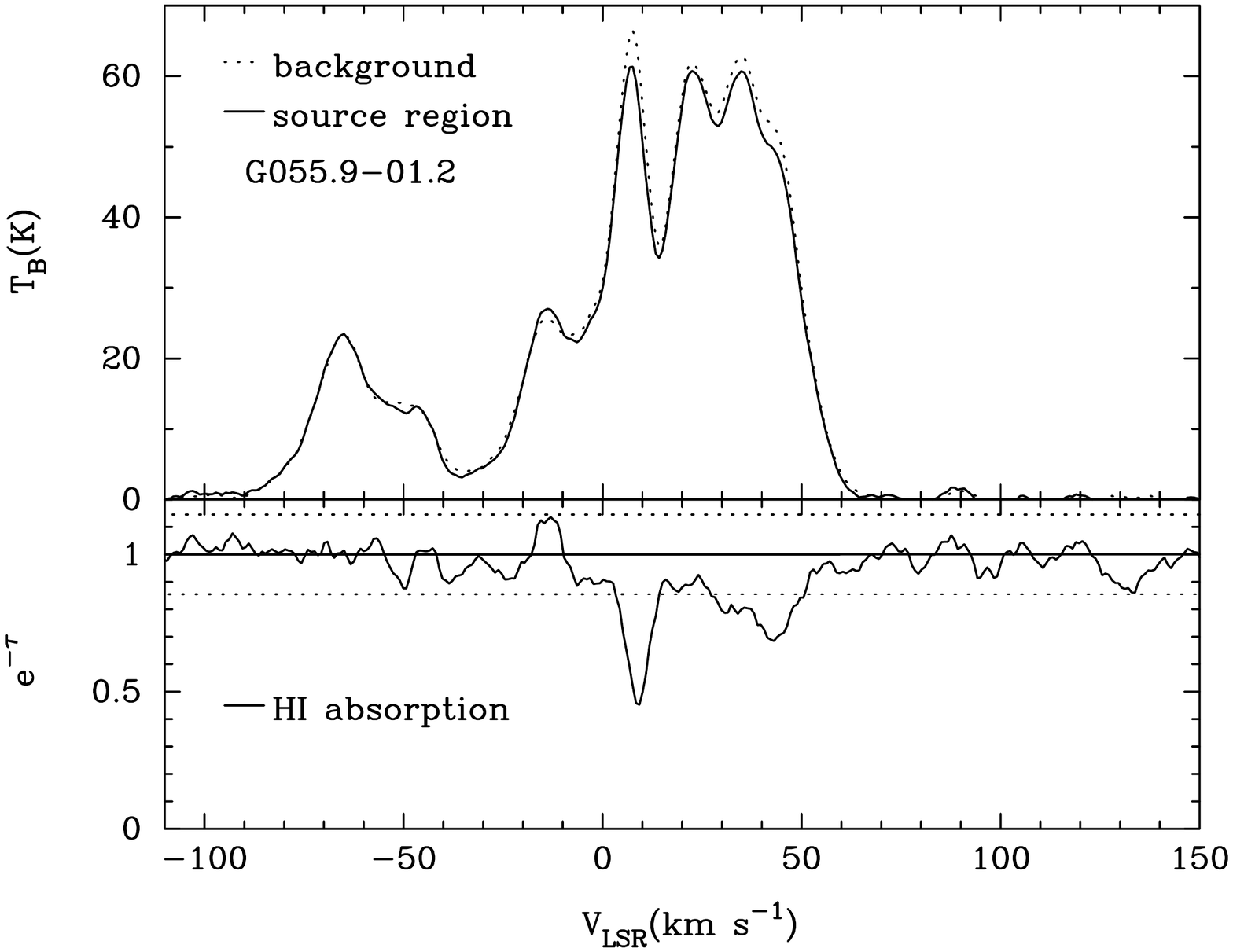}\\
   \end{tabular}
 \caption{1420\,MHz continuum image of PN G055.5$-$00.5 and its nearby background sources (top left),
 and the \HI~spectra of PN G055.5$-$00.5 (bottom left)
 , G055.5$-$01.1 (top right), and G055.9$-$01.2 (middle right). 
 The \HI~spectra of G055.7$-$00.2  is shown in appendix Fig.~\ref{fig21}.
 The map has superimposed contours (8,15,35\,K) of 1420\,MHz continuum emission.}
 \label{fig7}
\end{figure*}
\begin{figure*}
 \centering
 \begin{tabular}{cc}
    \includegraphics[width = 0.29\textwidth,height = 0.24\textwidth]{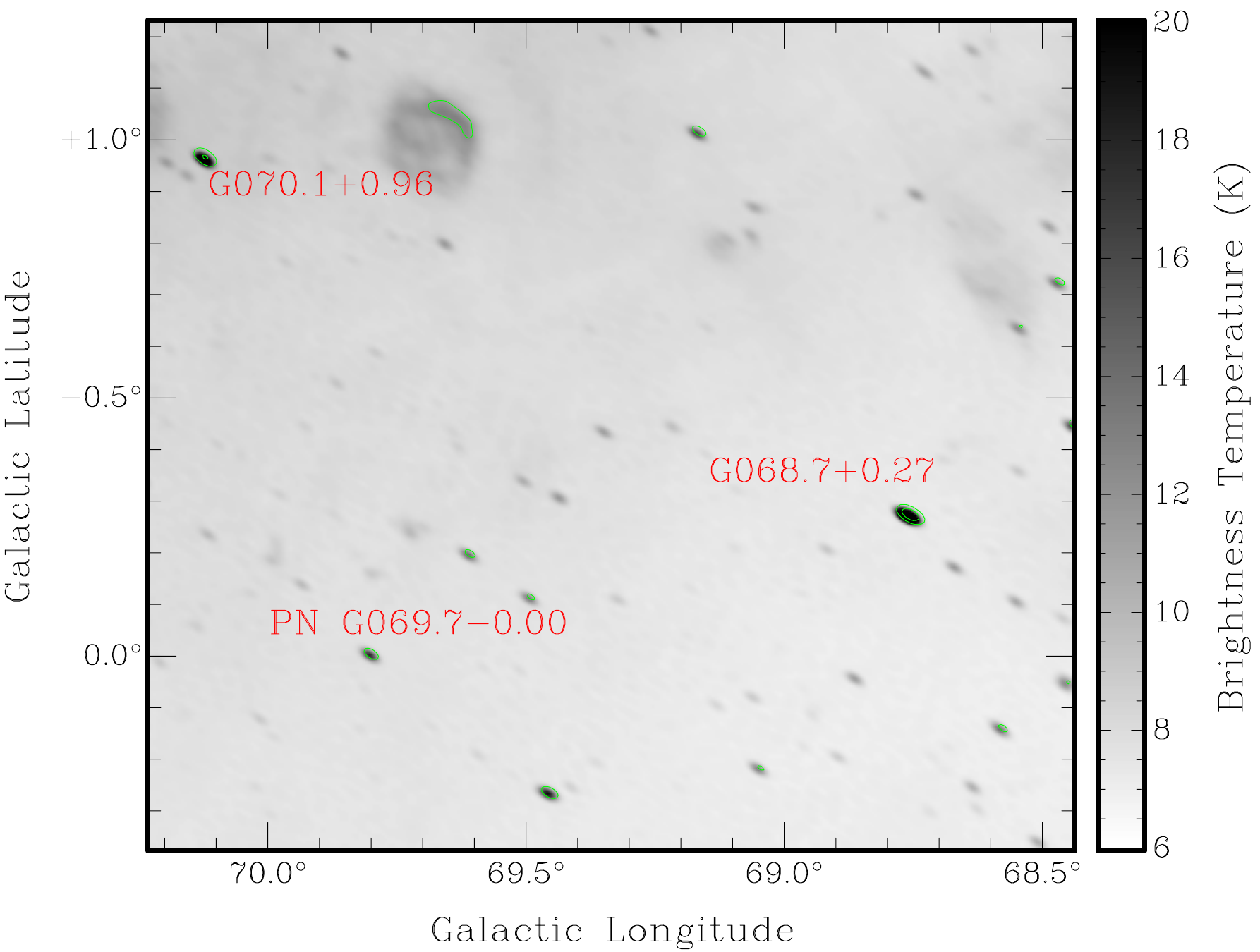}&
    \includegraphics[width = 0.3\textwidth]{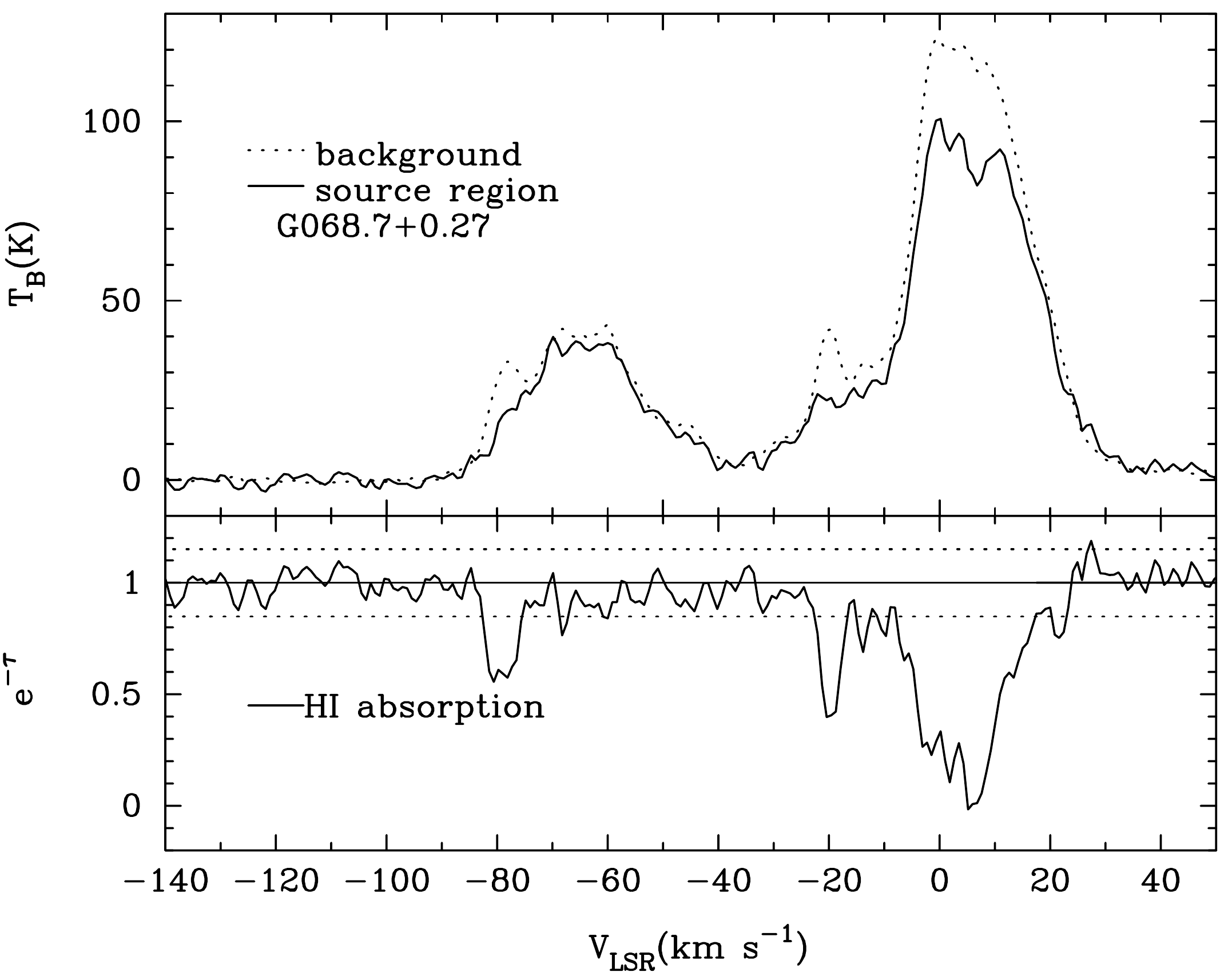}\\
    \includegraphics[width = 0.3\textwidth]{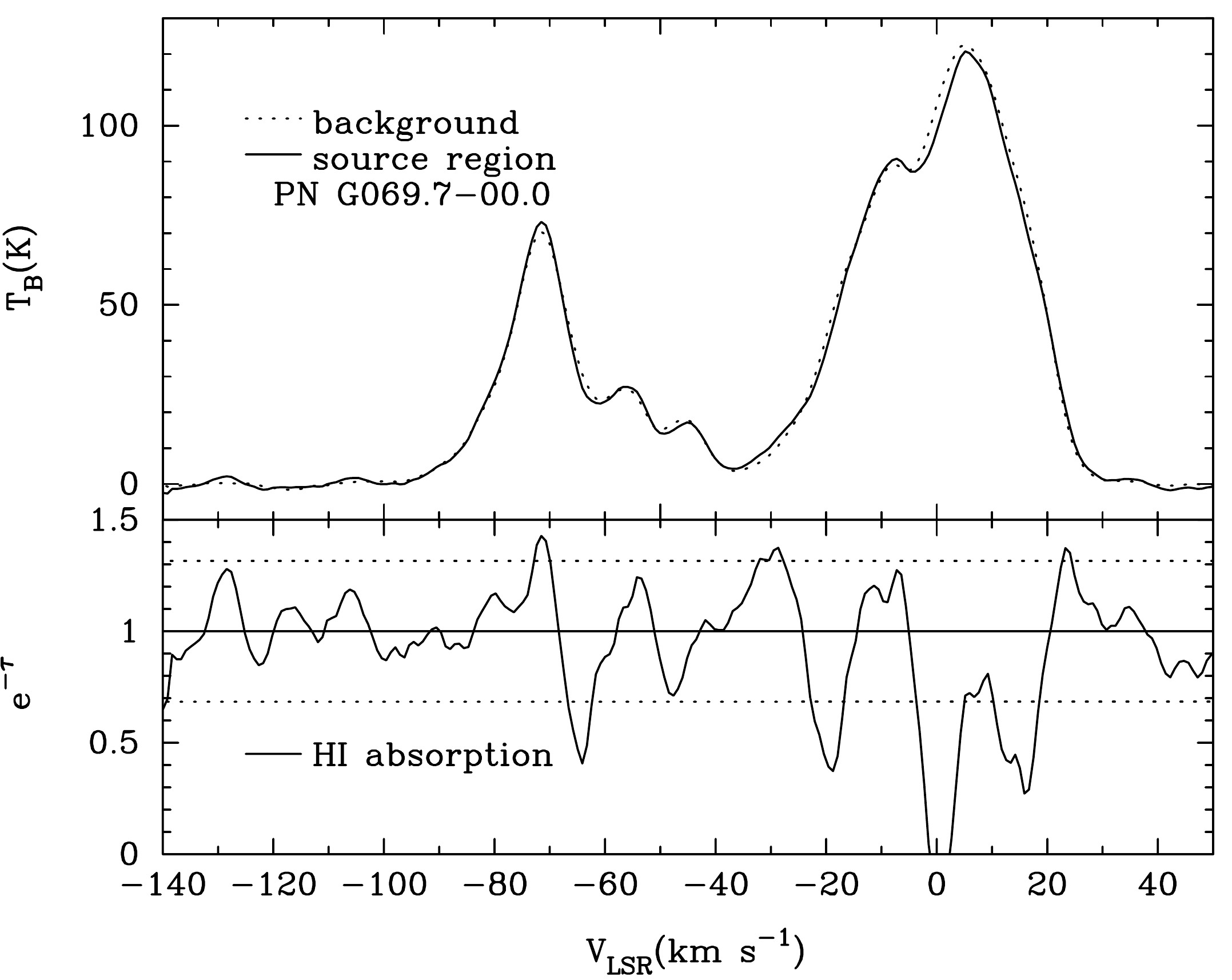}&
    \includegraphics[width = 0.3\textwidth]{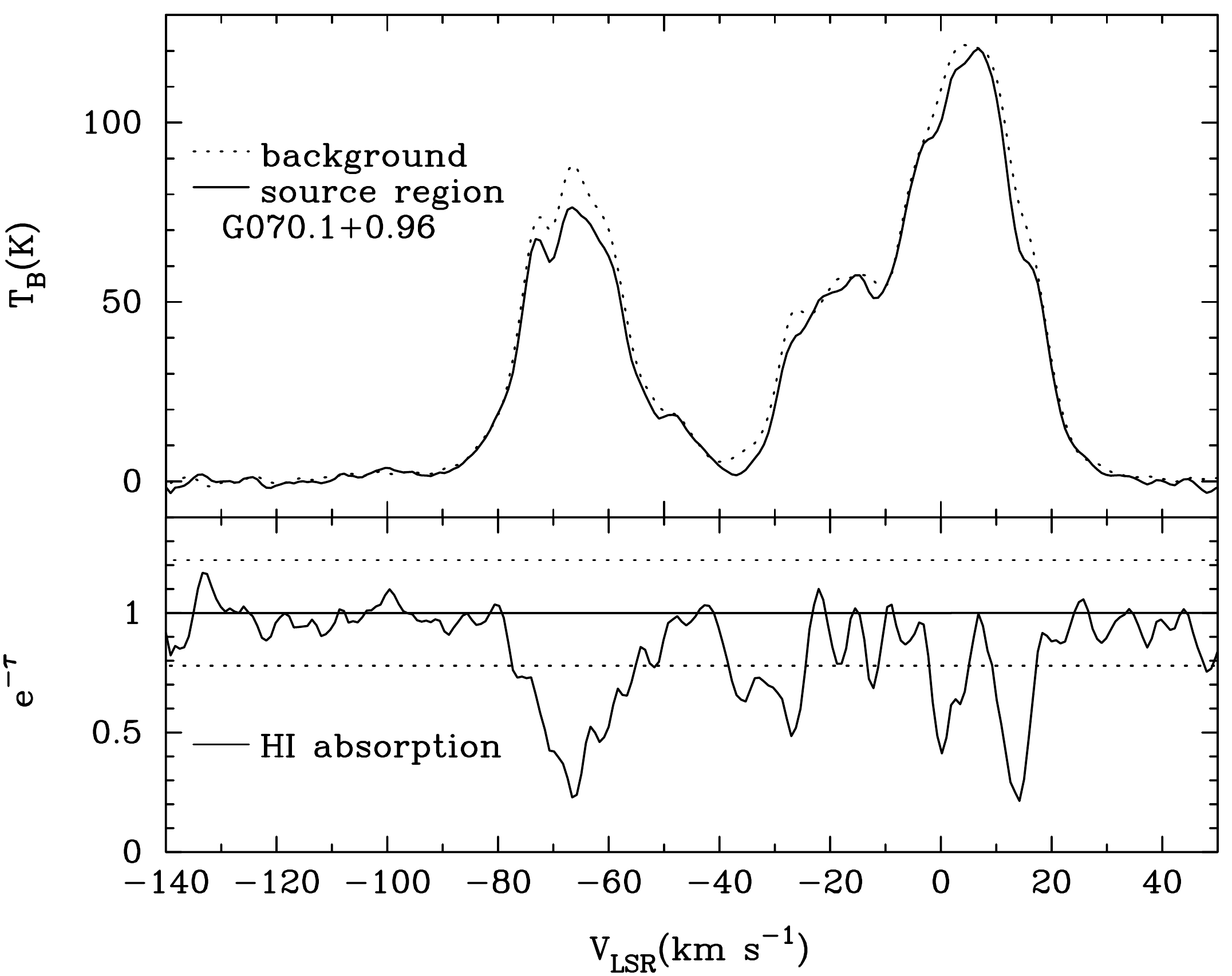}\\
   \end{tabular}
 \caption{1420\,MHz continuum image of PN G069.7$-$00.0, and its nearby background sources (top left),
 and the \HI~spectra of PN G069.7$-$00.0 (bottom left), G068.7$+$0.27 (top right) and G070.1$+$0.96 (bottom right).
 The map has superimposed contours (5,12,32\,K) of 1420\,MHz continuum emission.}
 \label{fig8}
\end{figure*}
\begin{figure*}
 \centering
 \begin{tabular}{cc}
    \includegraphics[width = 0.29\textwidth,height = 0.24\textwidth]{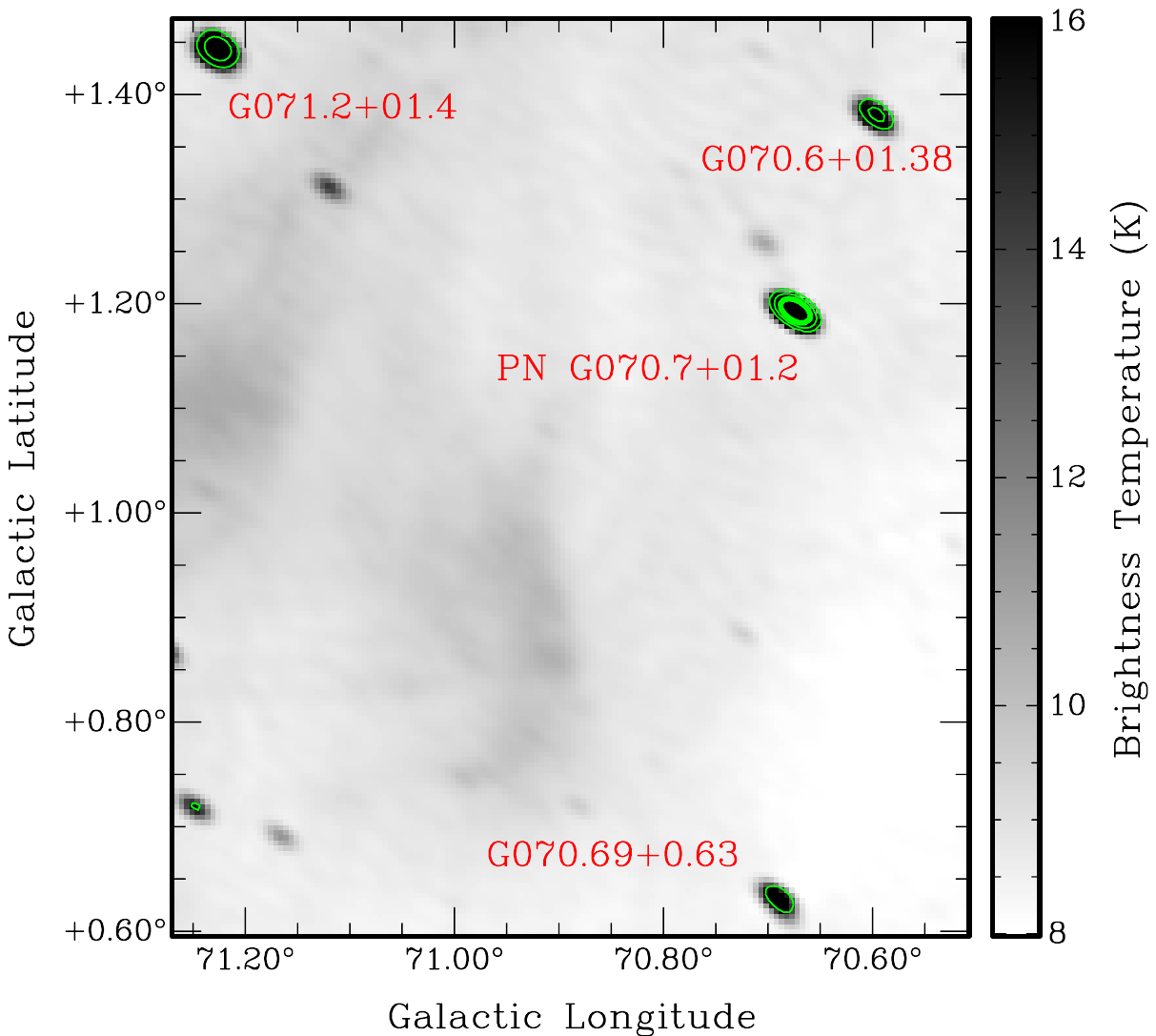}&
    \includegraphics[width = 0.3\textwidth]{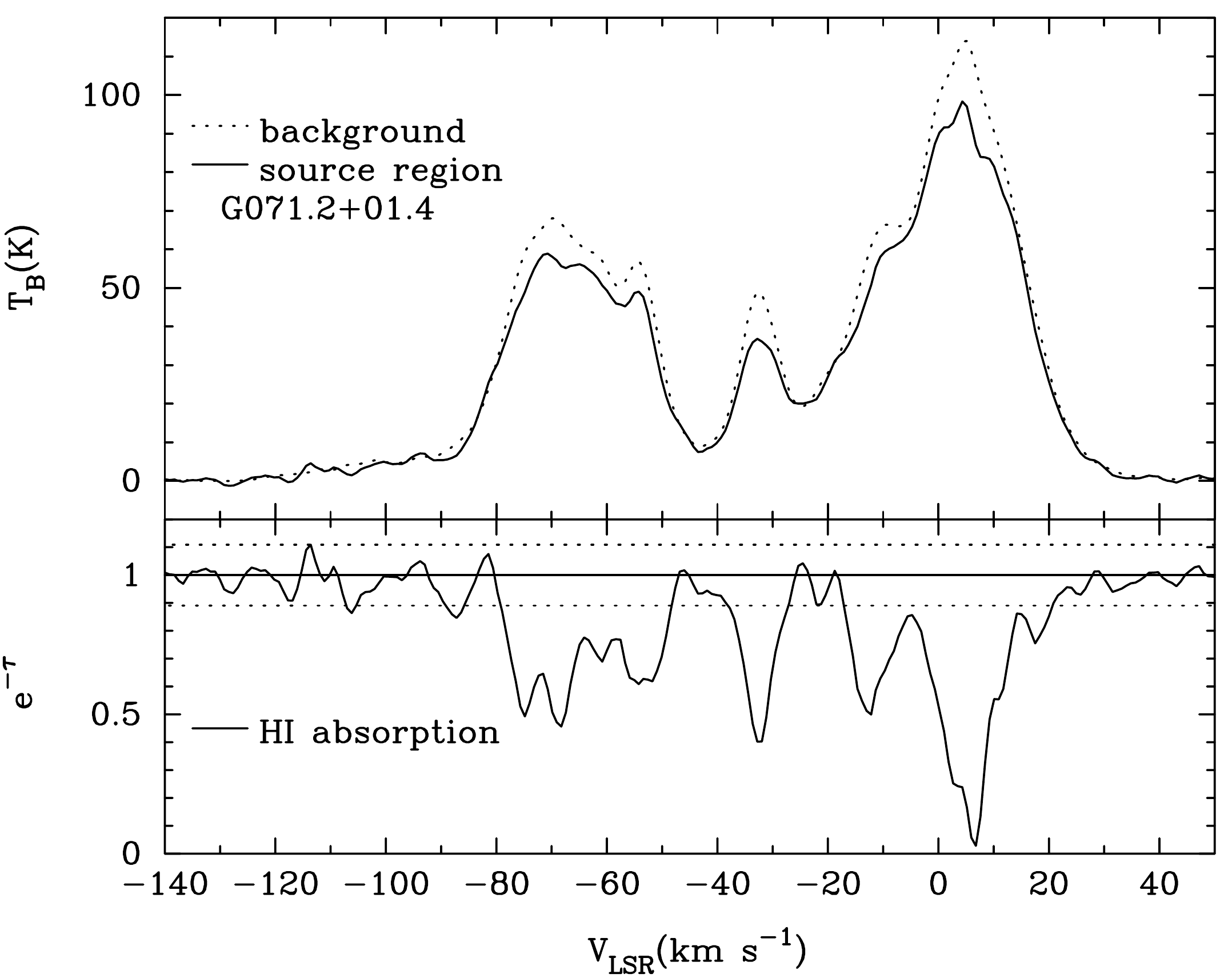}\\
    \includegraphics[width = 0.3\textwidth]{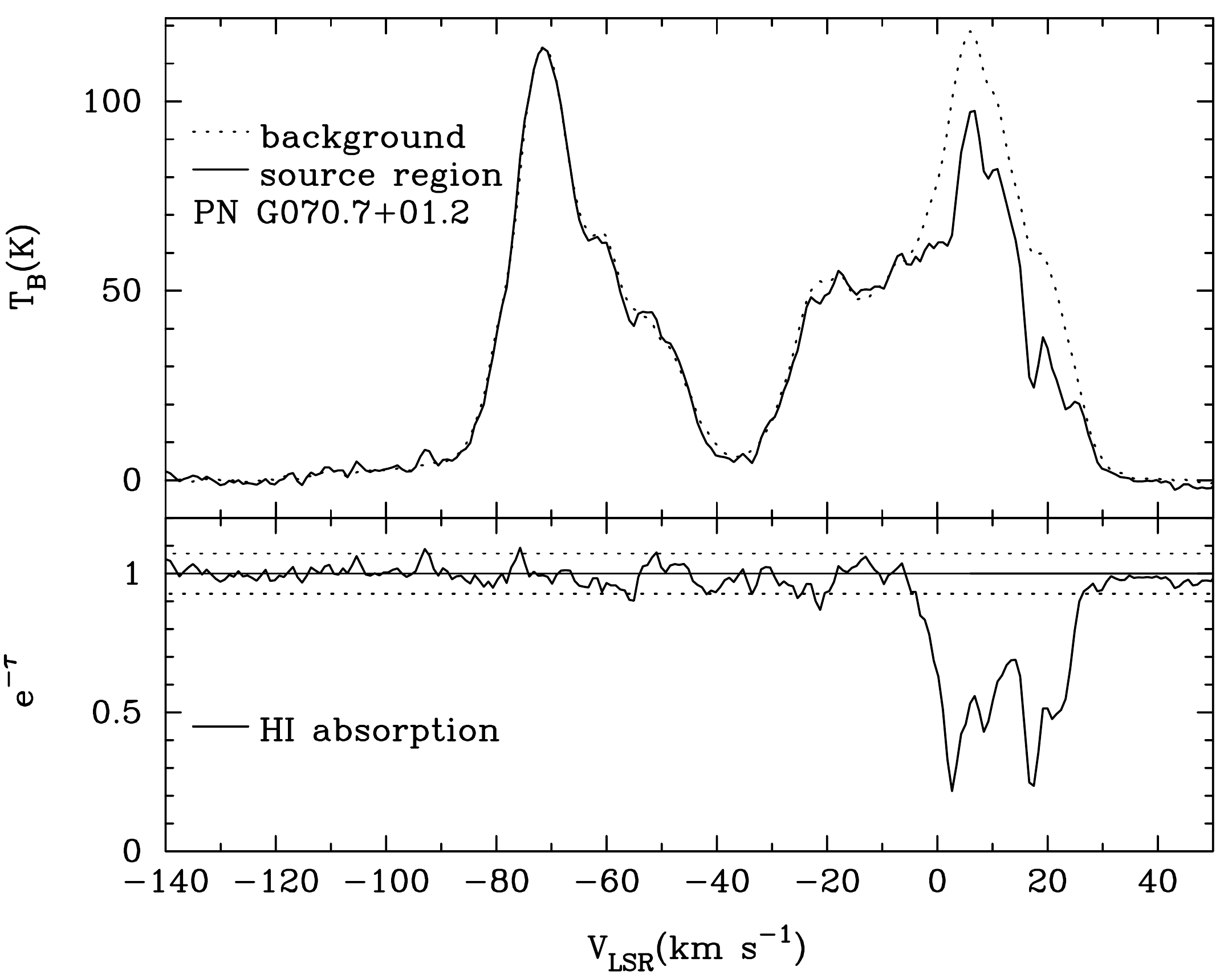}&
      \includegraphics[width = 0.3\textwidth]{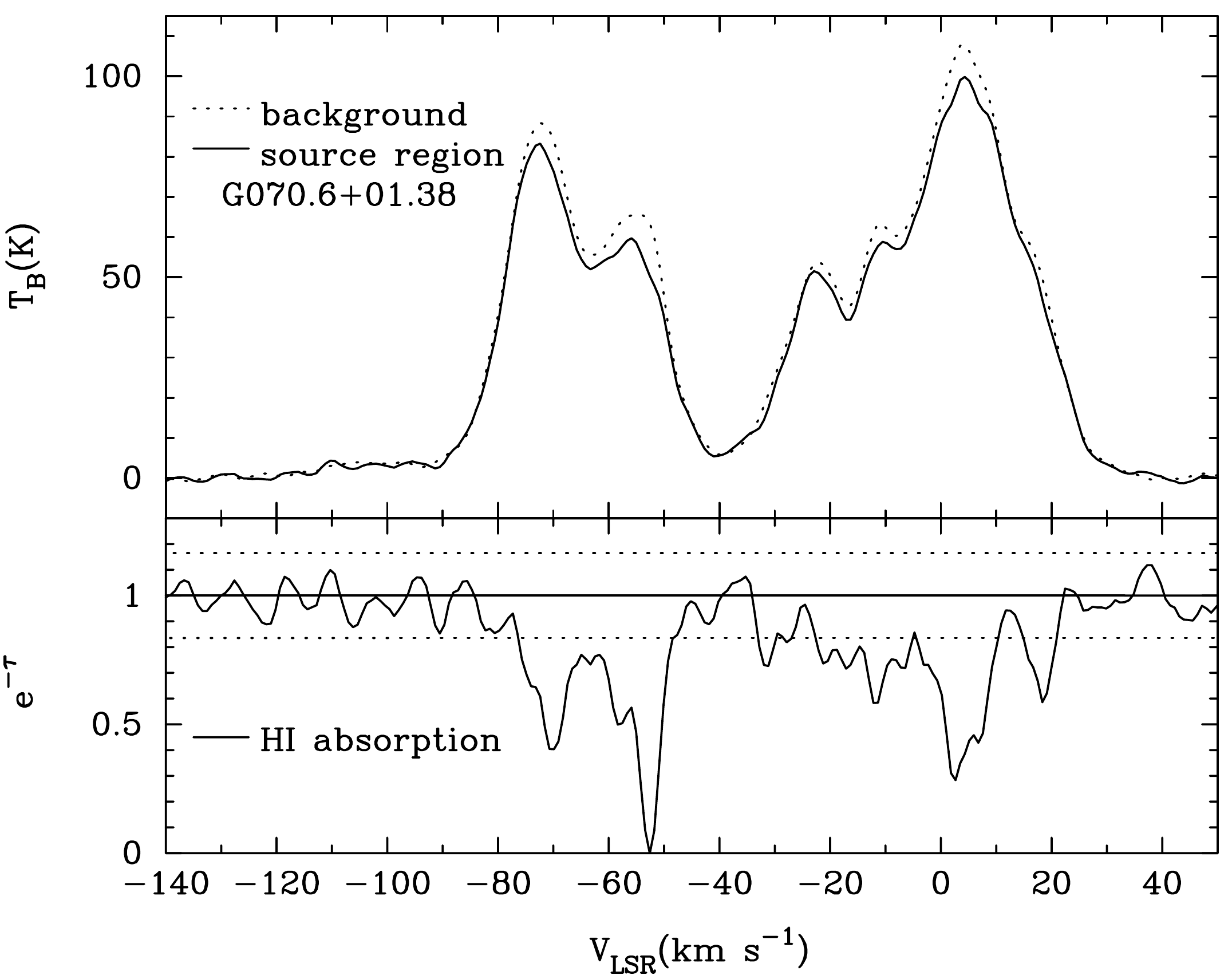}\\
   \end{tabular}
 \caption{1420\,MHz continuum image of PN G070.7$+$01.2, and its background sources (top left),
 and the \HI~spectra of PN G070.7$+$01.2 (bottom left), G071.2$+$01.4 (top right), and G070.6$+$01.38(bottom right).  The \HI~spectra of G070.69$+$0.63 is listed in appendix Fig.~\ref{fig21}.
 The map has superimposed contours (15,25,50\,K) of 1420\,MHz continuum emission.}
  \label{fig9}
\end{figure*}
\begin{figure*}
 \centering
 \begin{tabular}{cc}
    \includegraphics[width = 0.29\textwidth,height = 0.24\textwidth]{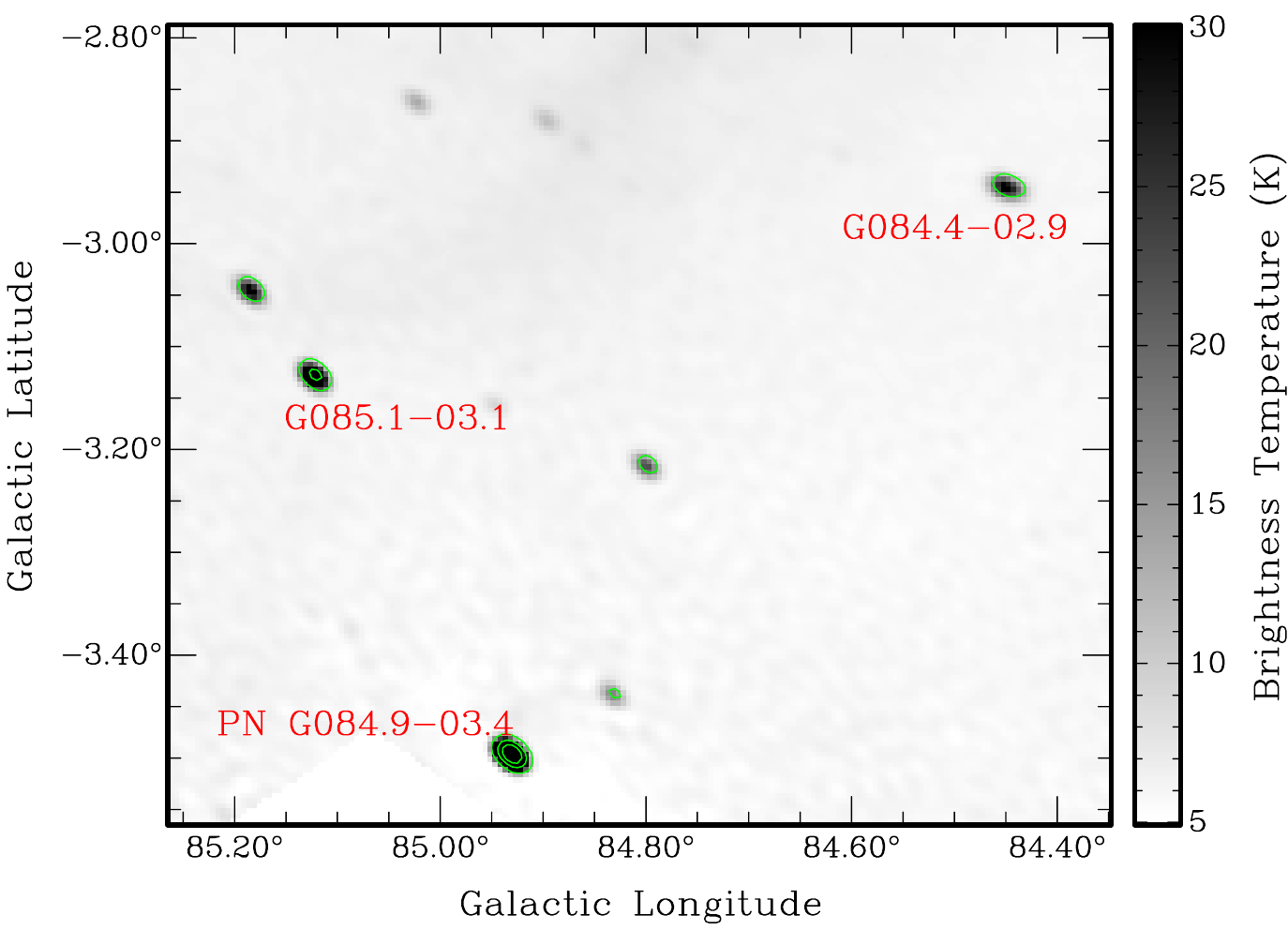}&
    \includegraphics[width = 0.3\textwidth]{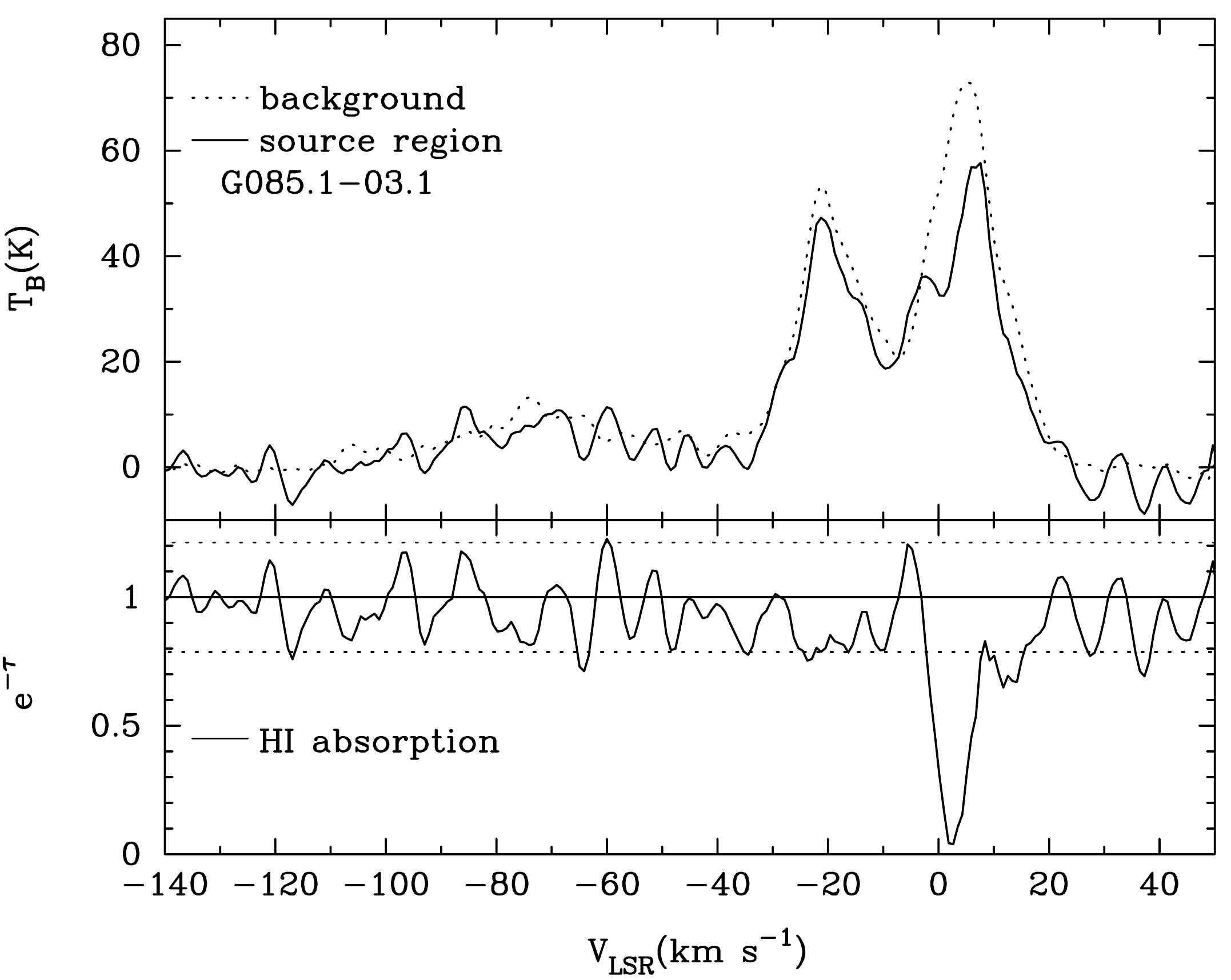}\\
    \includegraphics[width = 0.3\textwidth]{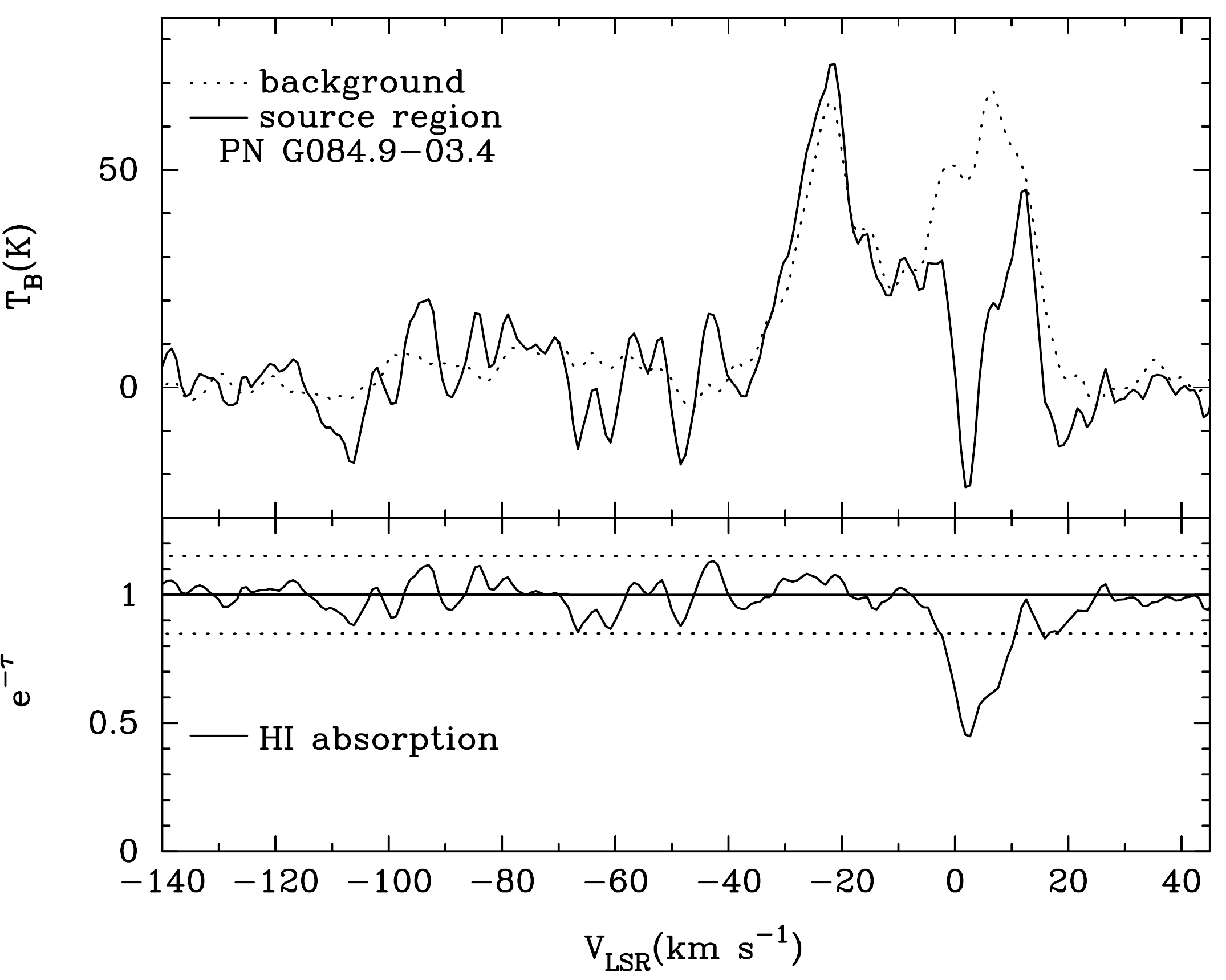}&
    \includegraphics[width = 0.3\textwidth]{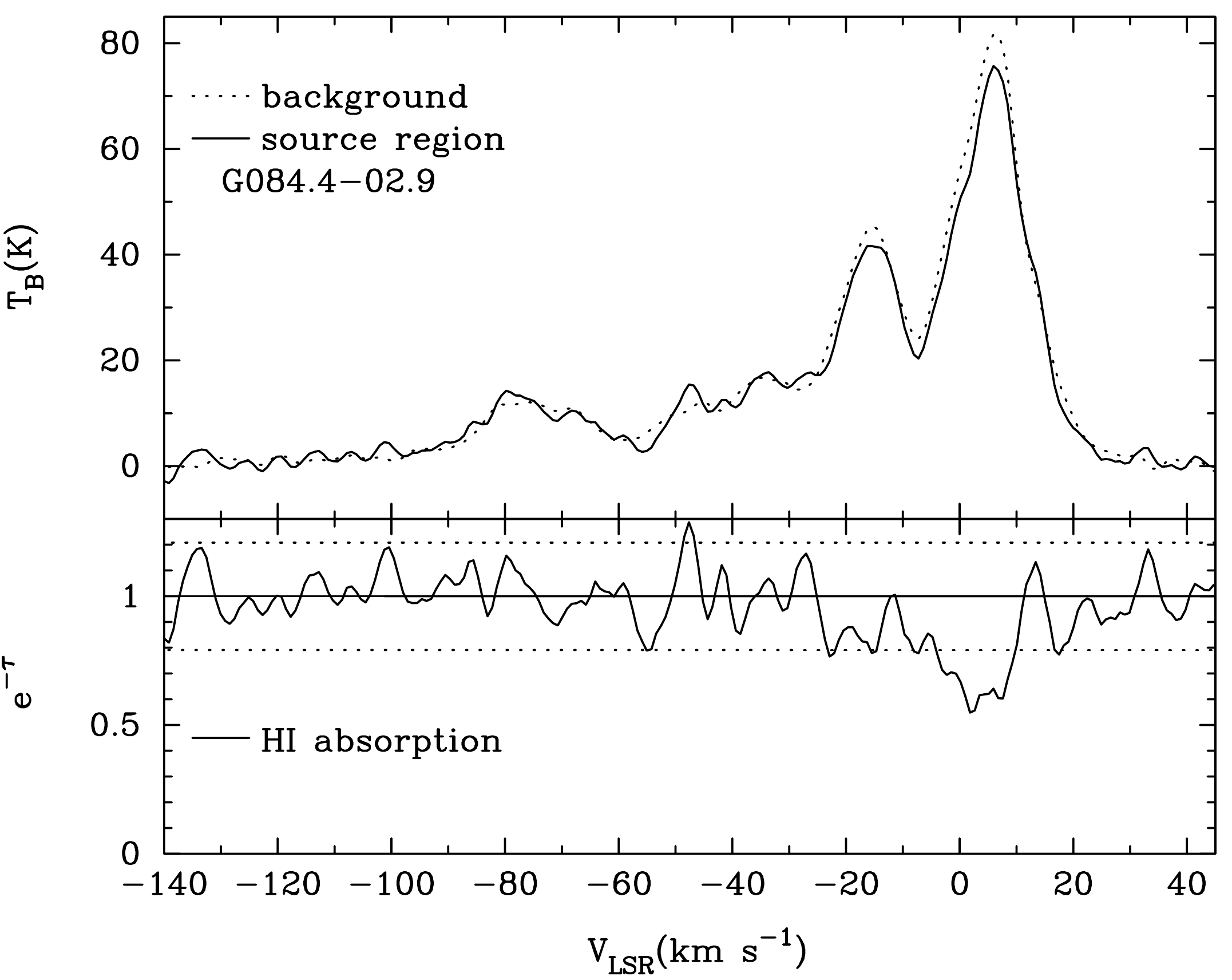}\\
   \end{tabular}
 \caption{1420\,MHz continuum image of PN G084.9$-$03.4 and it nearby background sources (top left),
  and the \HI~spectra of PN G084.9$-$03.4 (bottom left), G085.1$-$03.1 (top right), G084.4$-$02.9 (bottom right).
  The map has superimposed contours (15,50,100\,K) of 1420\,MHz continuum emission.}
 \label{fig10}
\end{figure*}
\begin{figure*}
 \centering
 \begin{tabular}{cc}
    \includegraphics[width = 0.29\textwidth,height = 0.24\textwidth]{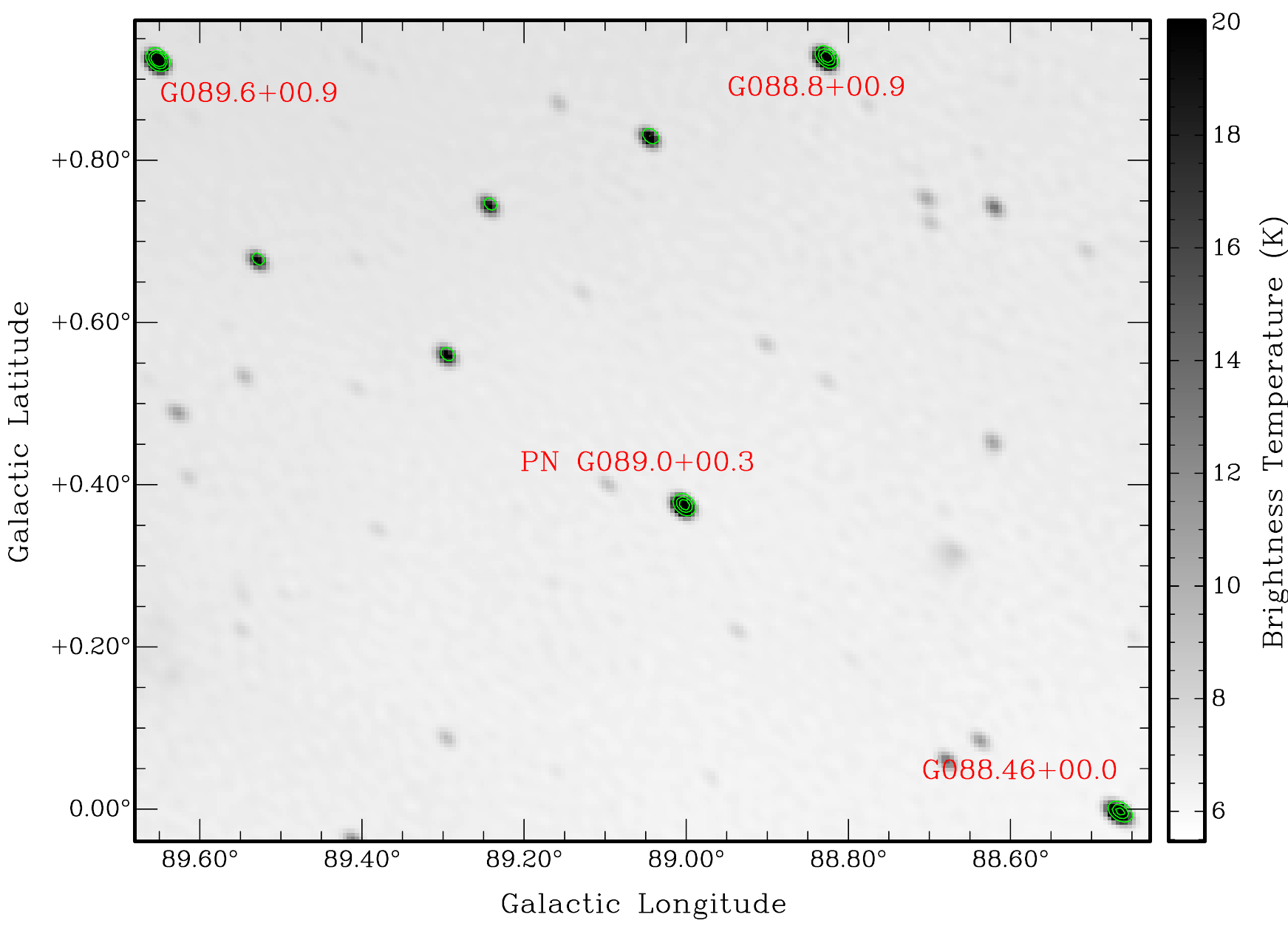}&
    \includegraphics[width = 0.3\textwidth]{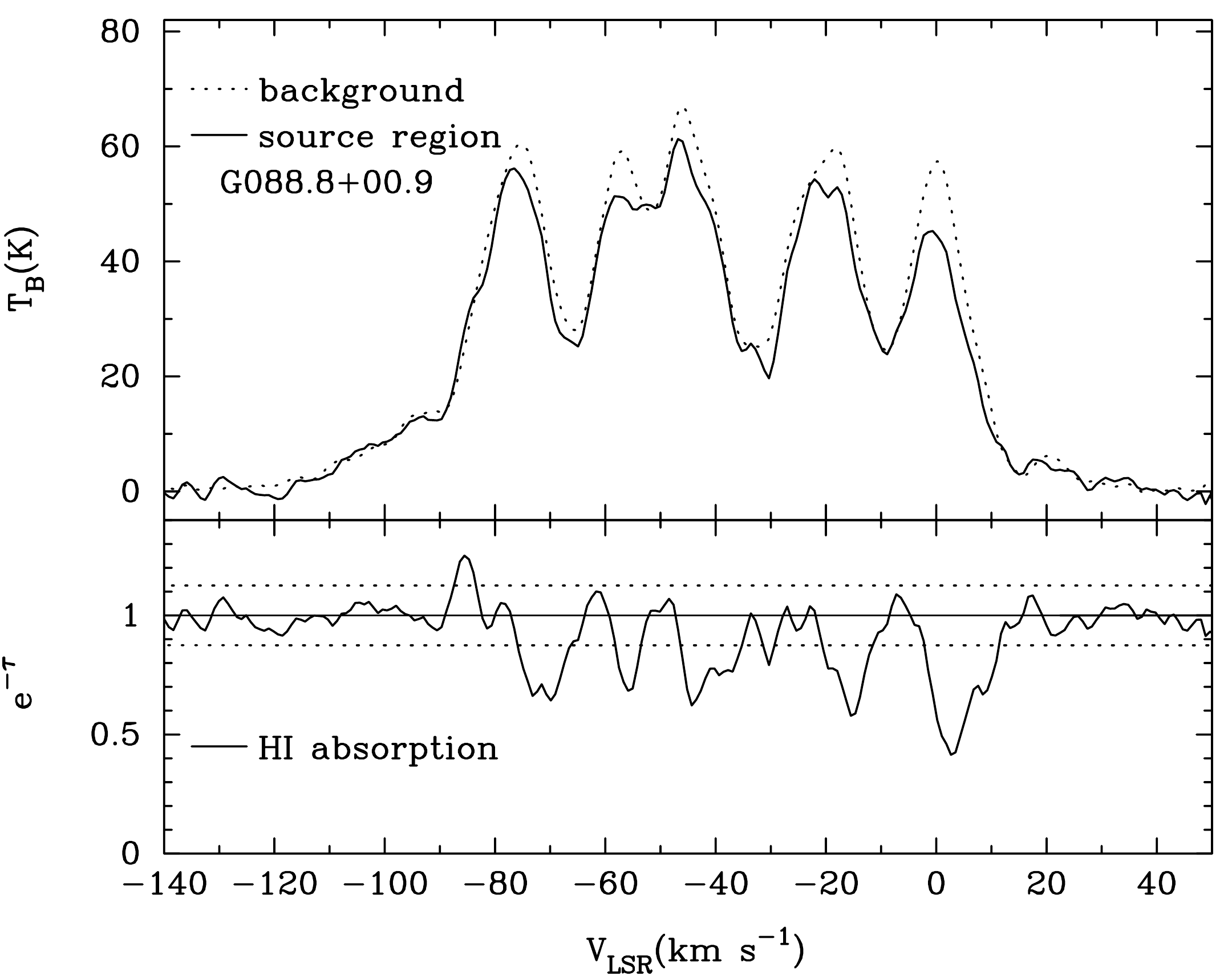}\\
    \includegraphics[width = 0.3\textwidth]{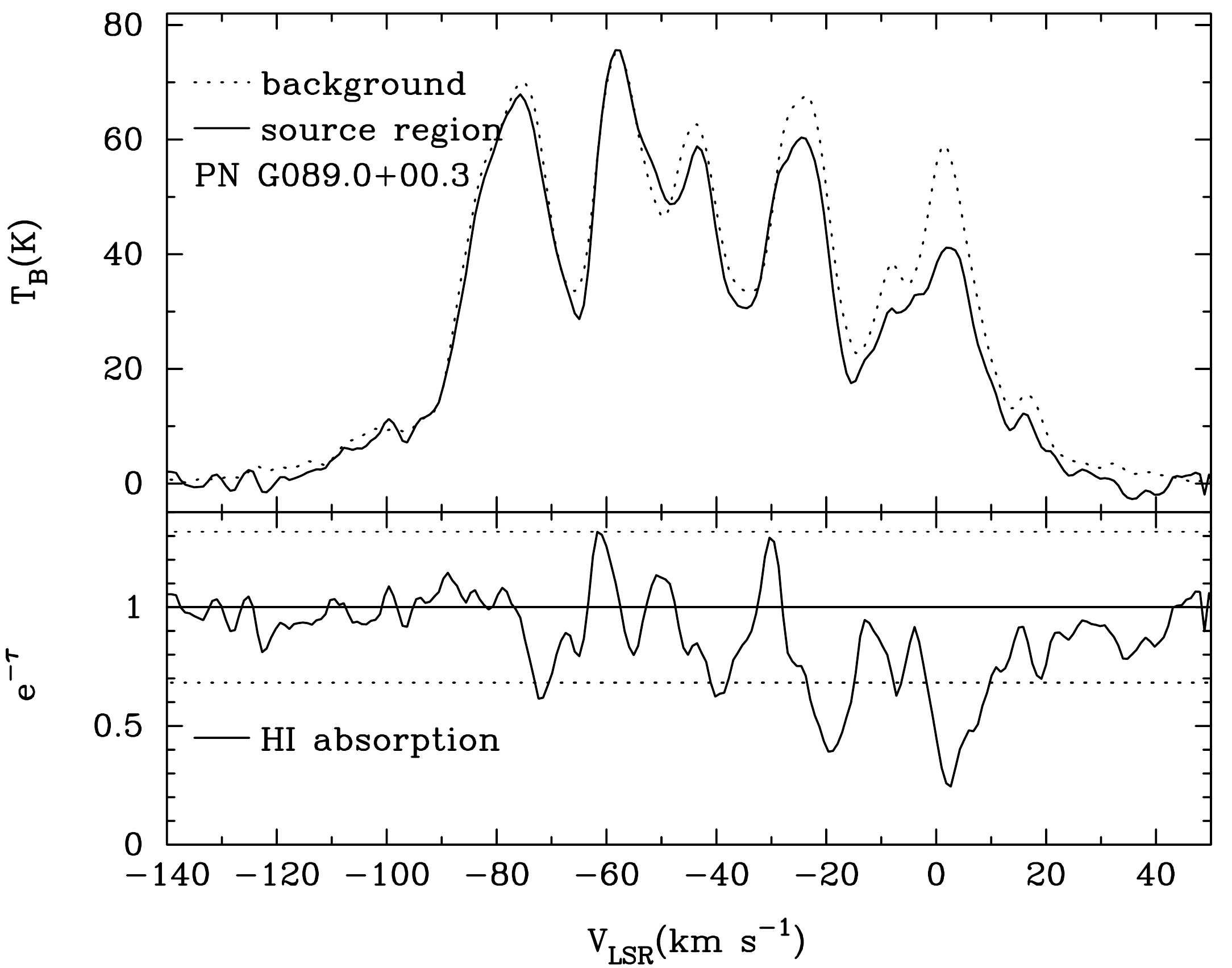}&
    \includegraphics[width = 0.3\textwidth]{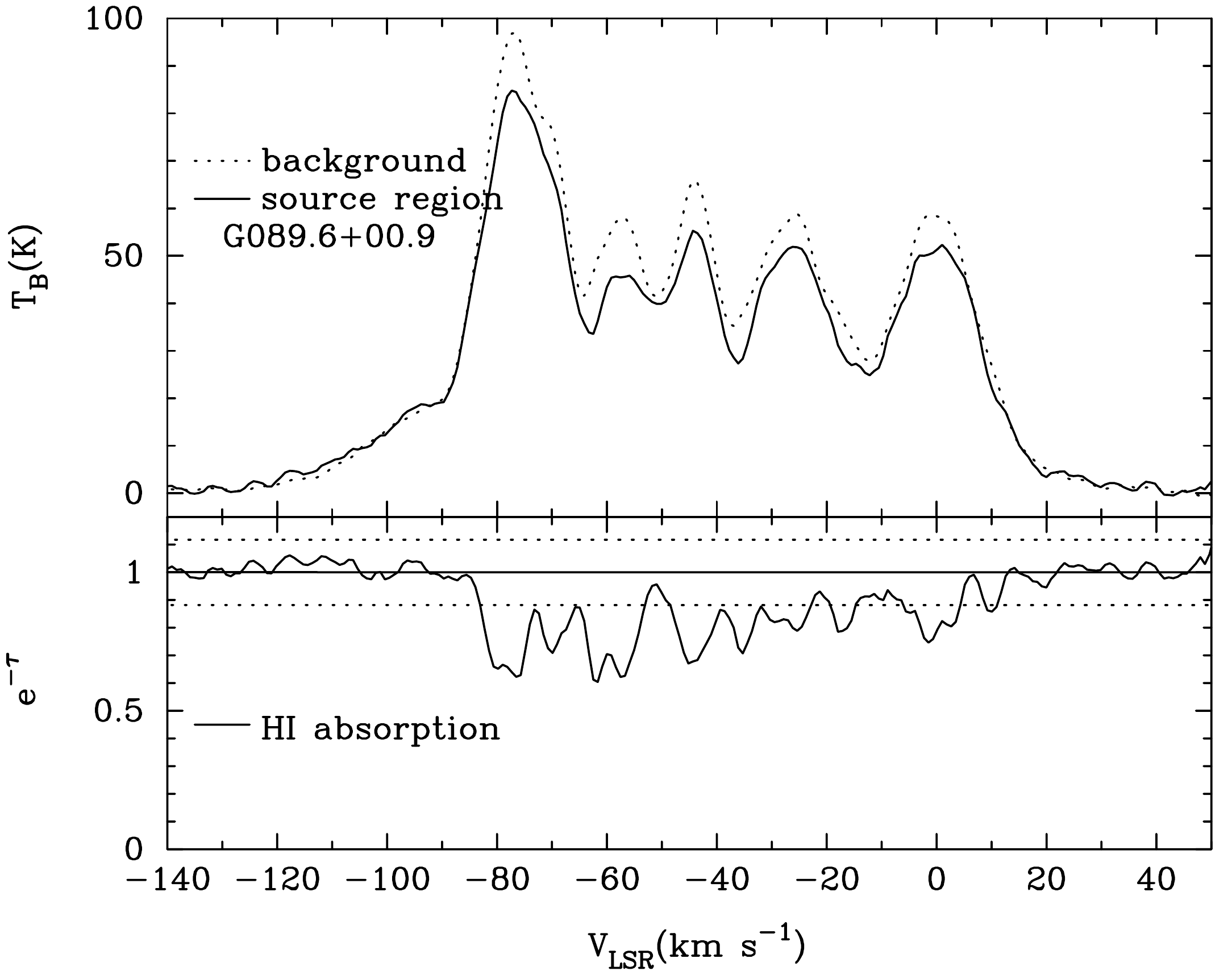}\\
   \end{tabular}
 \caption{1420\,MHz continuum image of PN G089.0$+$00.3 and its nearby background sources (top left), and the \HI~spectra of PN G089.0$+$00.3 (bottom left)
 , G088.8$+$00.9 (top right), and G089.6$+$00.9 (bottom right). 
 The  \HI~spectra of G088.46$+$00.0  is shown in appendix Fig.~\ref{fig21}.
 The map has superimposed contours (15,25,35\,K) of 1420\,MHz continuum emission.}
 \label{fig11}
\end{figure*}
\begin{figure*}
 \centering
 \begin{tabular}{cc}
    \includegraphics[width = 0.3\textwidth,height = 0.215\textwidth]{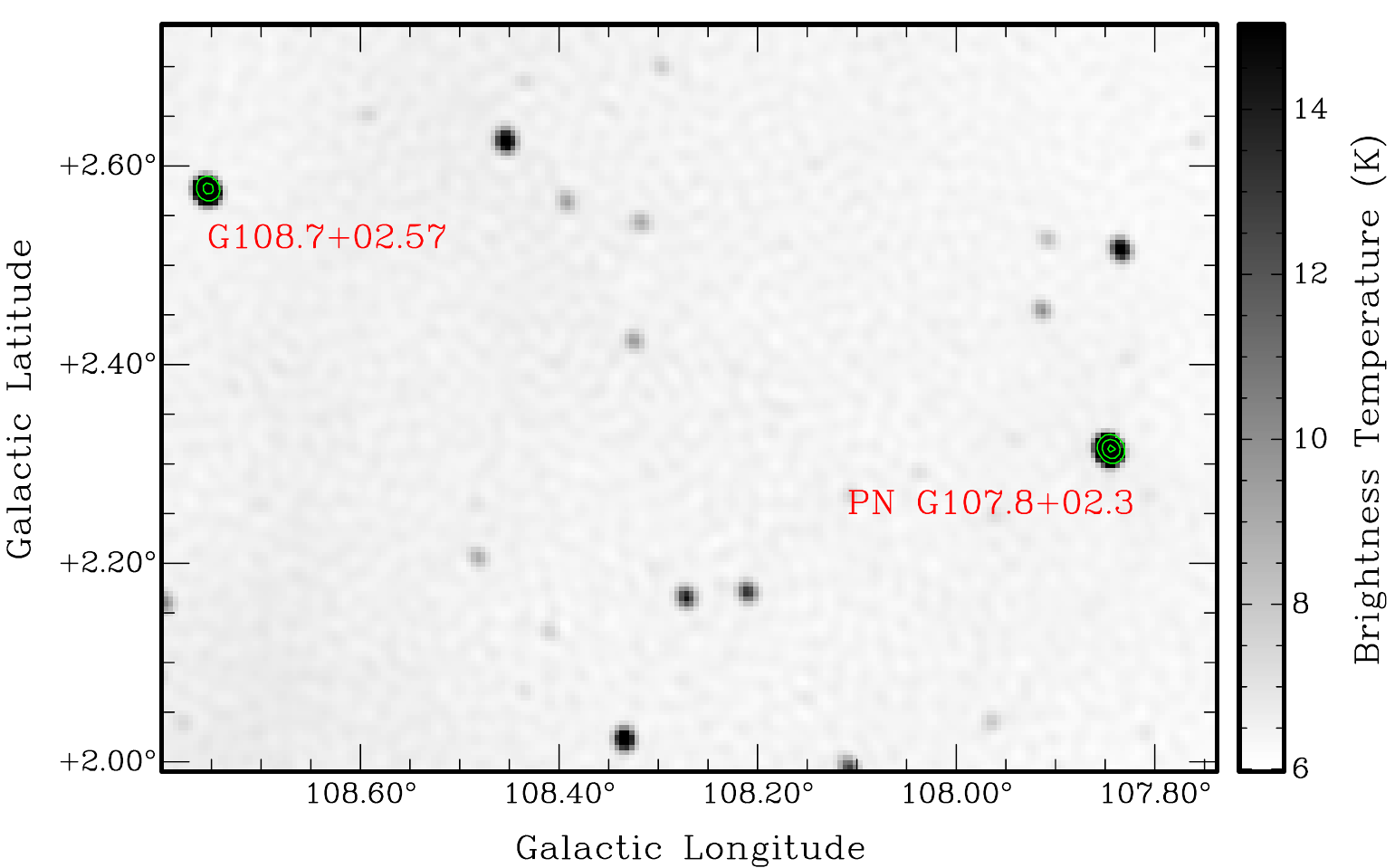}&
    \includegraphics[width = 0.31\textwidth]{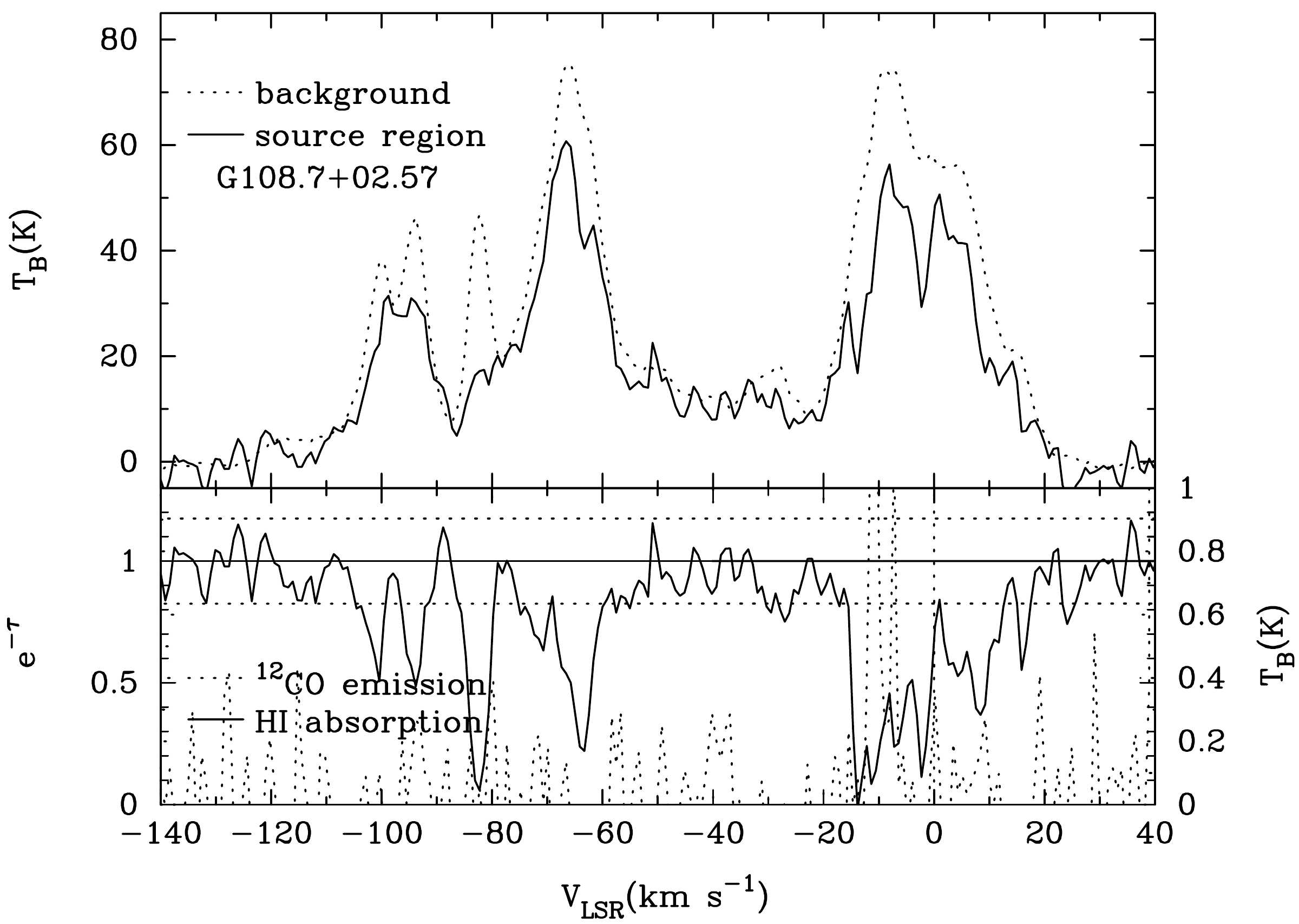}\\
    \includegraphics[width = 0.31\textwidth]{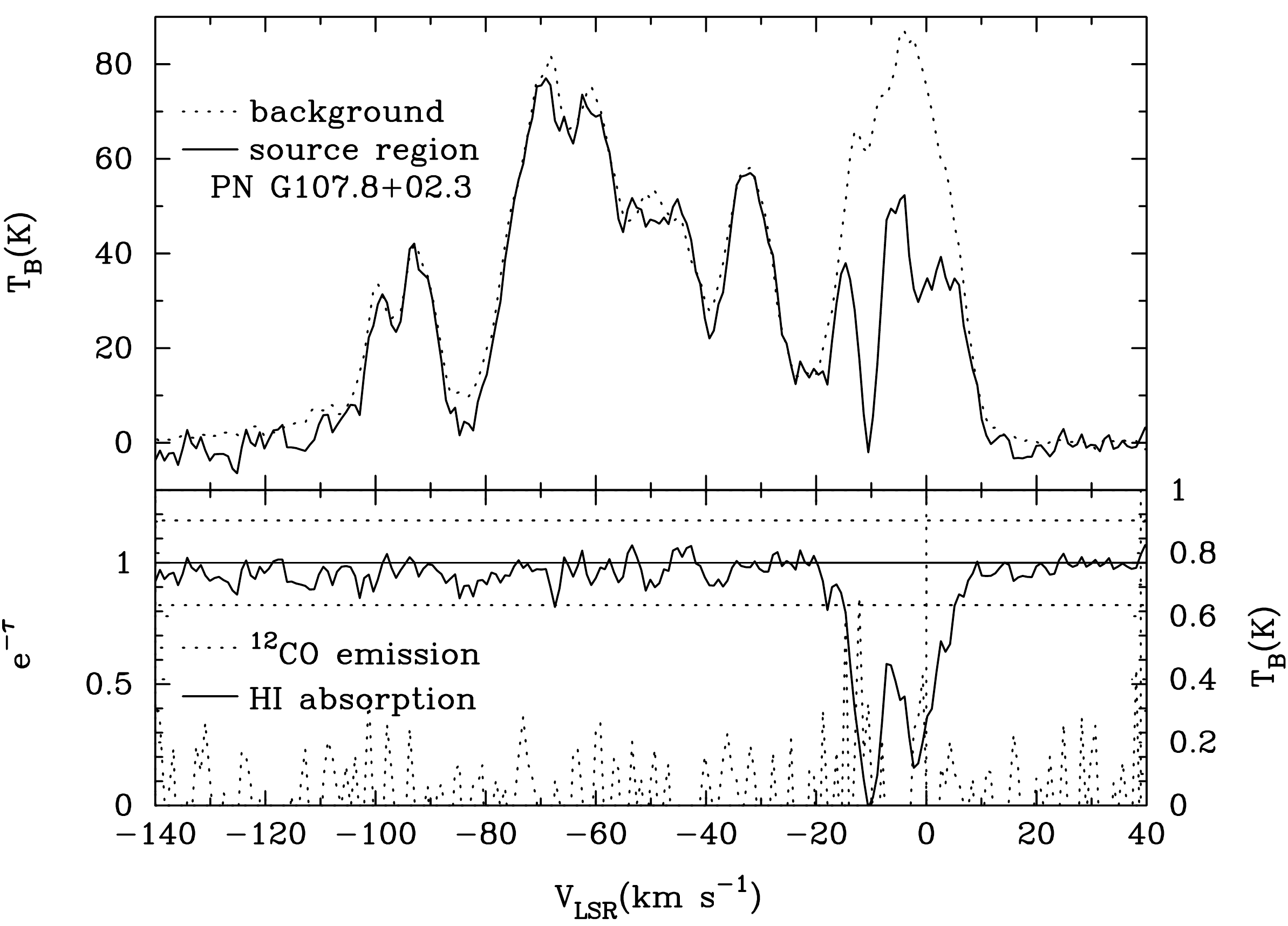}&\\
   \end{tabular}
 \caption{1420\,MHz continuum image of PN G107.8$+$02.3 and its nearby background sources(top right) ,
  and the \HI~spectra of PN G107.8$+$02.3 (bottom left), G108.7$+$02.57(top right).
  The map has superimposed contours (20,50,100\,K) of 1420\,MHz continuum emission.}
 \label{fig12}
\end{figure*}
\begin{figure*}
 \centering
 \begin{tabular}{cc}
    \includegraphics[width = 0.3\textwidth,height = 0.215\textwidth]{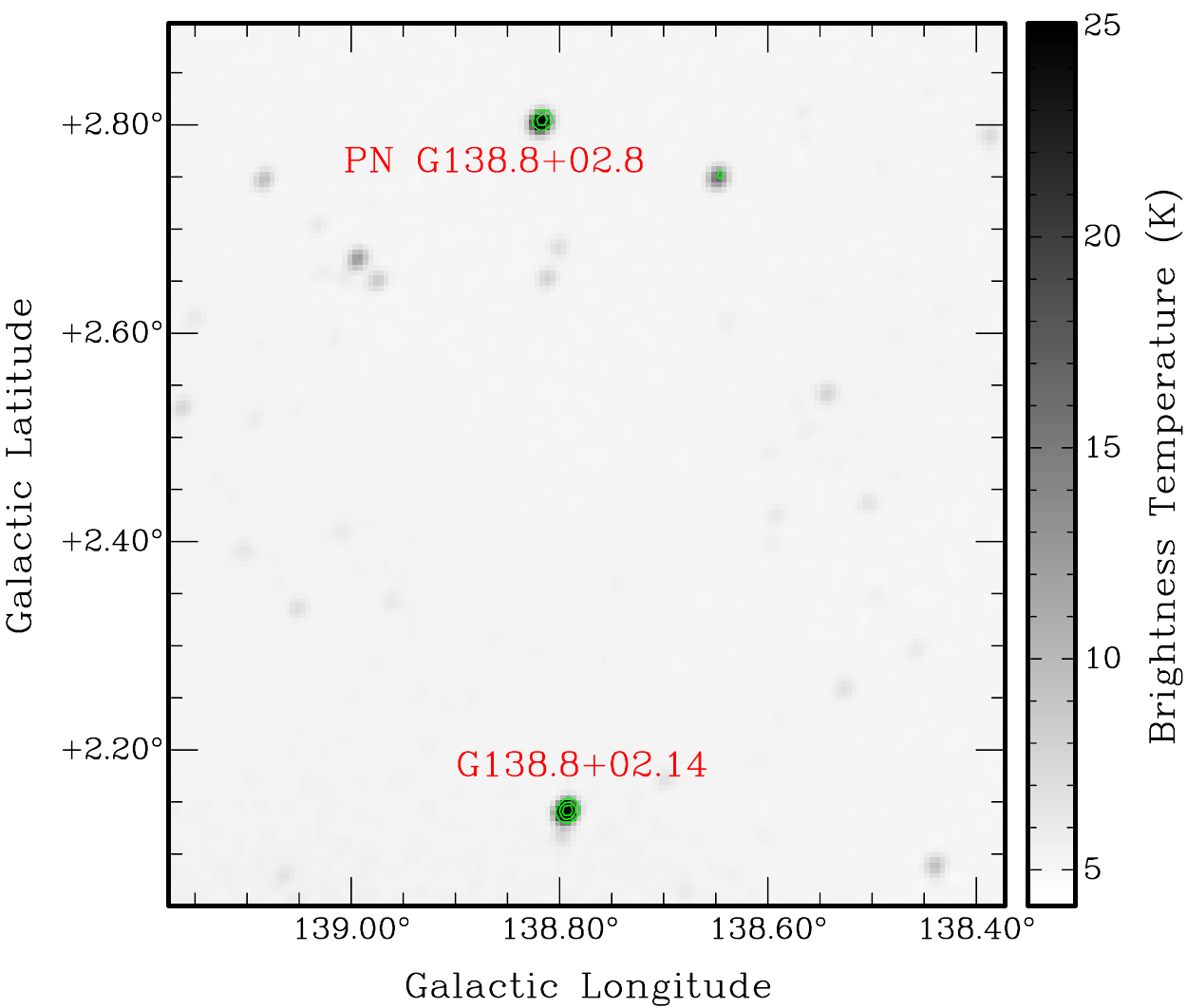}&
    \includegraphics[width = 0.31\textwidth]{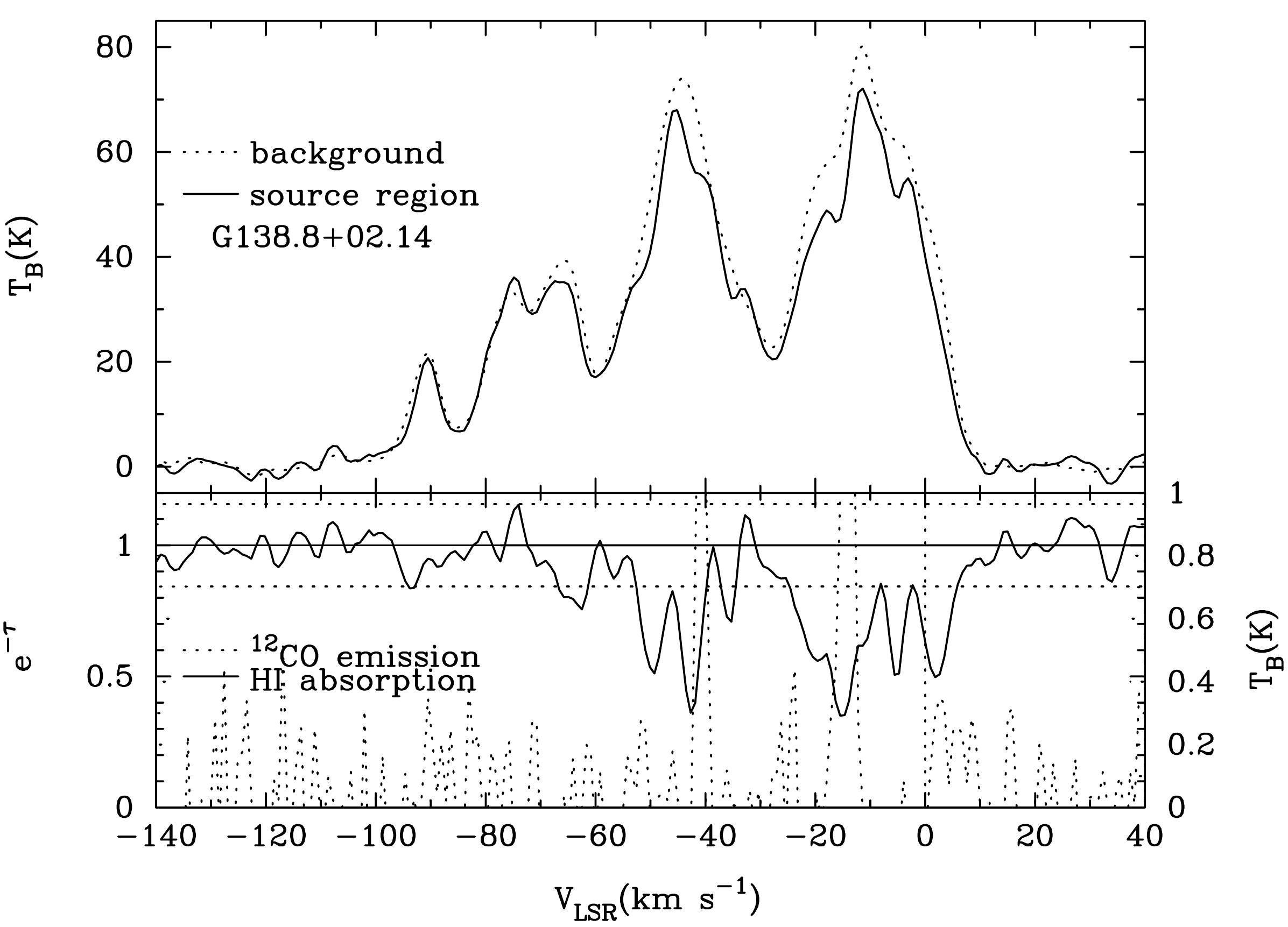}\\
    \includegraphics[width = 0.31\textwidth]{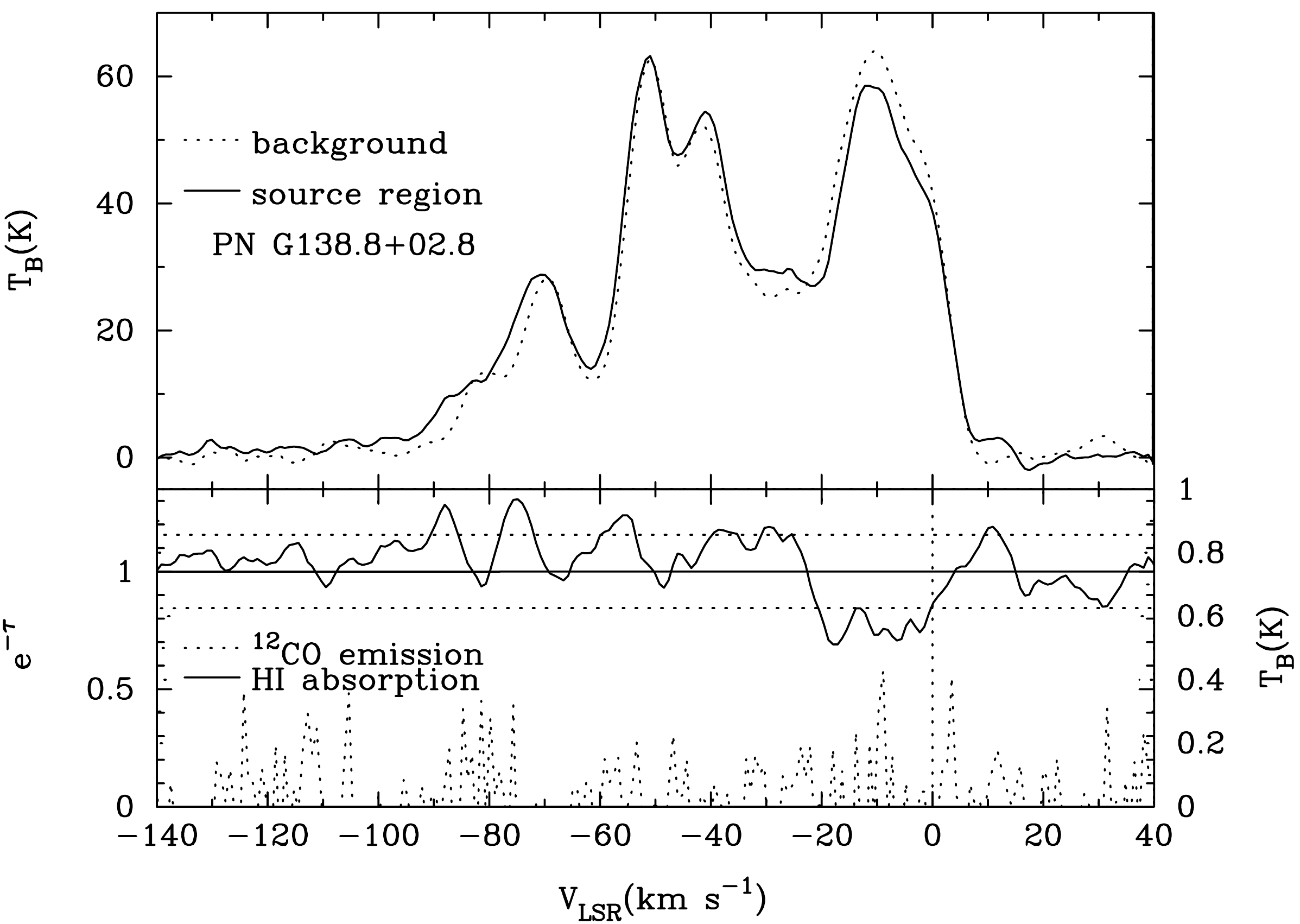}&
     \\
   \end{tabular}
 \caption{1420\,MHz continuum image of PN G138.8$+$02.8 and G138.8$+$02.4 (top left),
 and the \HI~spectra of PN G138.8$+$02.8 (bottom left), G138.8$+$02.14 (top right).
 The map has superimposed contours (15,25,35\,K) of 1420\,MHz continuum emission.}
 \label{fig13}
\end{figure*}
\begin{figure*}
 \centering
 \begin{tabular}{cc}
    \includegraphics[width = 0.29\textwidth,height = 0.24\textwidth]{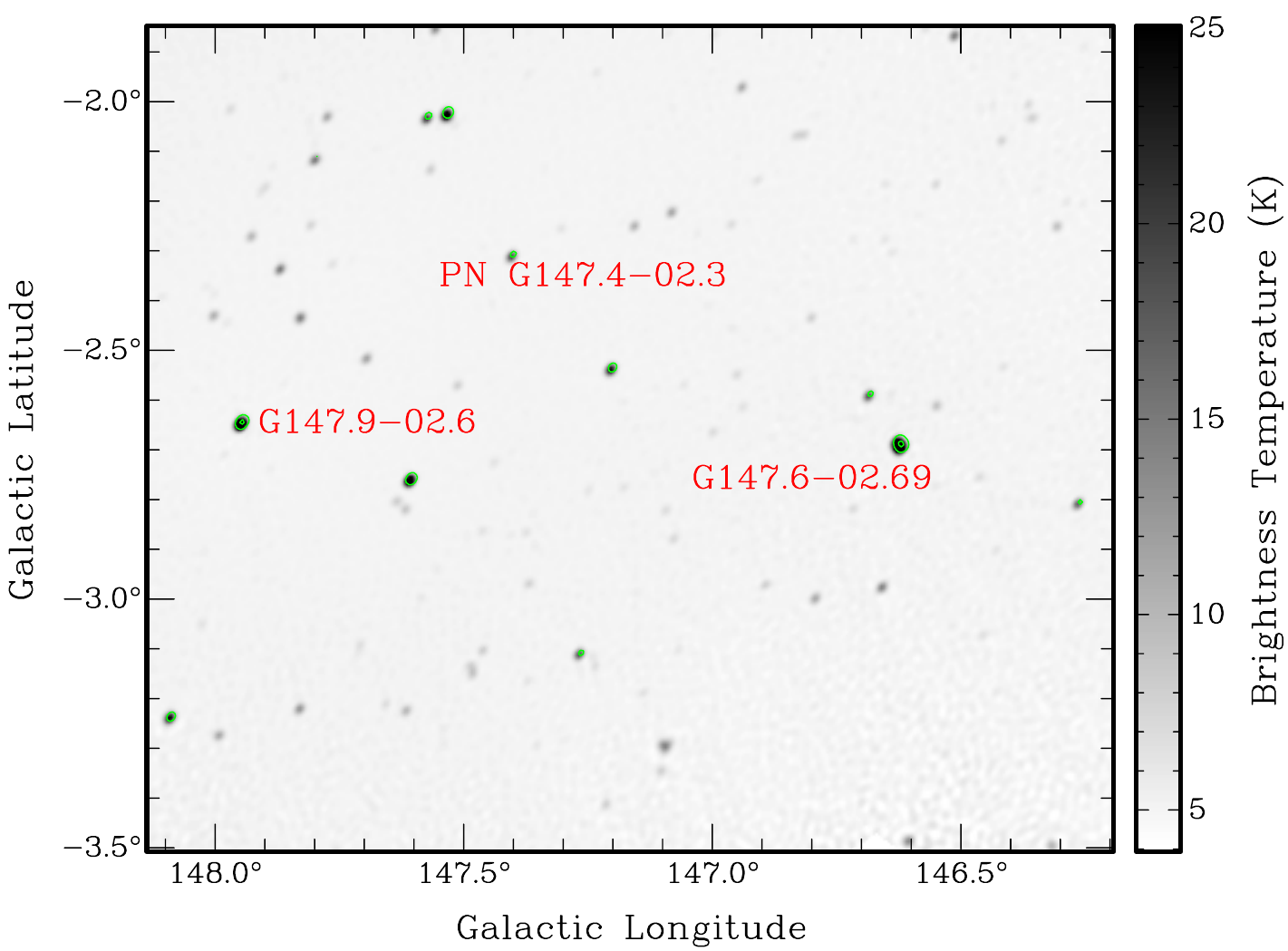}&
    \includegraphics[width = 0.3\textwidth]{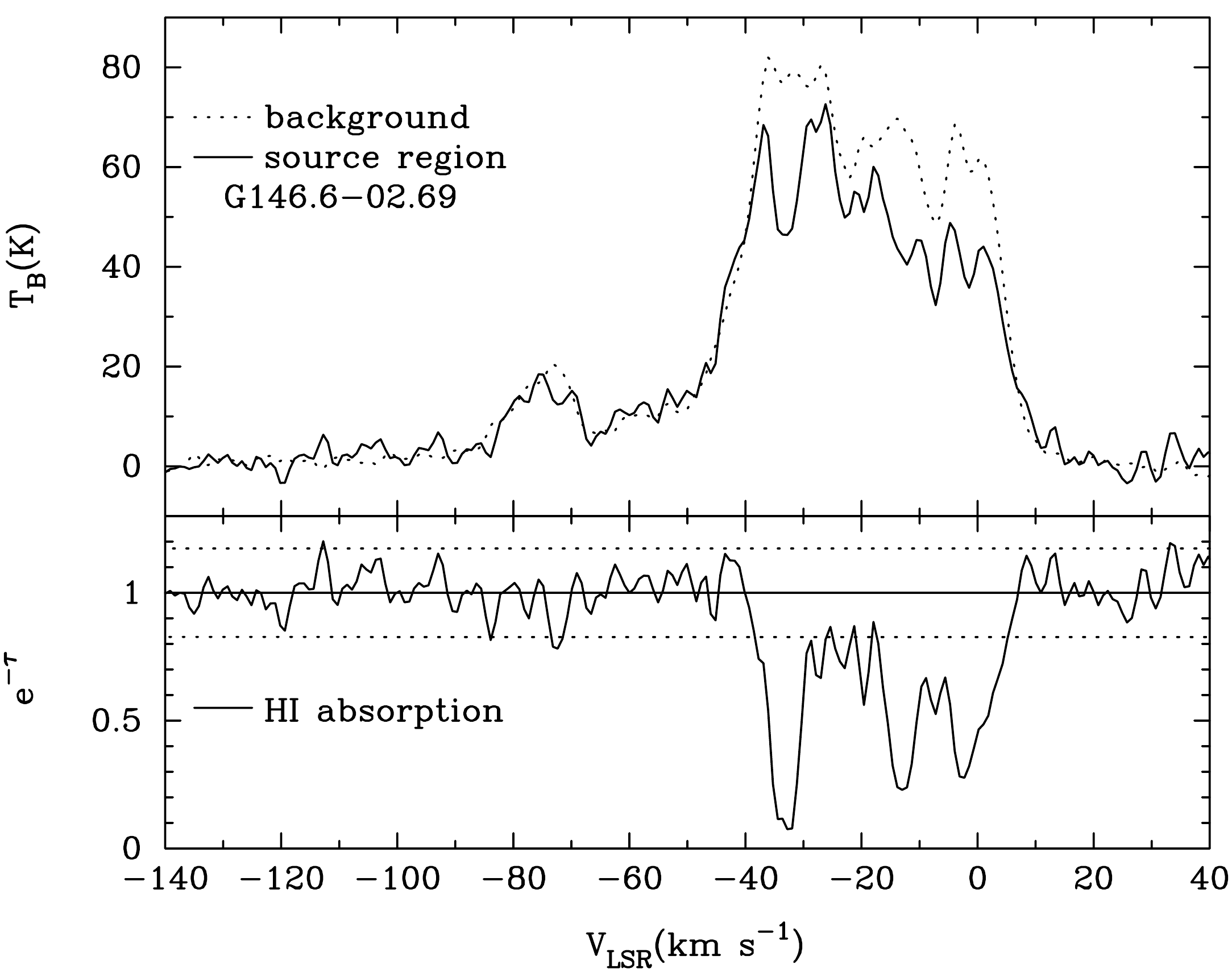}\\
    \includegraphics[width = 0.3\textwidth]{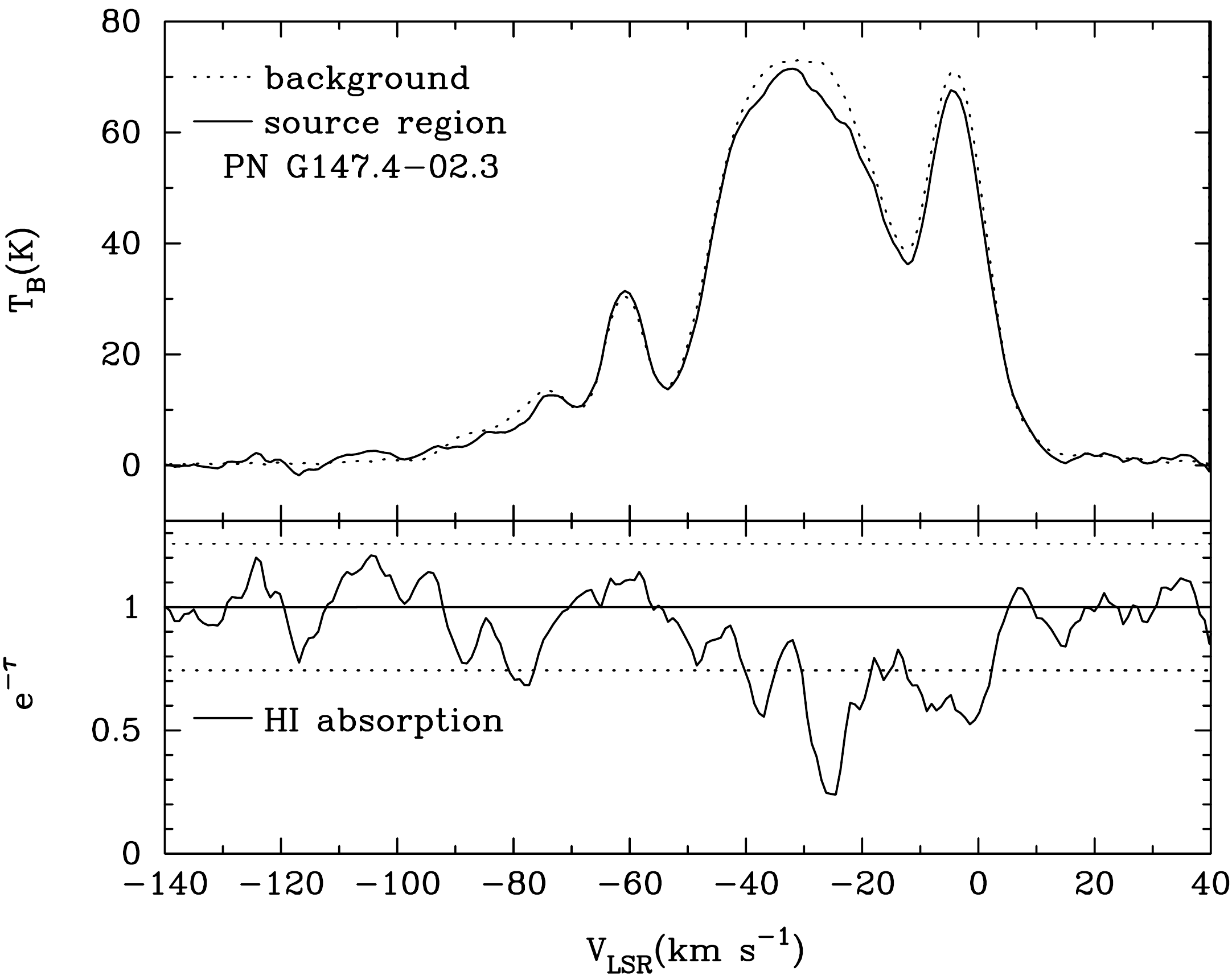}&
    \includegraphics[width = 0.3\textwidth]{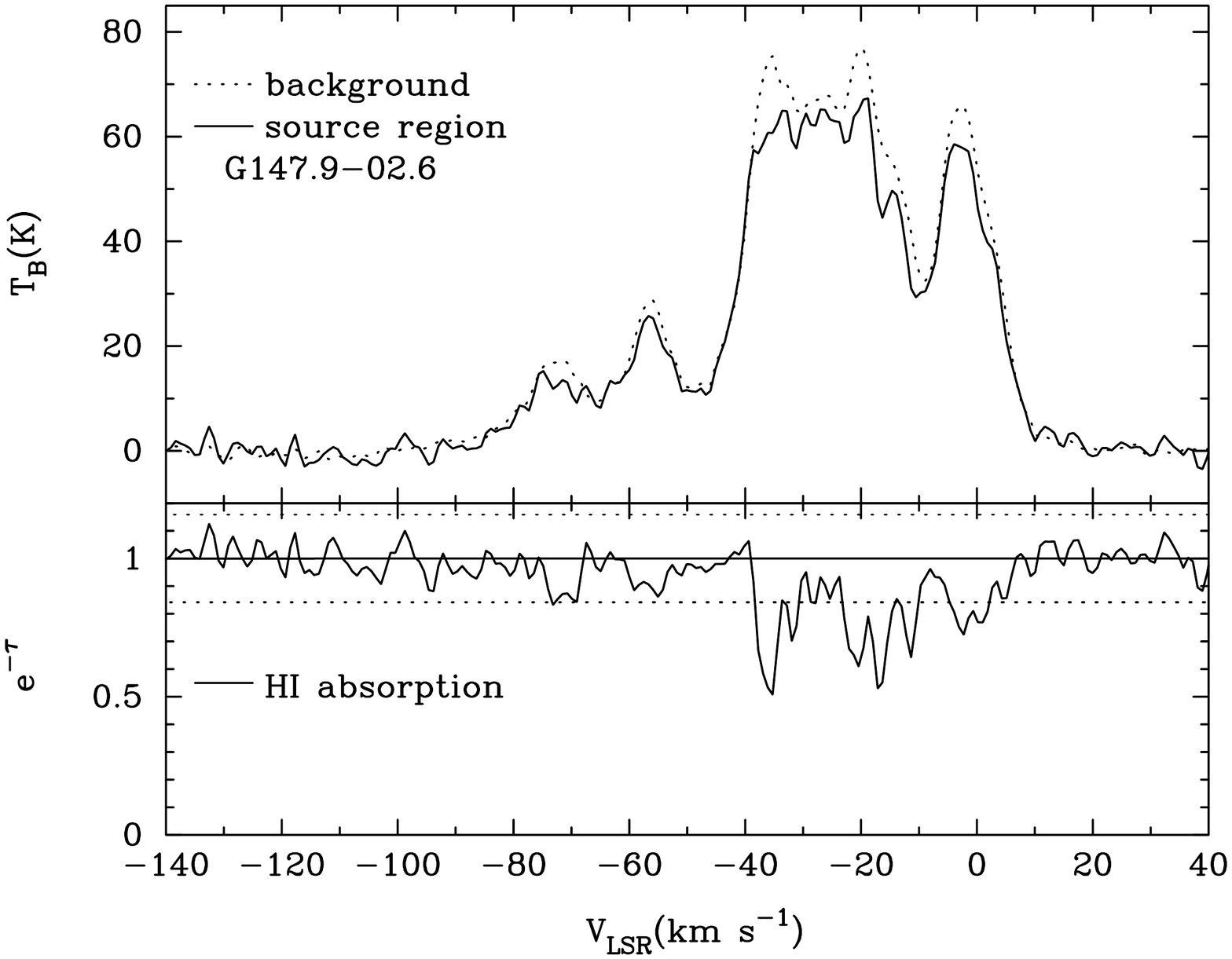}\\
   \end{tabular}
 \caption{1420\,MHz continuum image of PN G147.4$-$02.3, and its background sources (top left),
 and the \HI~spectra of PN G147.4$-$02.3 (bottom left), G146.6$-$2.69 (top right) and G147.9$-$02.6 (bottom right).
 The map has superimposed contours (15,55\,K) of 1420\,MHz continuum emission.}
 \label{fig14}
\end{figure*}
\begin{figure*}
 \centering
 \begin{tabular}{cc}
    \includegraphics[width = 0.29\textwidth,height = 0.24\textwidth]{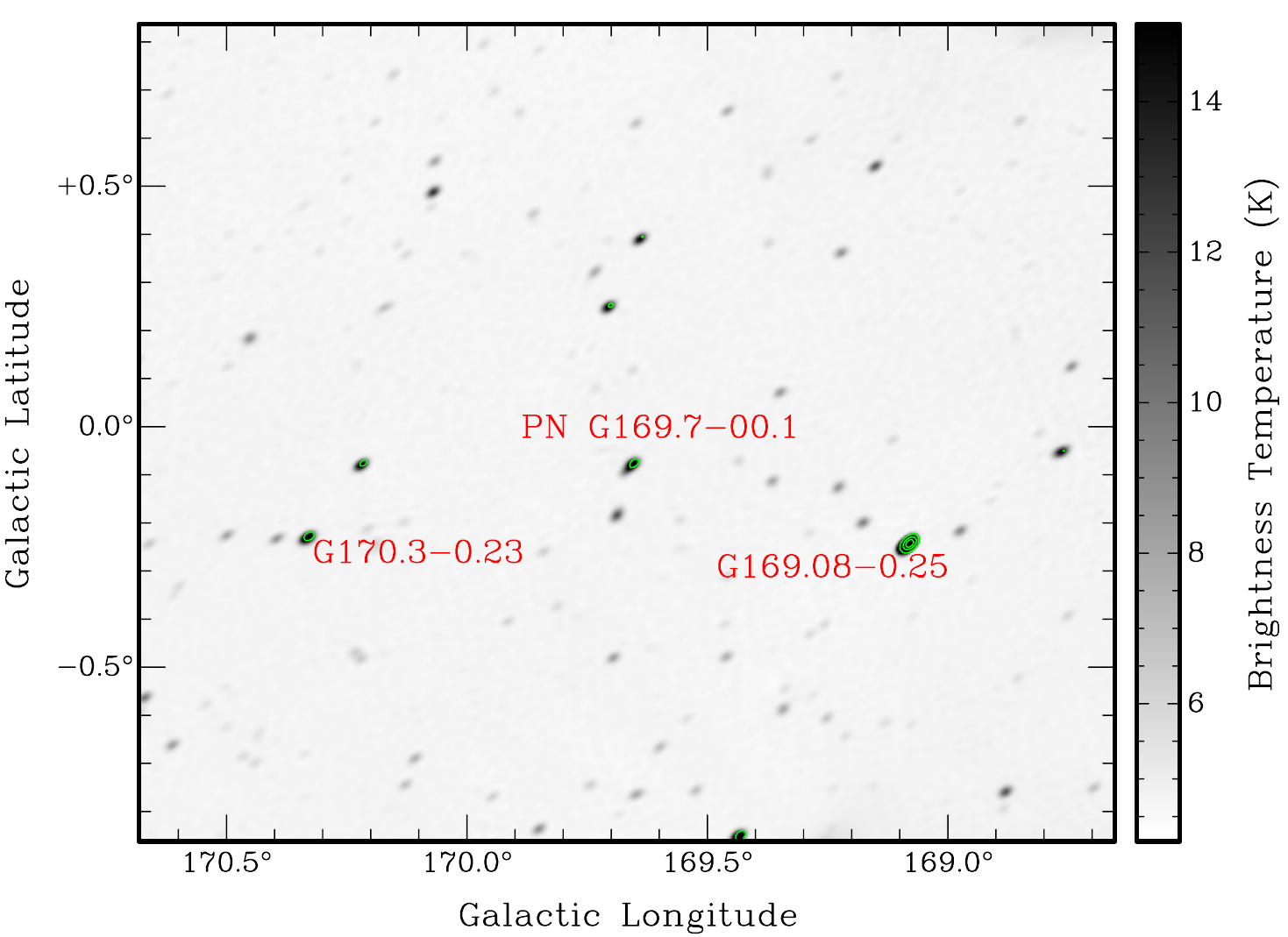}&
    \includegraphics[width = 0.3\textwidth]{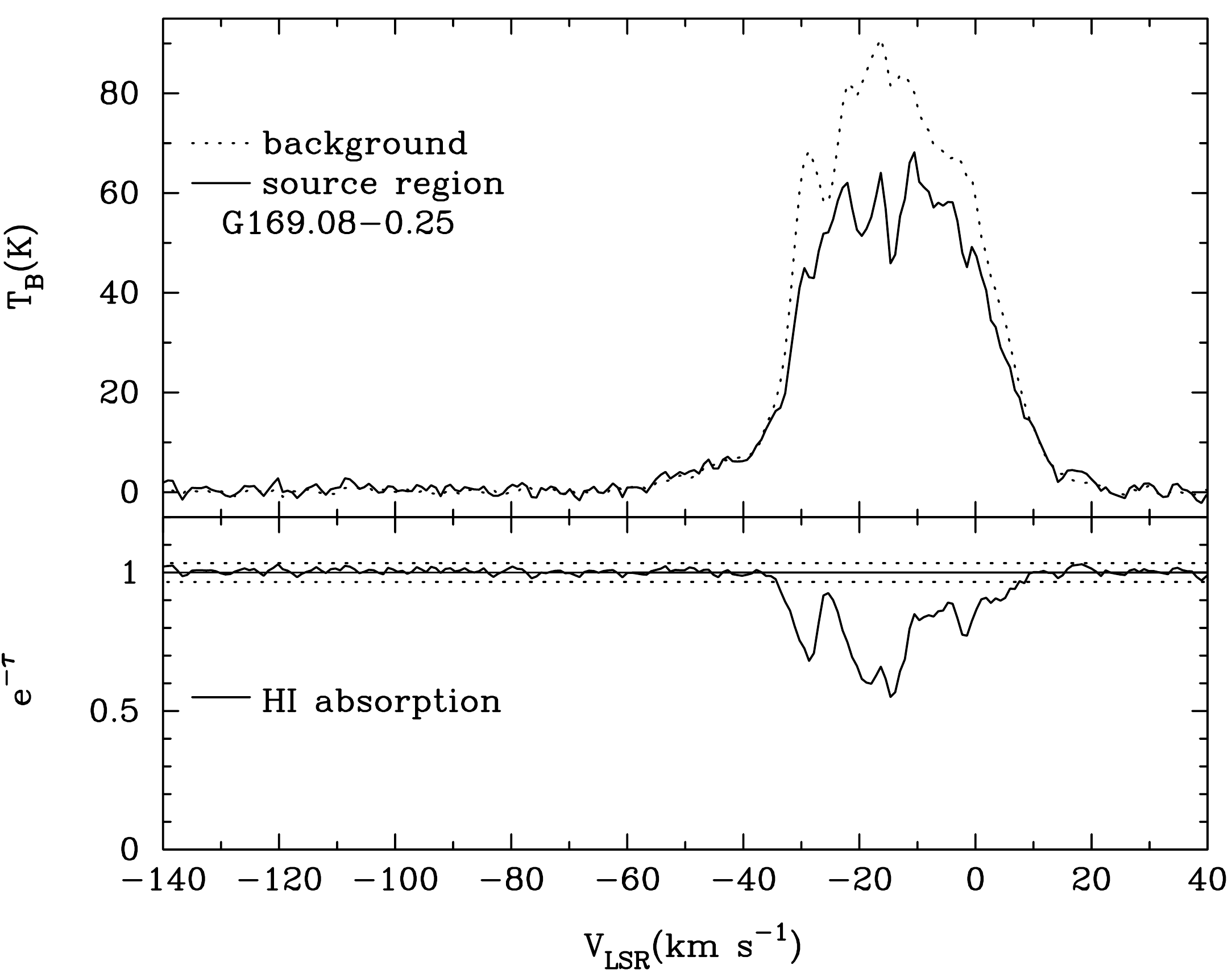}\\
    \includegraphics[width = 0.3\textwidth]{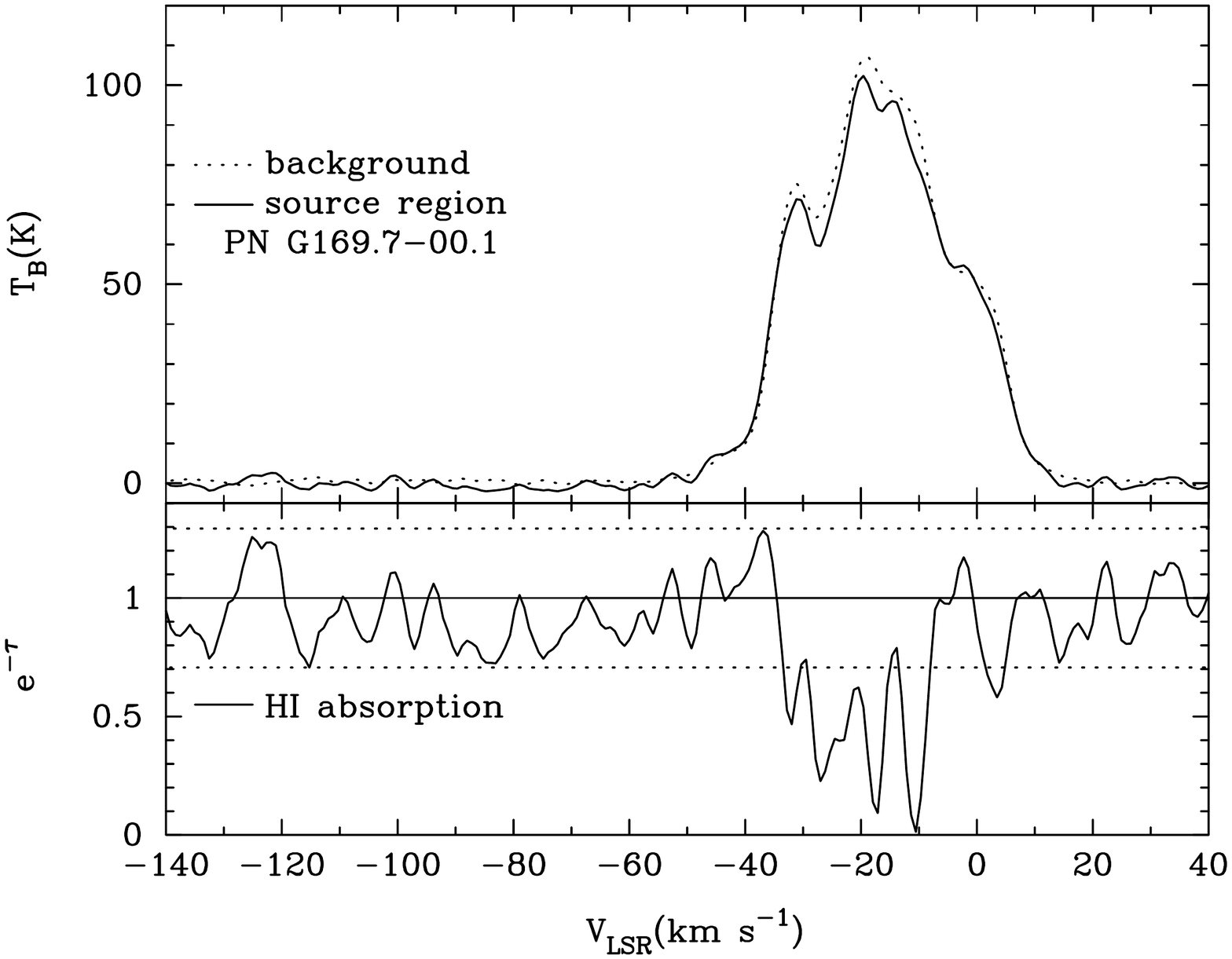}&
    \includegraphics[width = 0.3\textwidth]{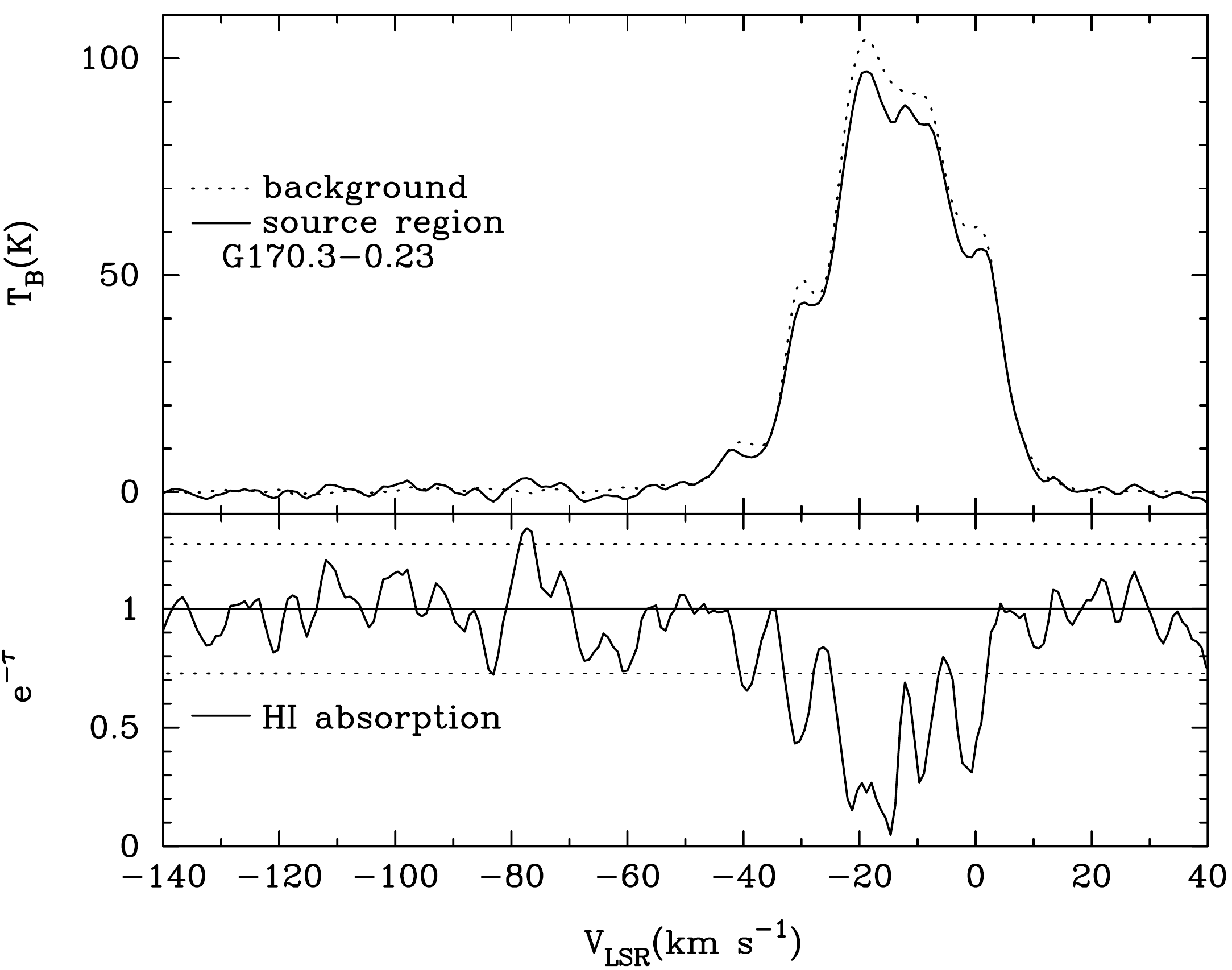}\\
   \end{tabular}
 \caption{1420\,MHz continuum image of PN G169.7$-$00.1 and its background sources (top left),
 and the \HI~spectra of PN G169$-$00.1 (bottom left), G169.08$-$0.25 (top right) and G170.3$-$0.23 (bottom right).
 The map has superimposed contours (15,50,100\,K) of 1420\,MHz continuum emission.}
 \label{fig15}
\end{figure*}
\begin{figure*}
 \centering
 \begin{tabular}{cc}
    \includegraphics[width = 0.29\textwidth,height = 0.24\textwidth]{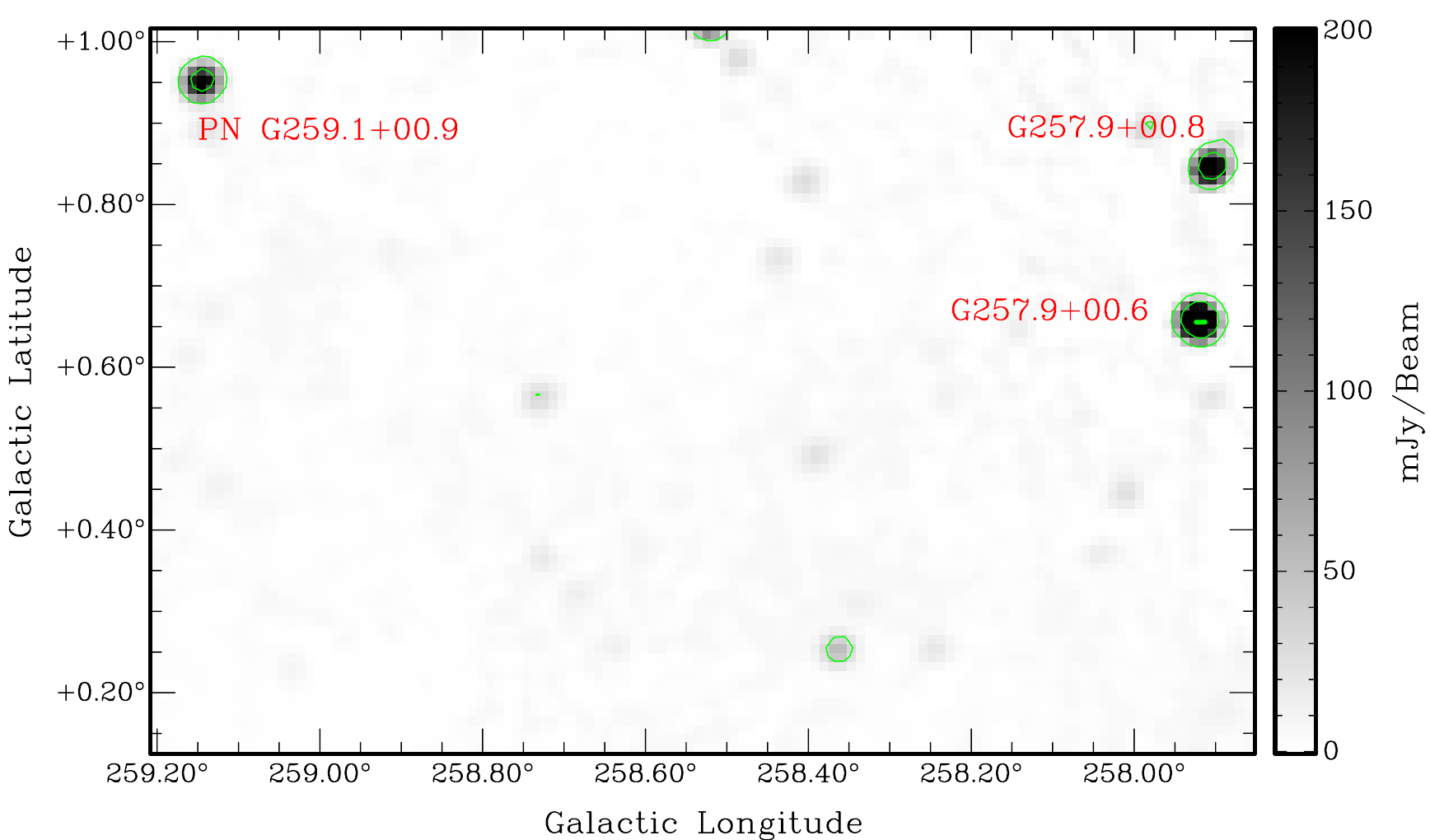}&
    \includegraphics[width = 0.3\textwidth]{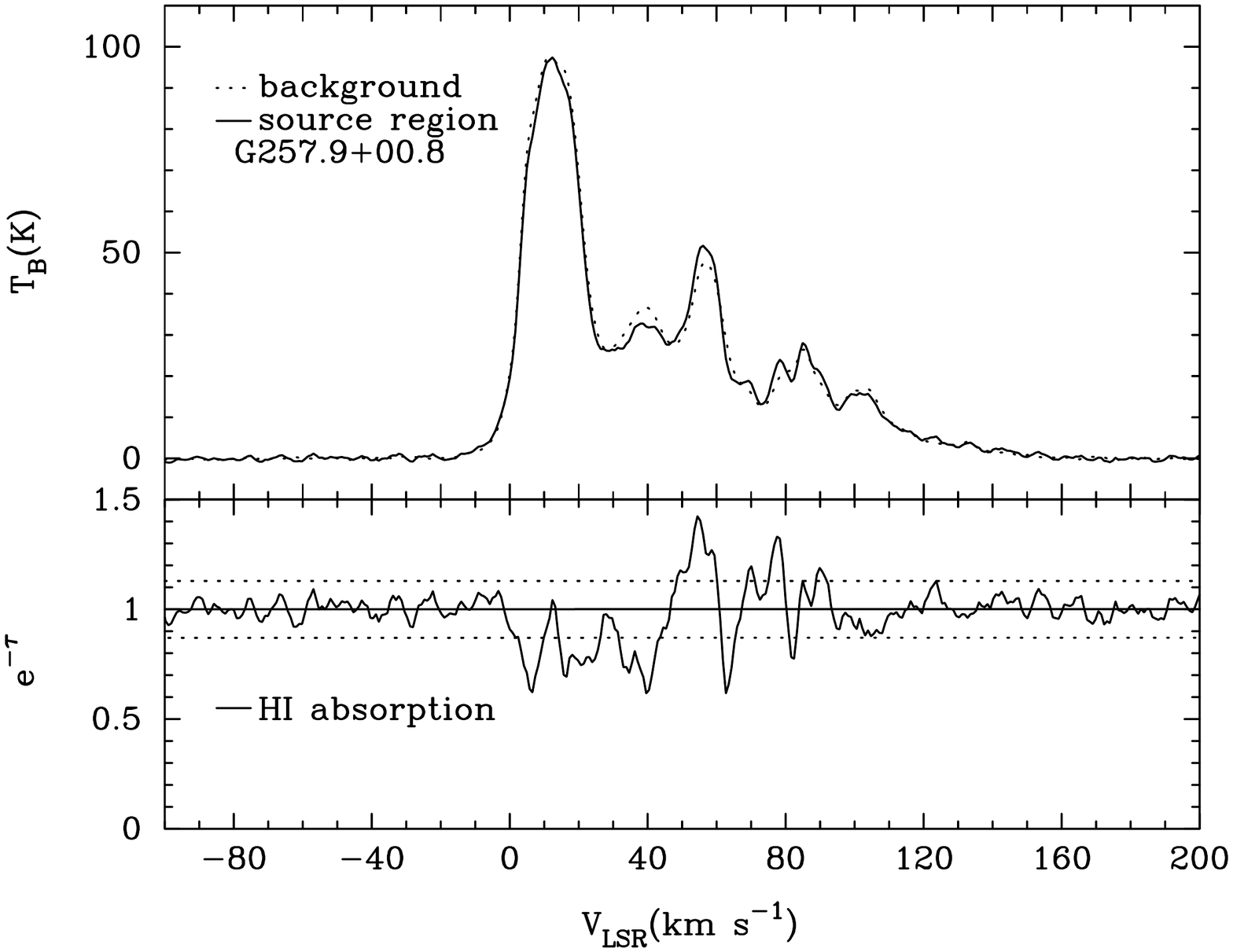}\\
    \includegraphics[width = 0.3\textwidth]{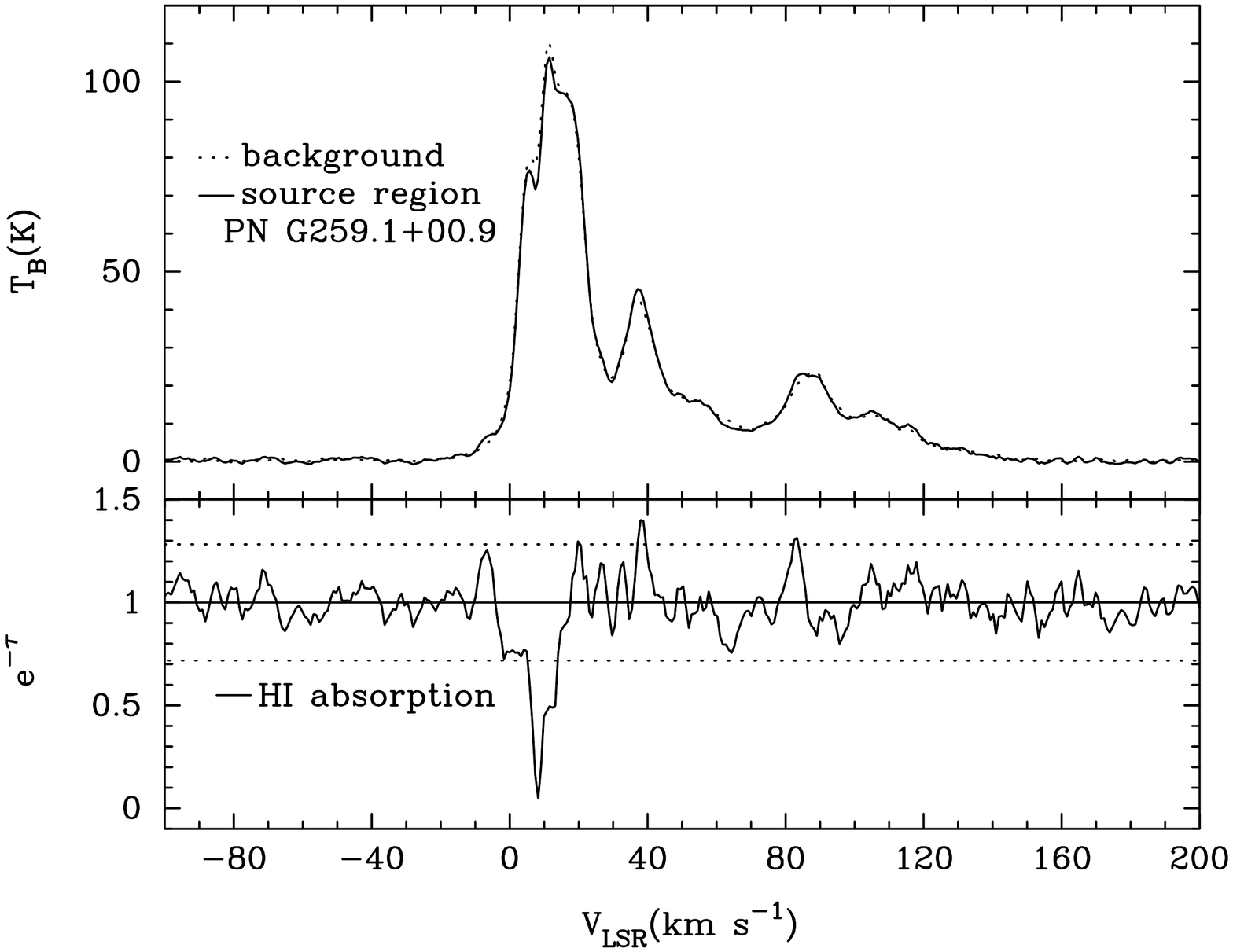}&
    \includegraphics[width = 0.3\textwidth]{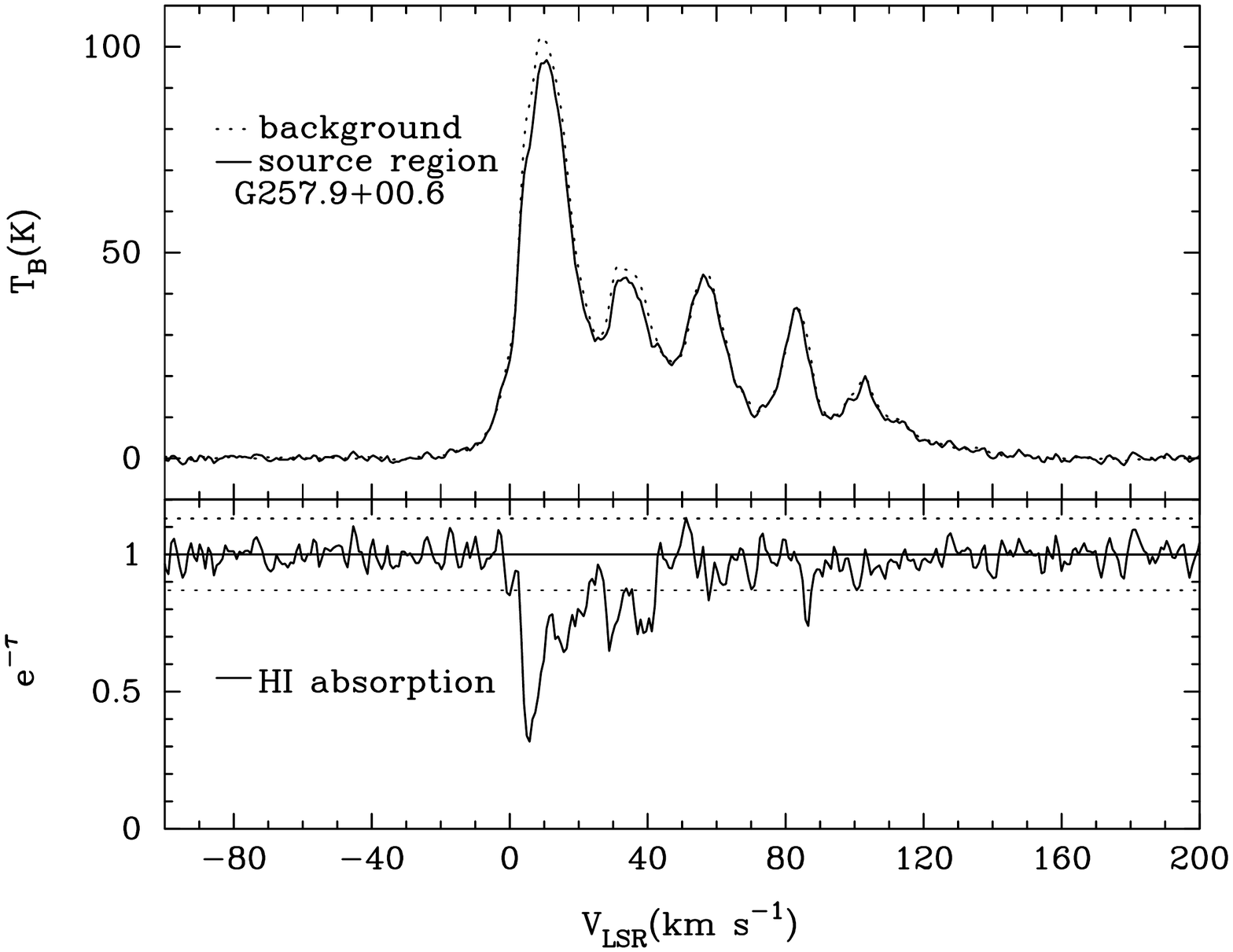}\\
   \end{tabular}
 \caption{1420\,MHz continuum image of PN G259.1$+$00.9, and its nearby background sources (top left),
 and the \HI~spectra of PN G259.1$+$00.9 (bottom left), G257.9$+$00.8 (top right) and G257.9$+$00.6 (bottom right).
 The map has superimposed contours (0.03,0.15,0.5\,Jy) of 1420\,MHz continuum emission.}
 \label{fig16}
\end{figure*}
\begin{figure*}[!hbt]
 \centering
 \begin{tabular}{cc}
    \includegraphics[width = 0.29\textwidth,height = 0.24\textwidth]{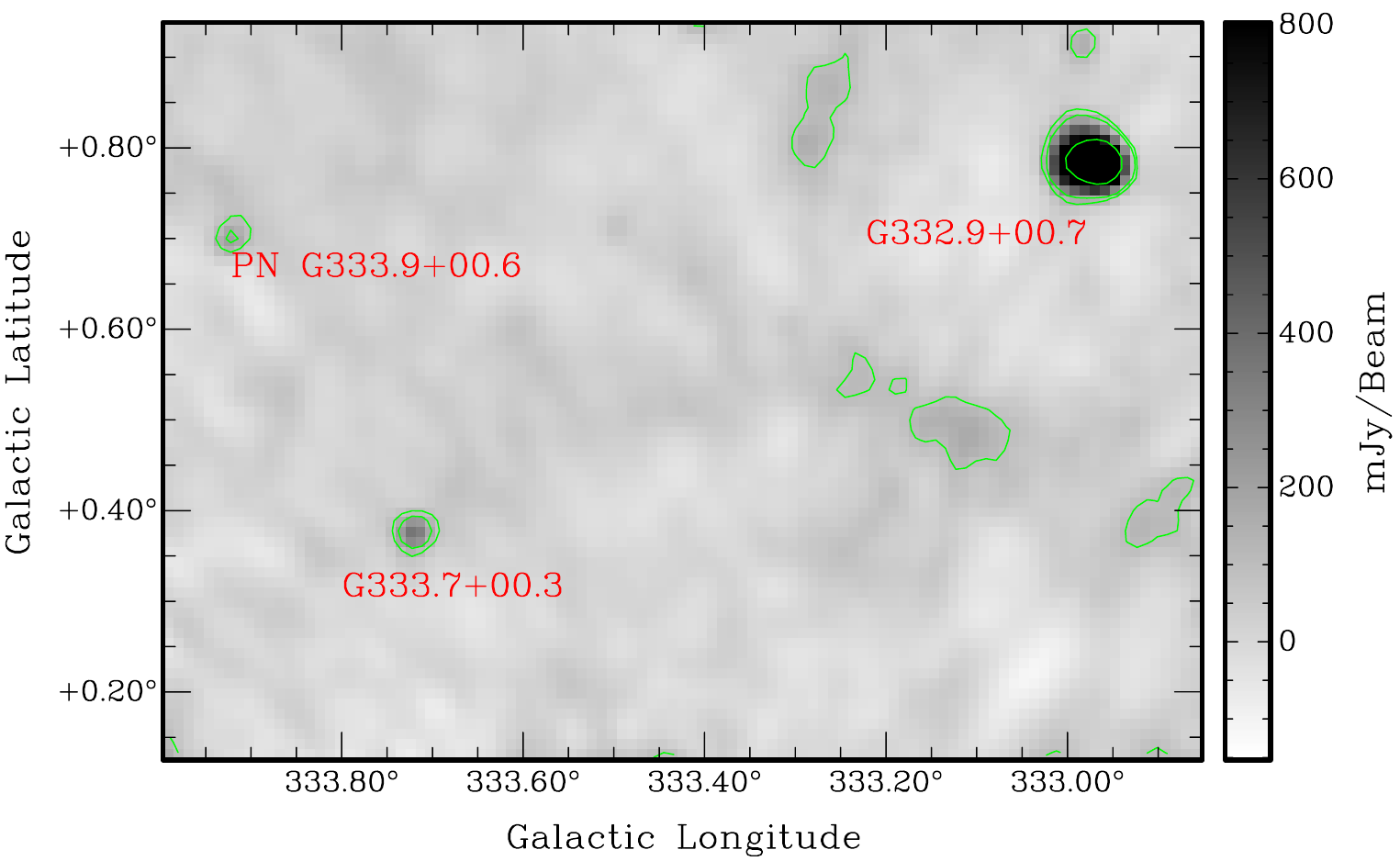}&
    \includegraphics[width = 0.3\textwidth]{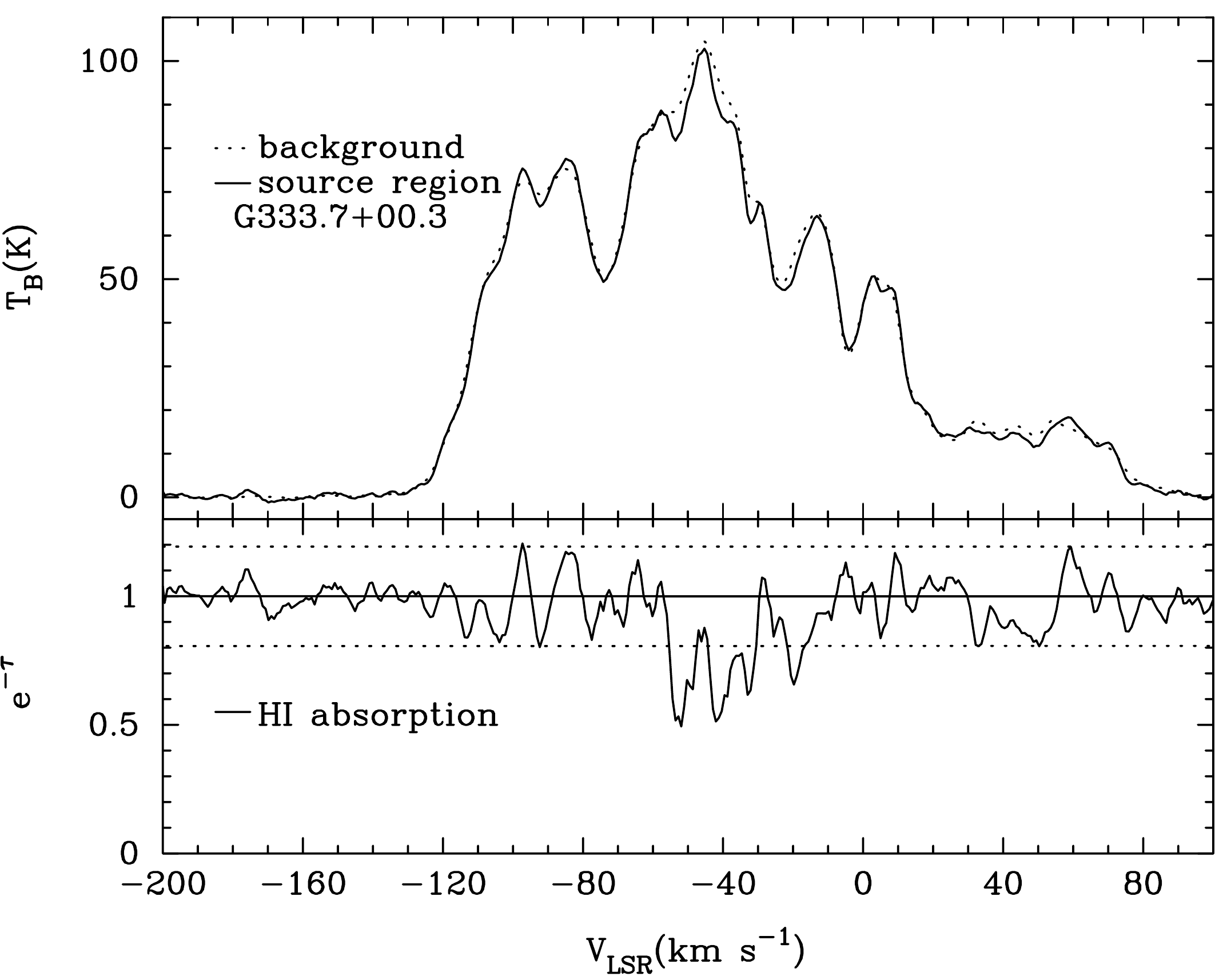}\\
    \includegraphics[width = 0.3\textwidth]{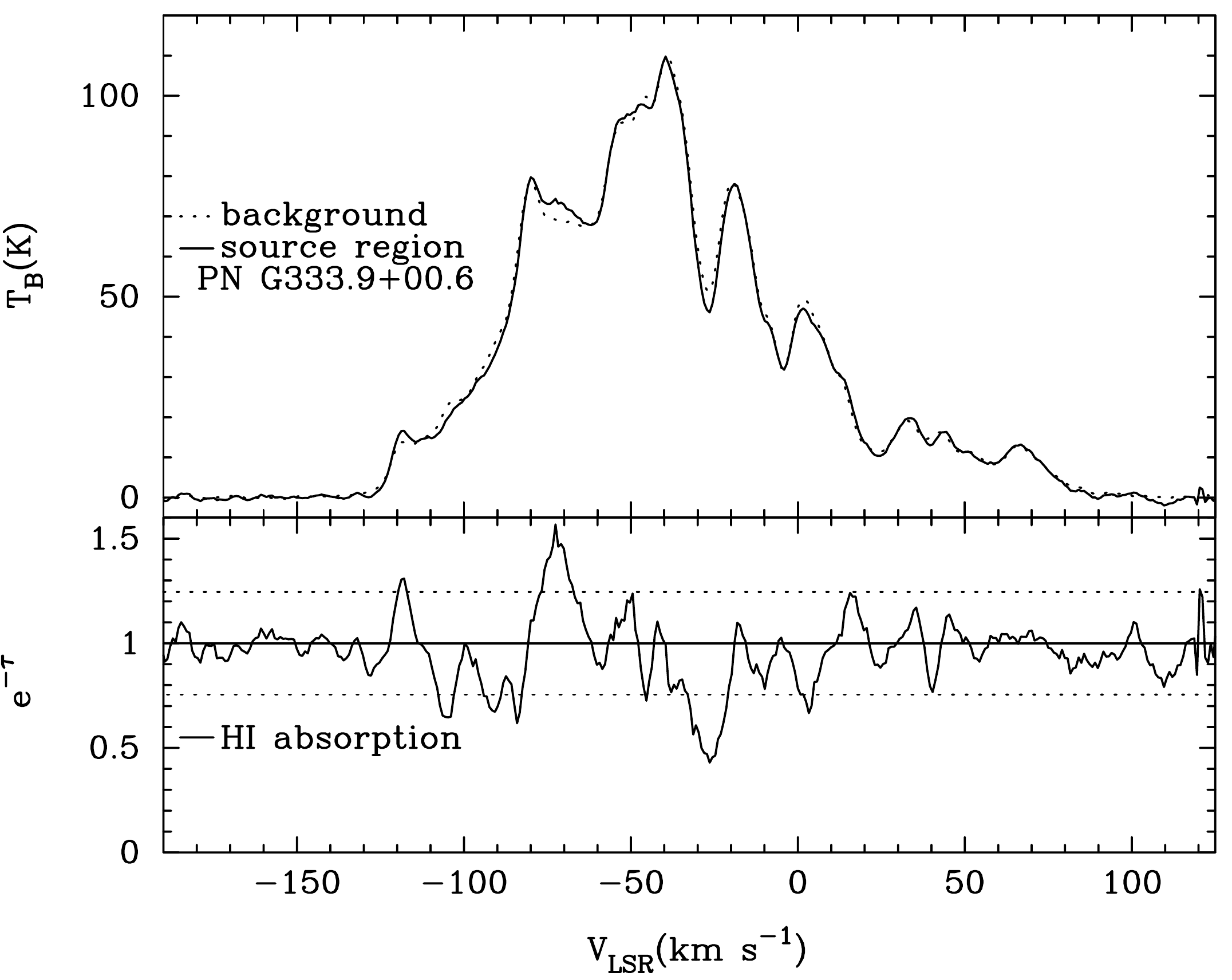}&
    \includegraphics[width = 0.3\textwidth]{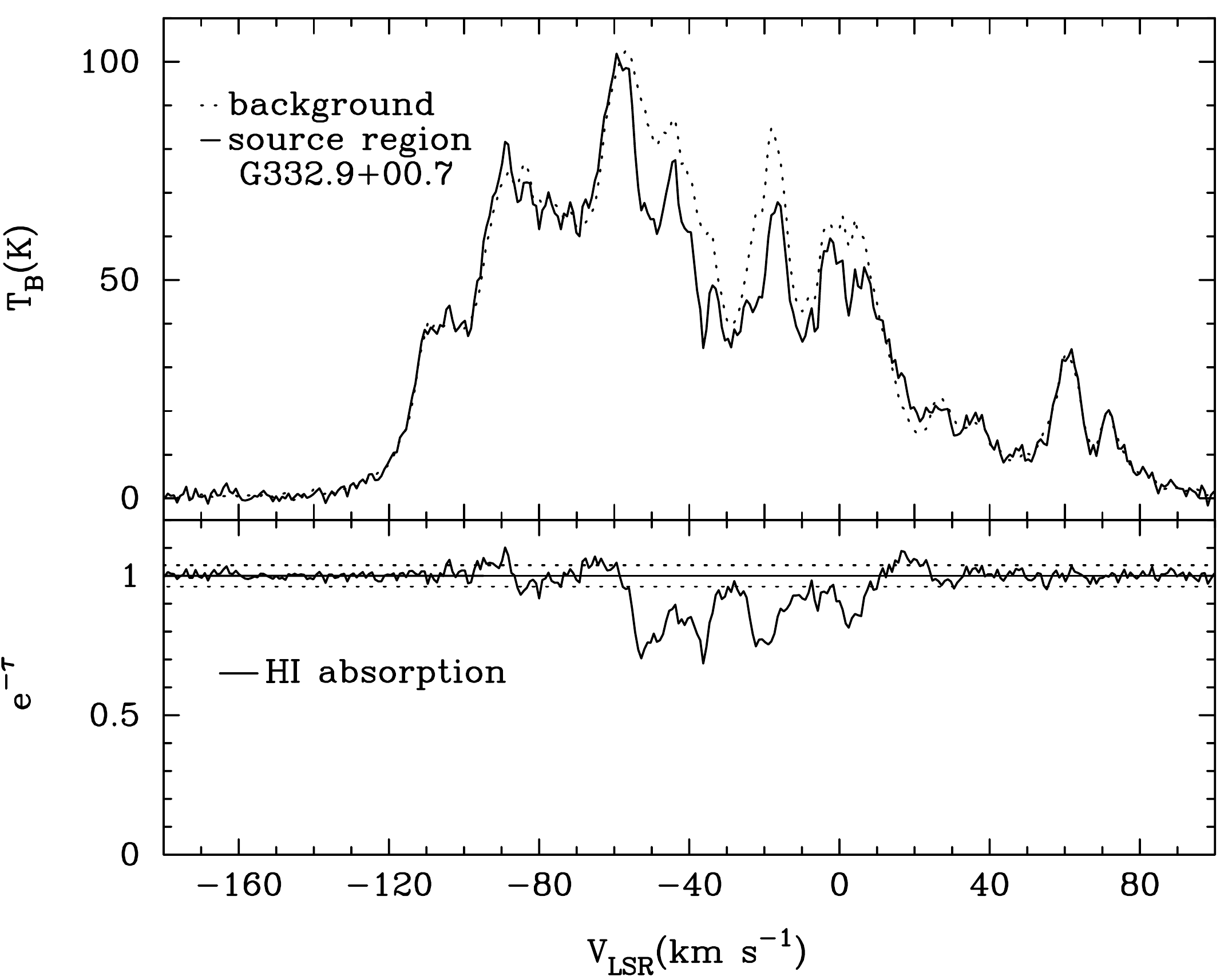}\\
   \end{tabular}
 \caption{1420\,MHz continuum image of PN G333.9$+$00.6, and its background sources (top left),
 and the \HI~spectra of PN G333.9$+$00.6 (bottom left), G333.7$+$00.3 (top right) and G332.9$+$00.7 (bottom right). 
 The map has superimposed contours (0.1,0.18,1.0\,Jy) of 1420\,MHz continuum emission.}
 \label{fig17}
 \end{figure*}
 
\clearpage
\begin{figure*}
 \centering
 \begin{tabular}{cc}
    \includegraphics[width = 0.29\textwidth,height = 0.24\textwidth]{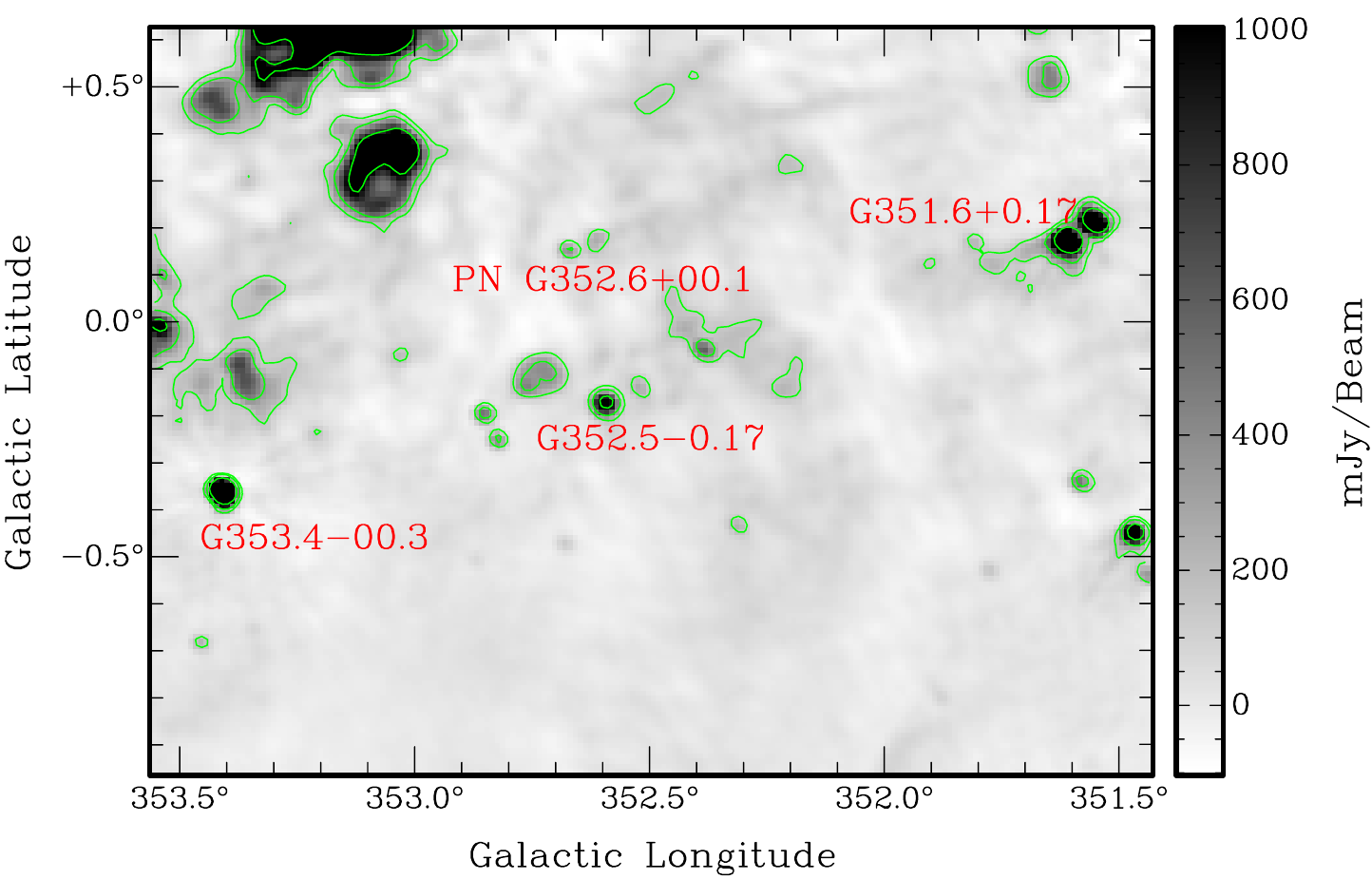}&
    \includegraphics[width = 0.3\textwidth]{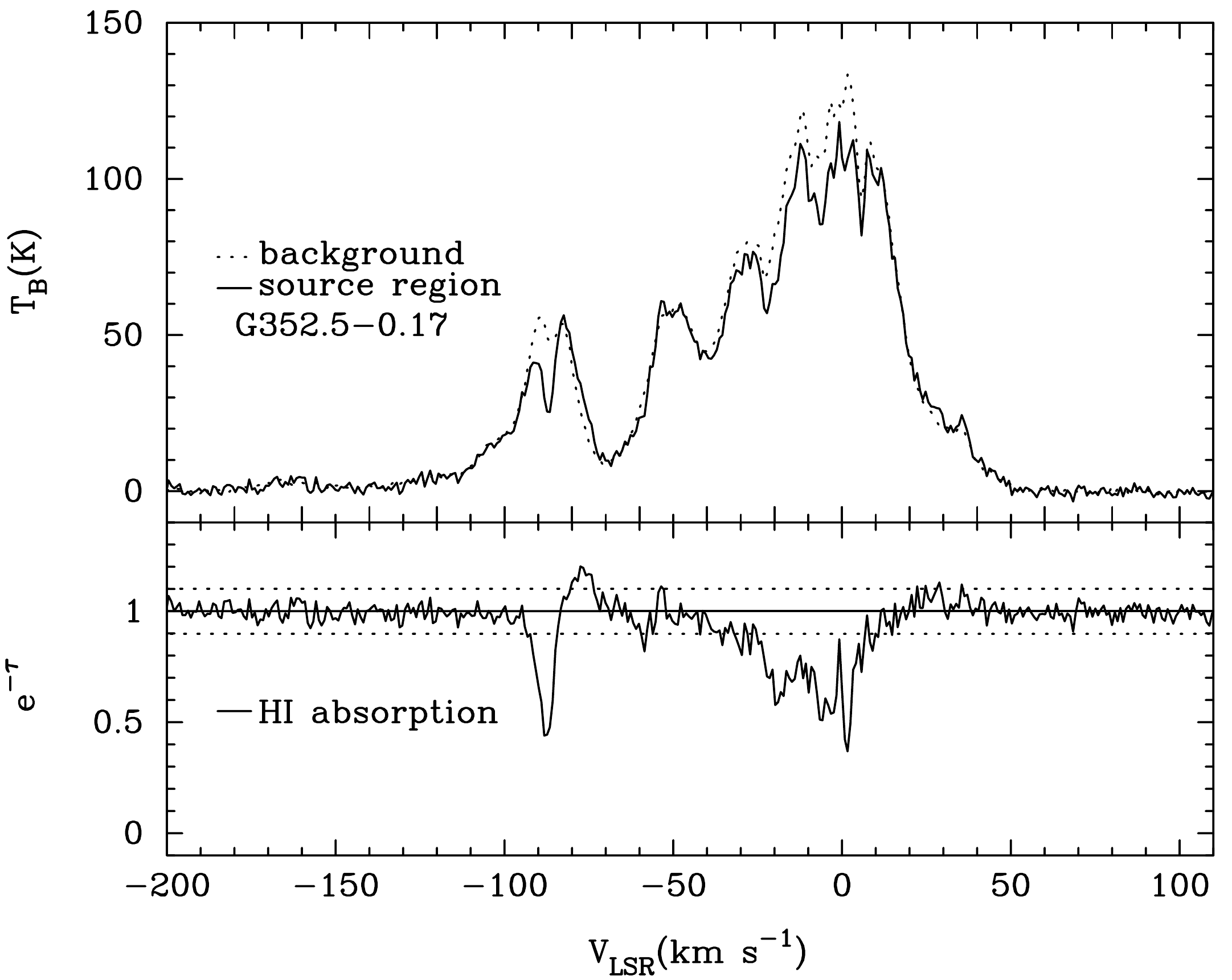}\\
    \includegraphics[width = 0.3\textwidth]{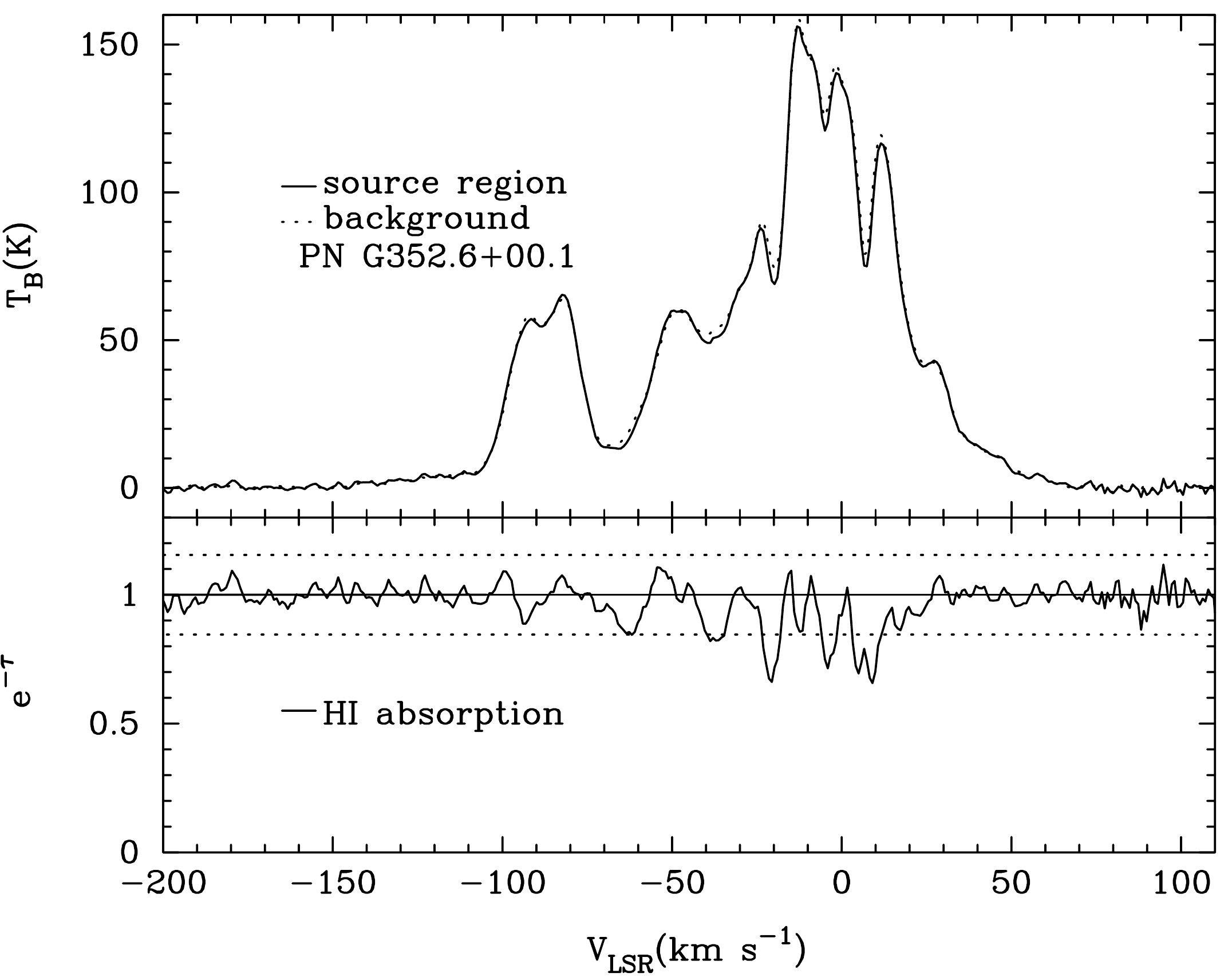}&
    \includegraphics[width = 0.3\textwidth]{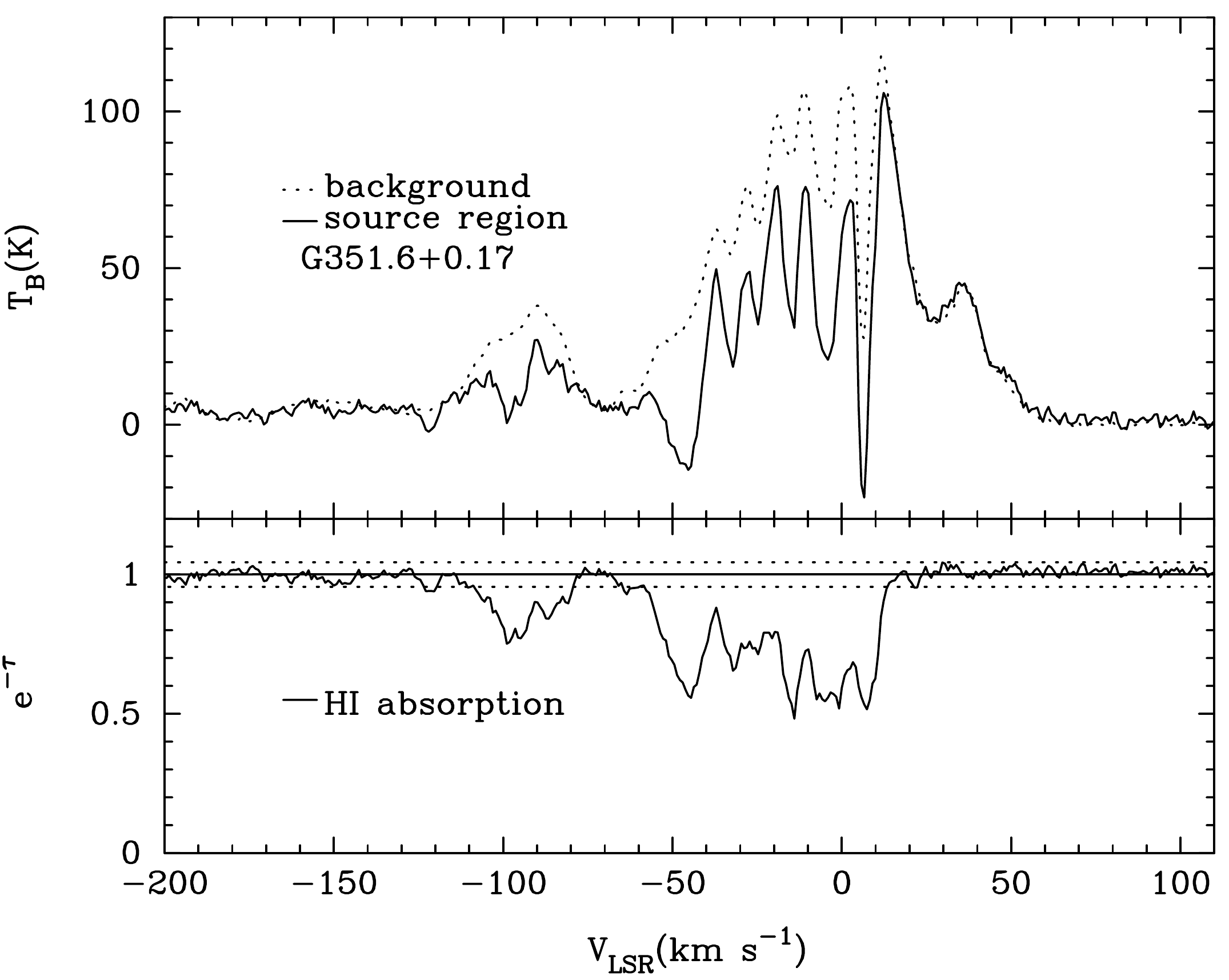}\\
   \end{tabular}
 \caption{1420\,MHz continuum image of PN G352.6$+$00.1 and its nearby background sources (top left), 
 and the \HI~spectra of PN G352.6$+$00.1 (bottom left)
 , G352.5$-$0.17 (top right), and G351.6$+$0.17 (bottom right). 
 The \HI~spectra of G353.4$-$00.3 is shown in Fig.~\ref{fig19}.
 The map has superimposed contours (0.15,0.35,1.0\,Jy) of 1420\,MHz continuum emission.}
 \label{fig18}
\end{figure*}

\begin{figure*}
 \centering
 \begin{tabular}{cc}
    \includegraphics[width = 0.29\textwidth,height = 0.24\textwidth]{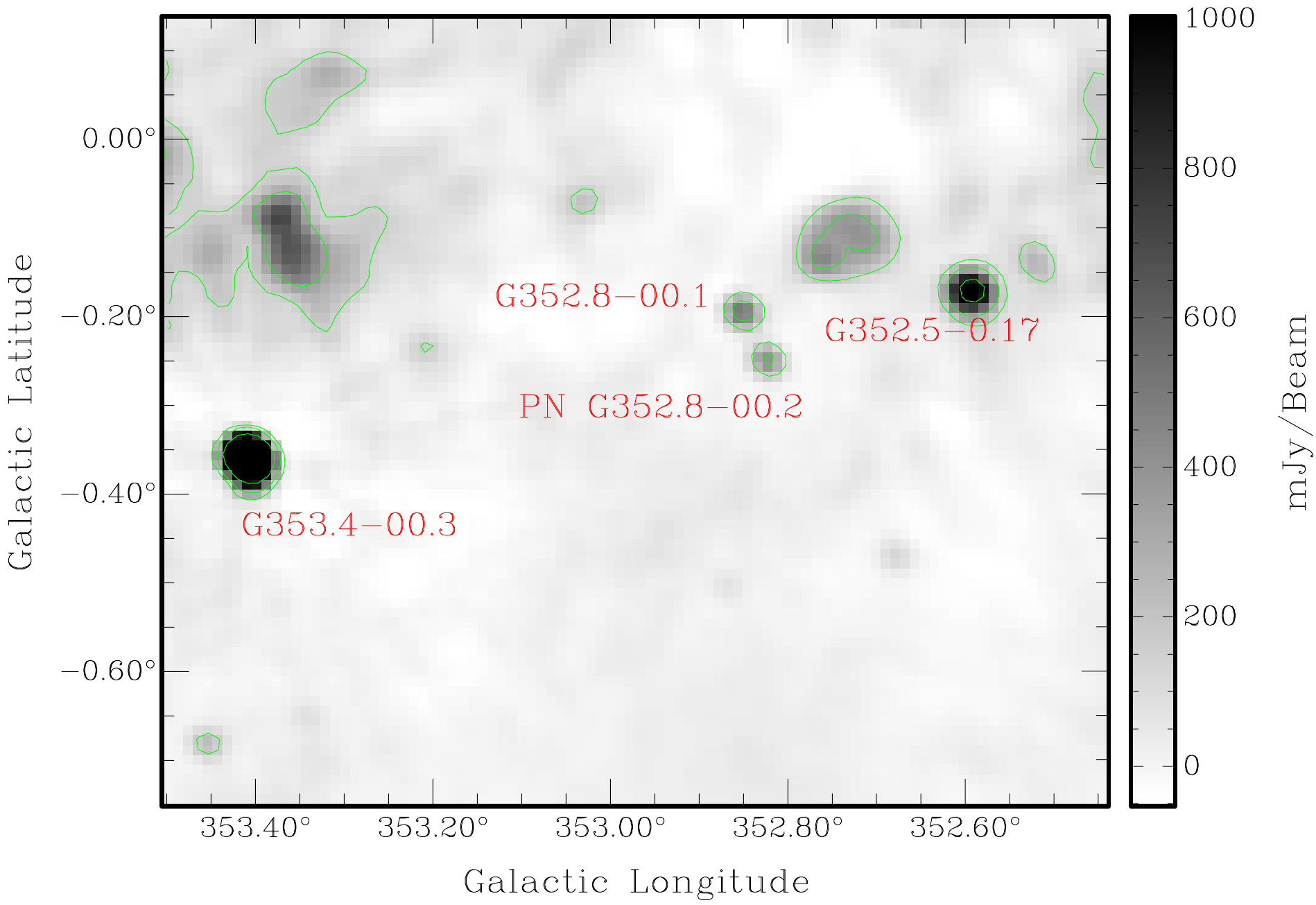}&
    \includegraphics[width = 0.3\textwidth]{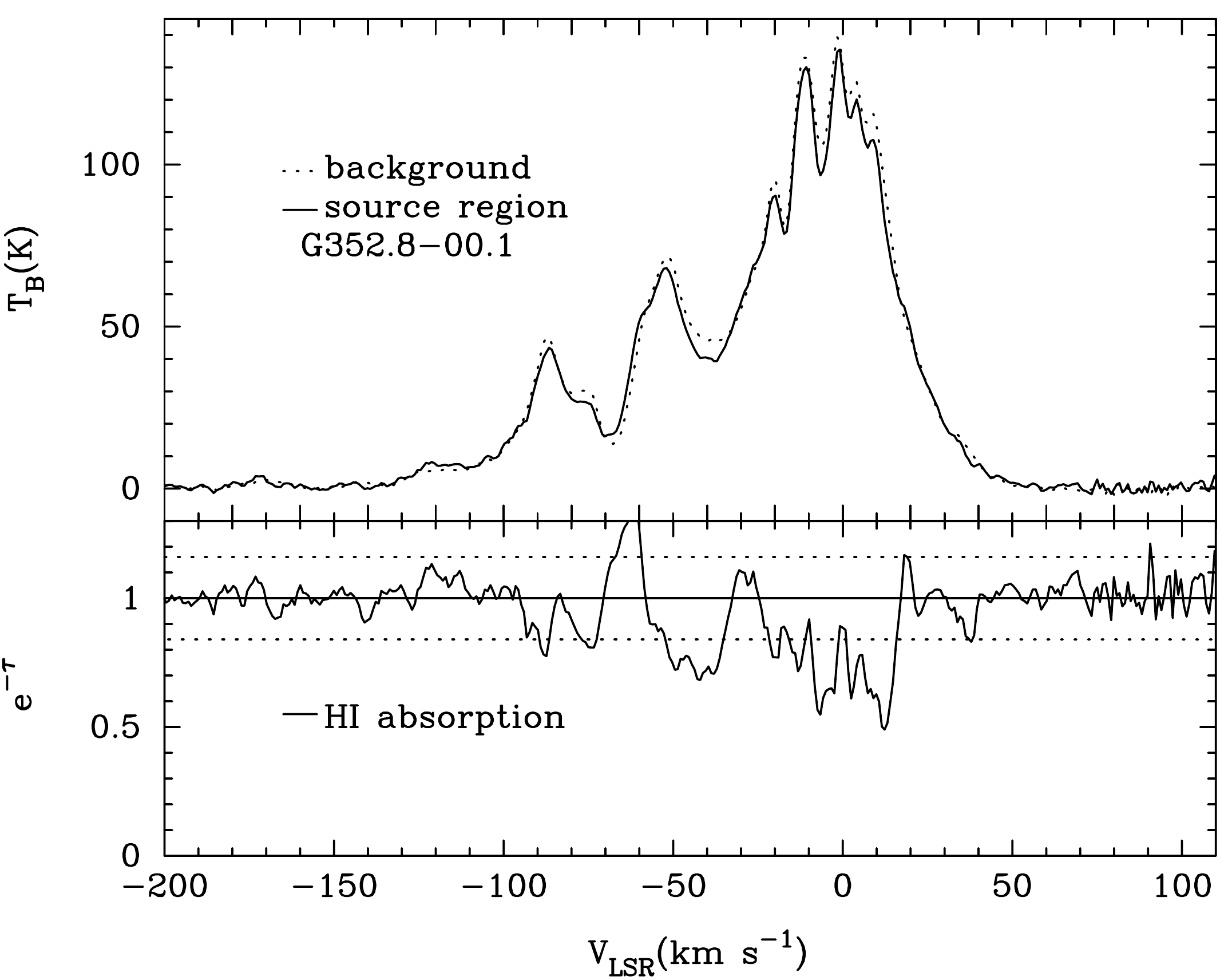}\\
    \includegraphics[width = 0.3\textwidth]{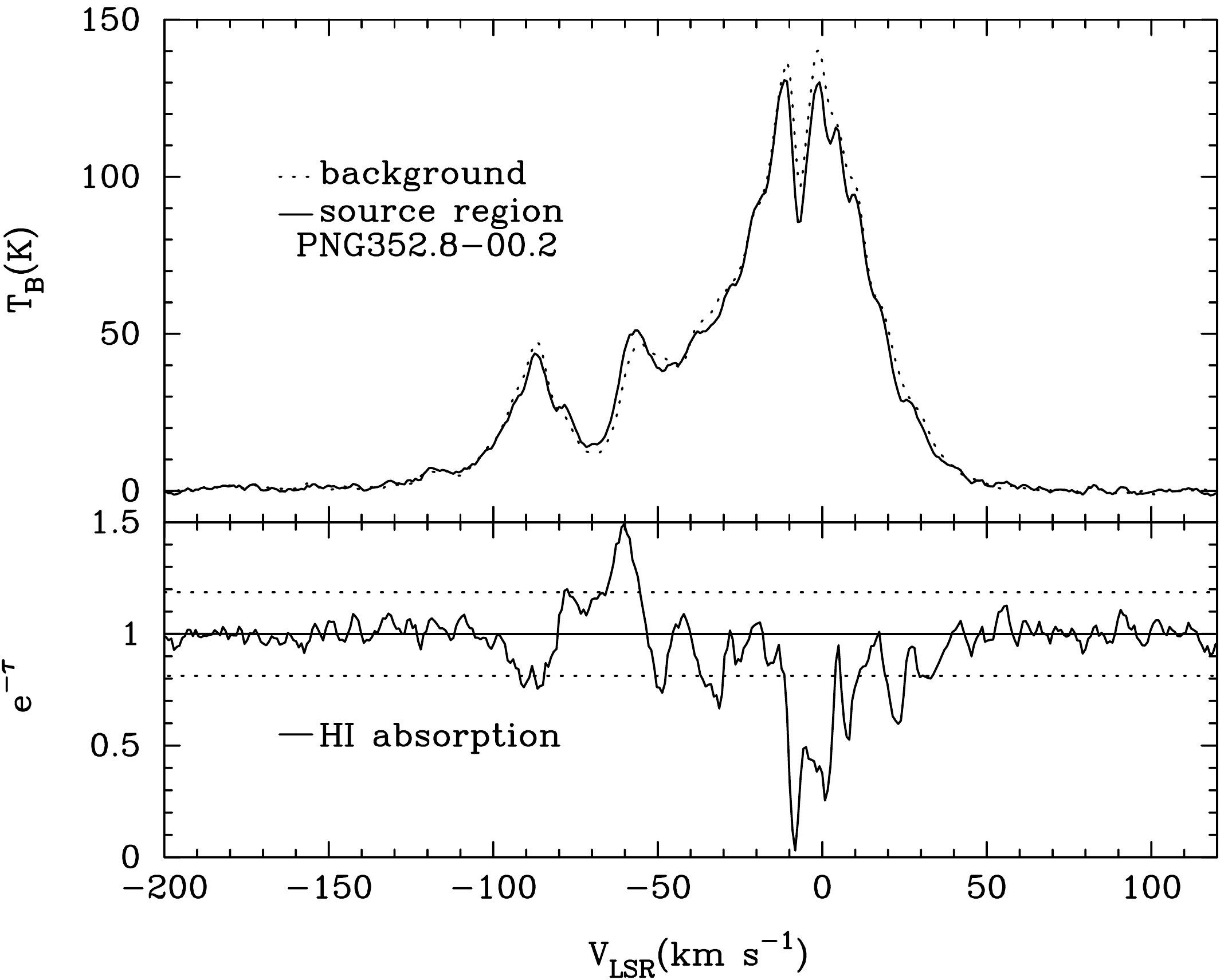}&
    \includegraphics[width = 0.3\textwidth]{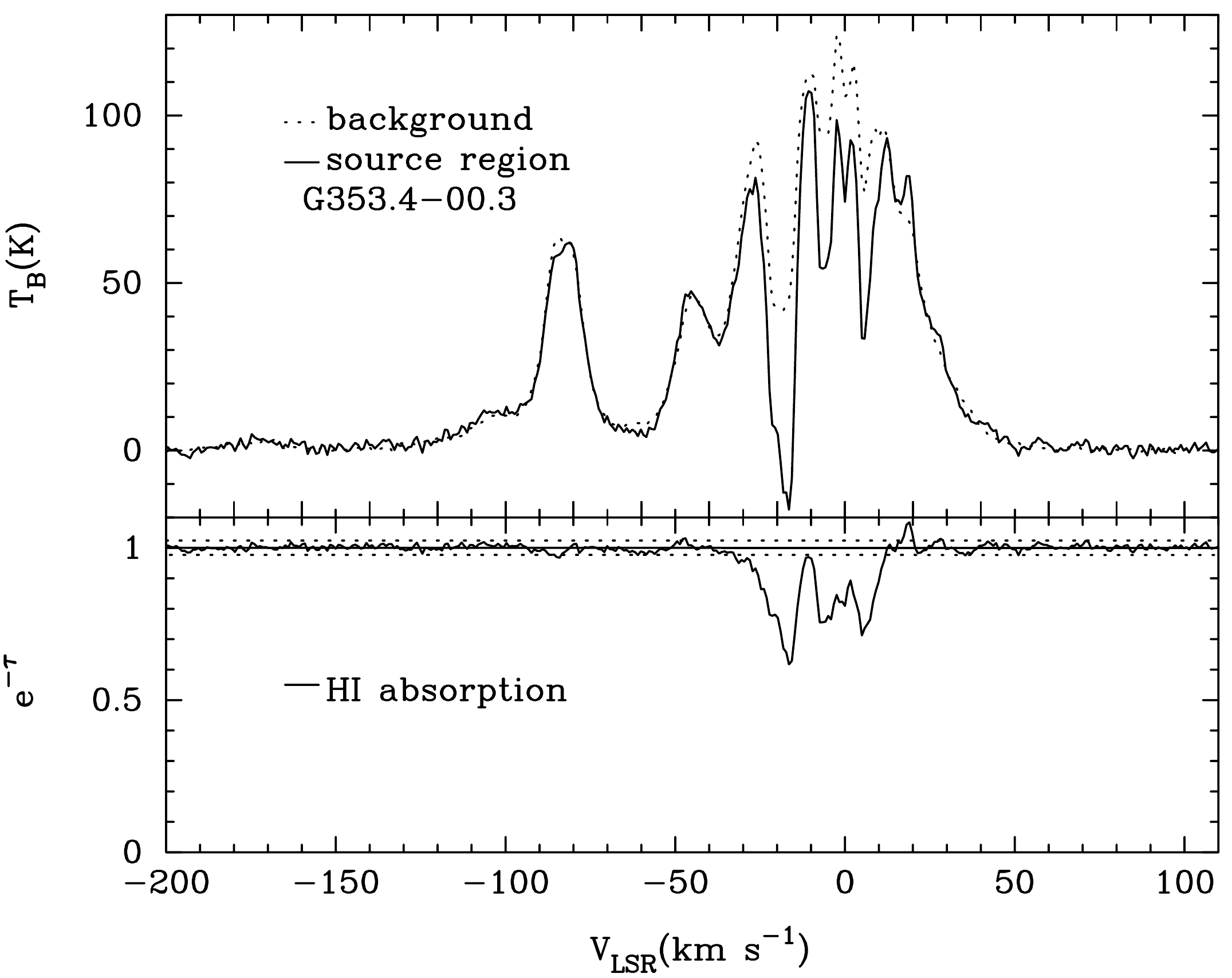}\\
   \end{tabular}
 \caption{1420\,MHz continuum image of PN G352.8$-$00.2 and its nearby background sources (top left), and the \HI~spectra of PN G352.8$-$00.2 (bottom left)
 , G352.8$-$00.1 (top right), and G353.4$-$00.3 (bottom right). 
  The \HI~spectra of G352.5$-$0.17 is shown in Fig.~\ref{fig18}.
 The map has superimposed contours (0.15,0.35,1.0\,Jy) of 1420\,MHz continuum emission.}
 \label{fig19}
\end{figure*}


\begin{figure*}
 \centering
 \begin{tabular}{cc}
    \includegraphics[width = 0.28\textwidth]{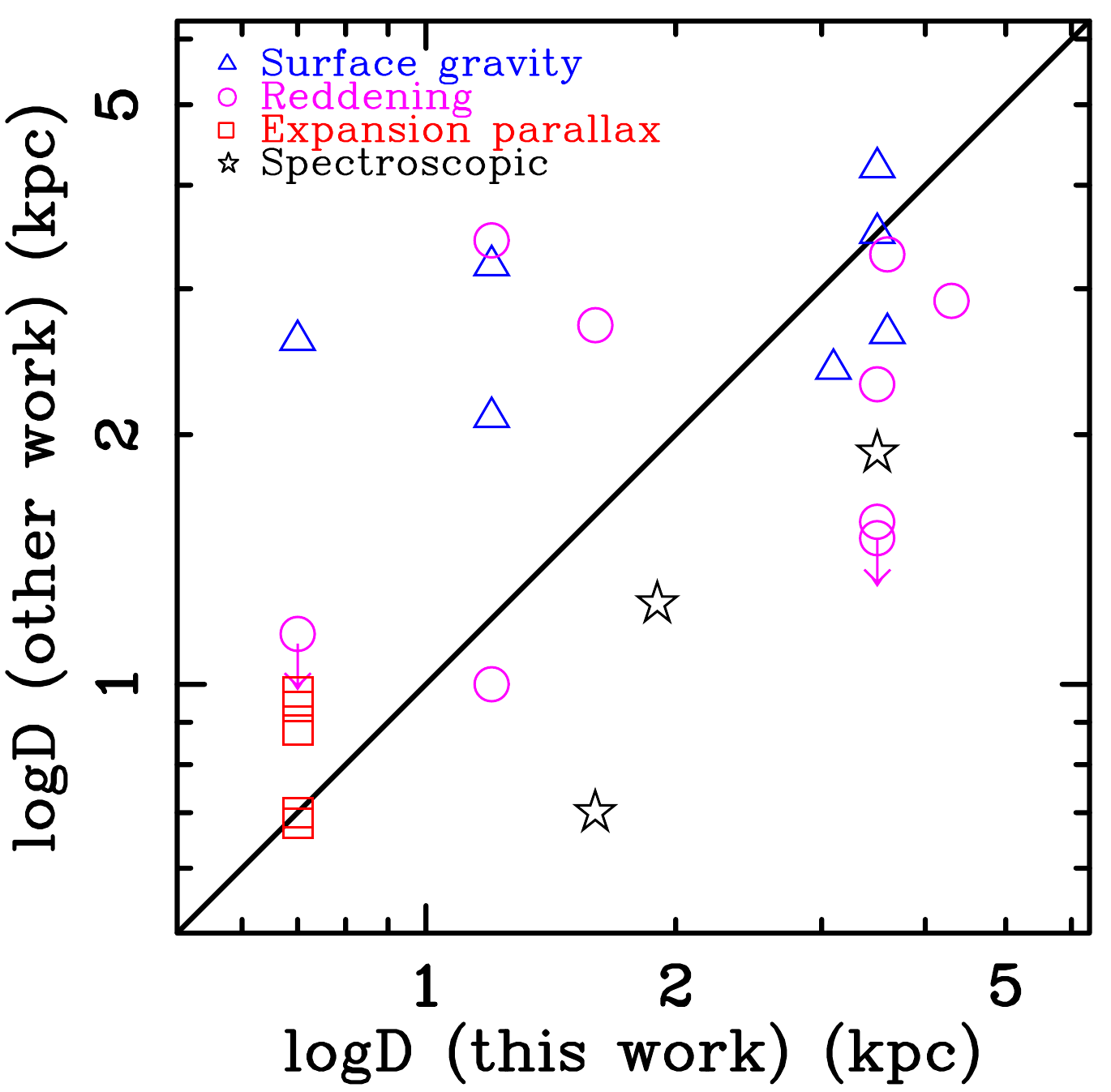}&
       \includegraphics[width = 0.28\textwidth]{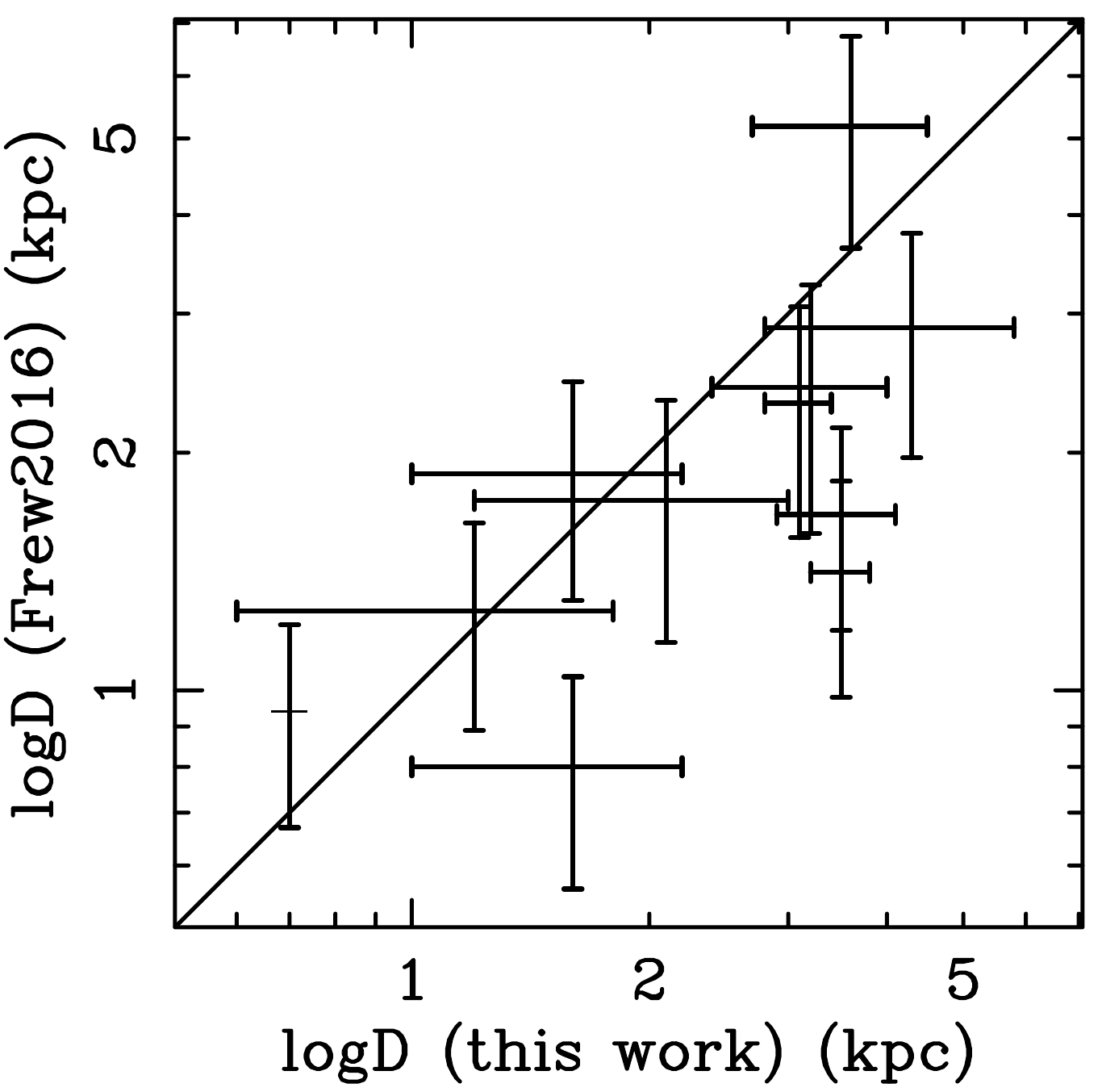} 
    \includegraphics[width = 0.28\textwidth]{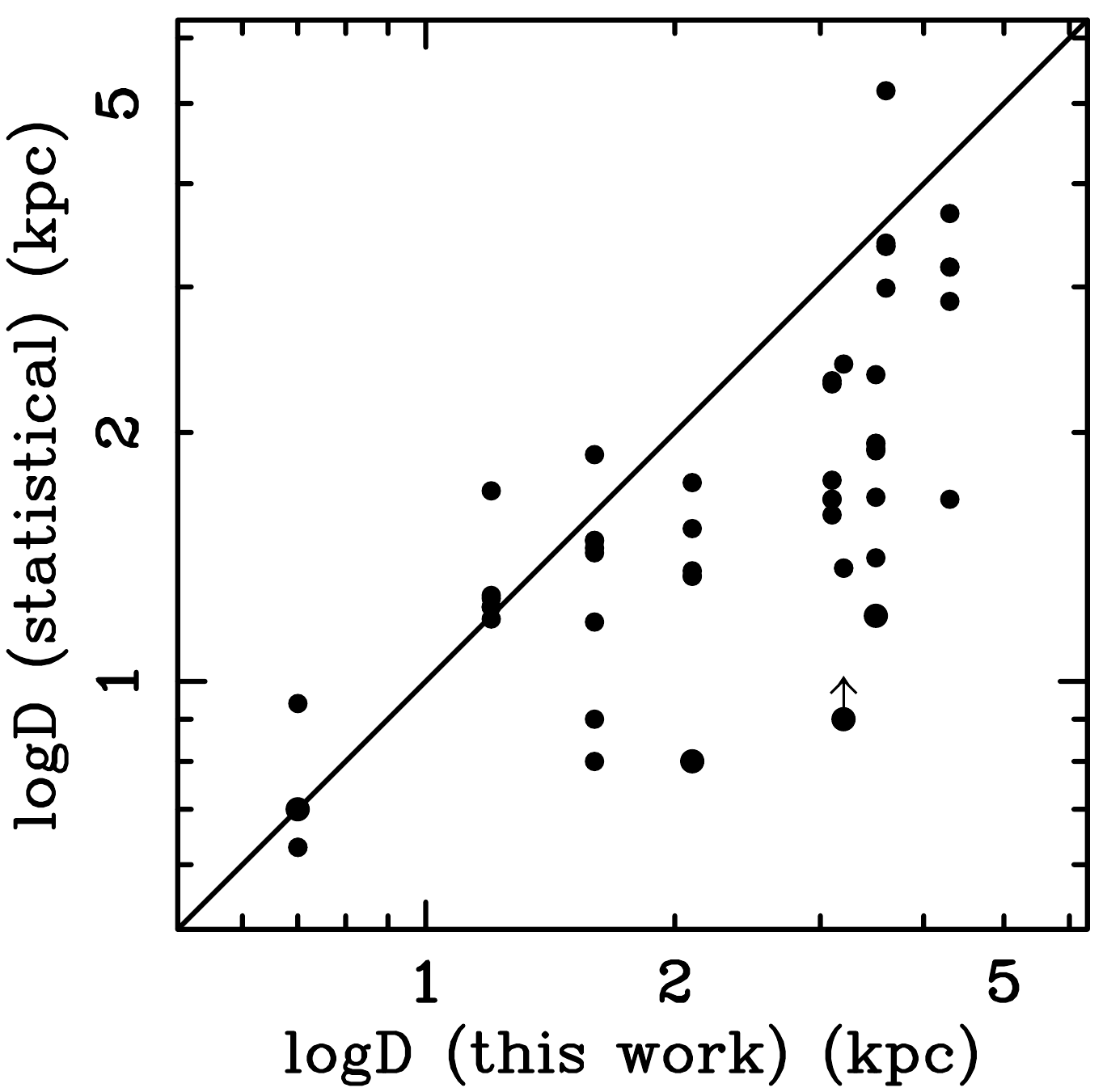}\\
   \end{tabular}
 \caption{Left panel: the correlation between our work and all other work (except
statistical method). Middle panel: the correlation between our work and \cite{Frew2016}. 
Right panel: the correlation between our work and all statistical results.}
 \label{fig20}
\end{figure*}

\clearpage

\appendix
\begin{figure*}[!hbt]
 \centering
 \begin{tabular}{cc}
    ~~\includegraphics[width = 0.33\textwidth]{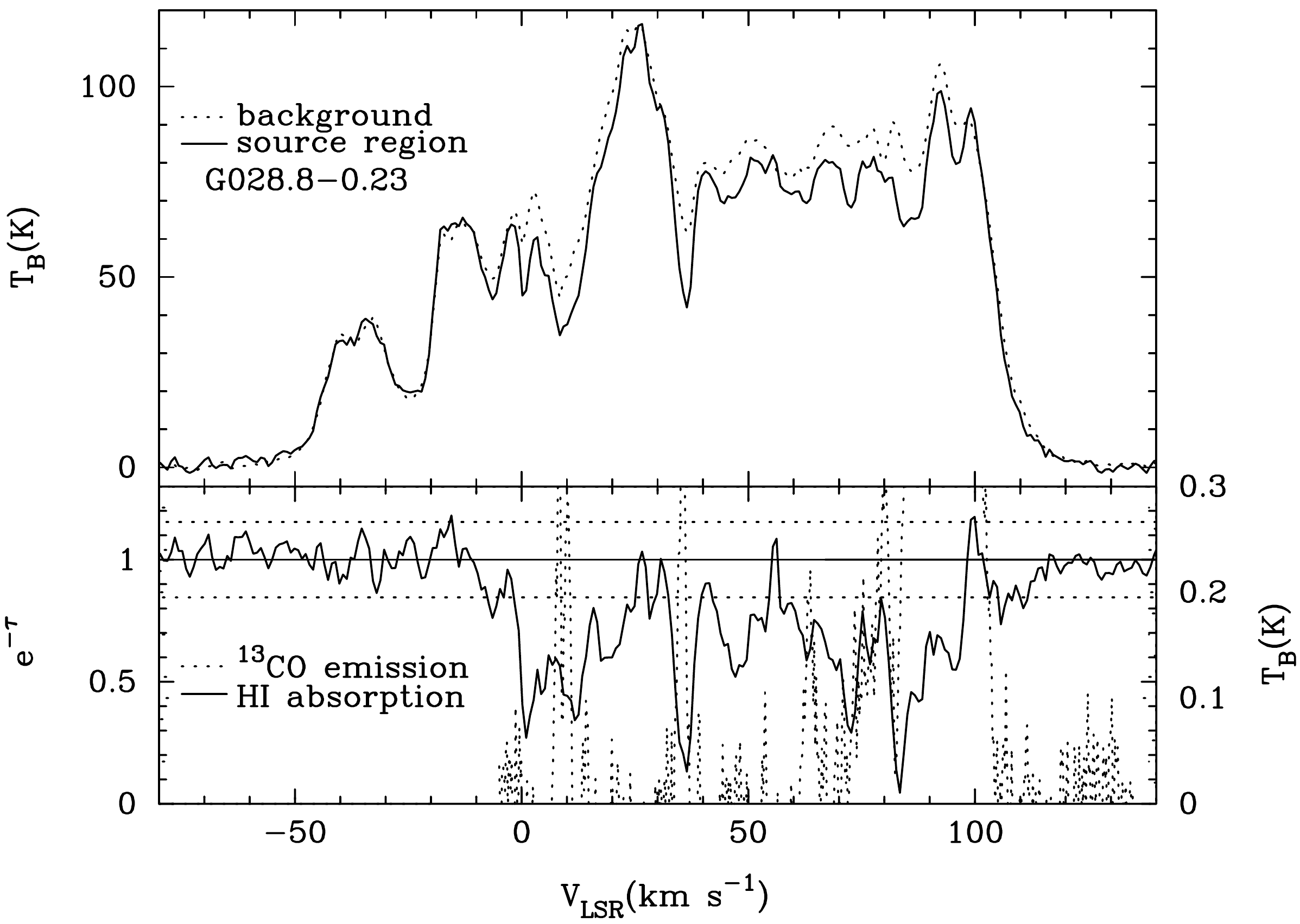}&
    ~~\includegraphics[width = 0.33\textwidth]{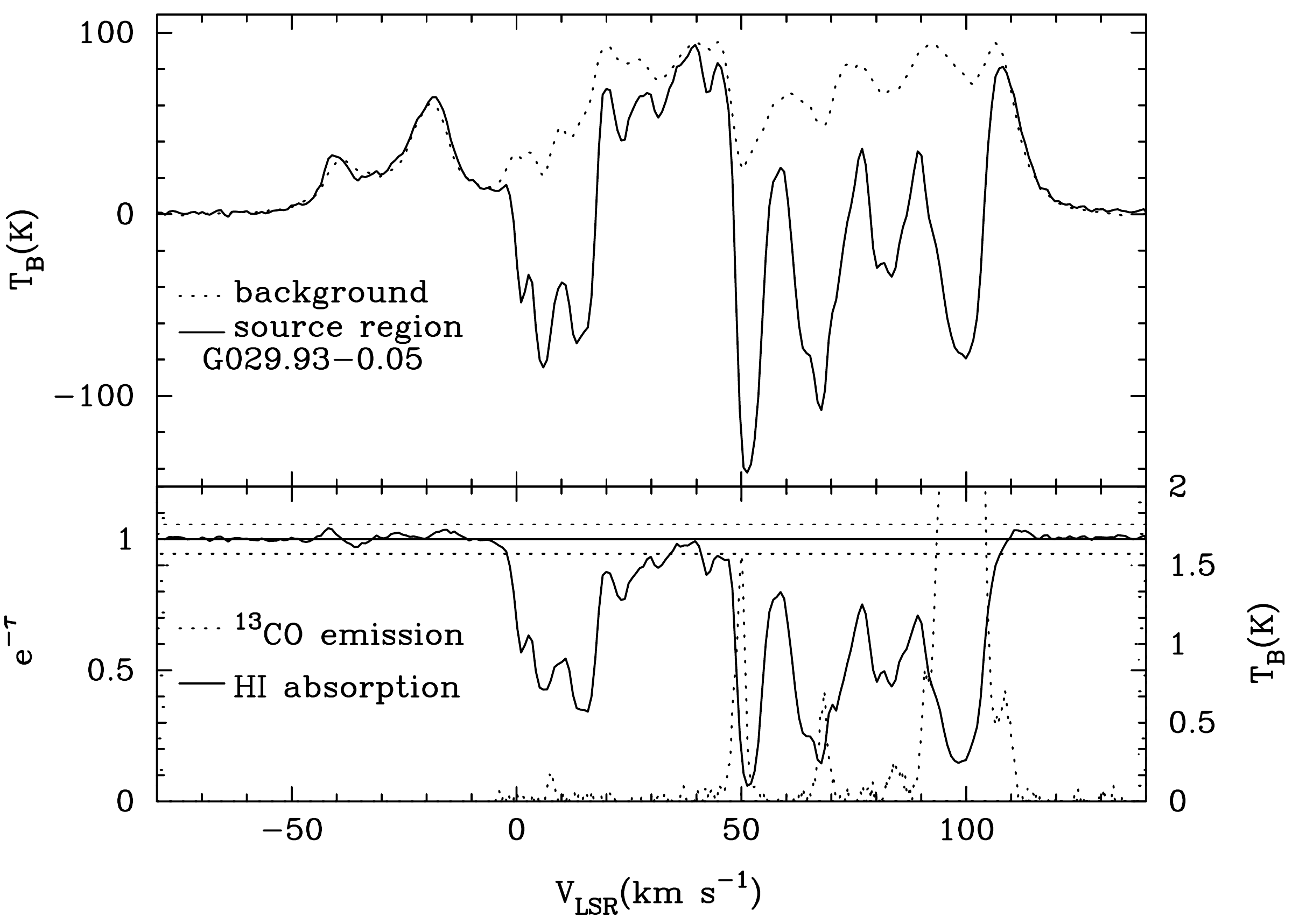}\\
   ~~\includegraphics[width = 0.33\textwidth]{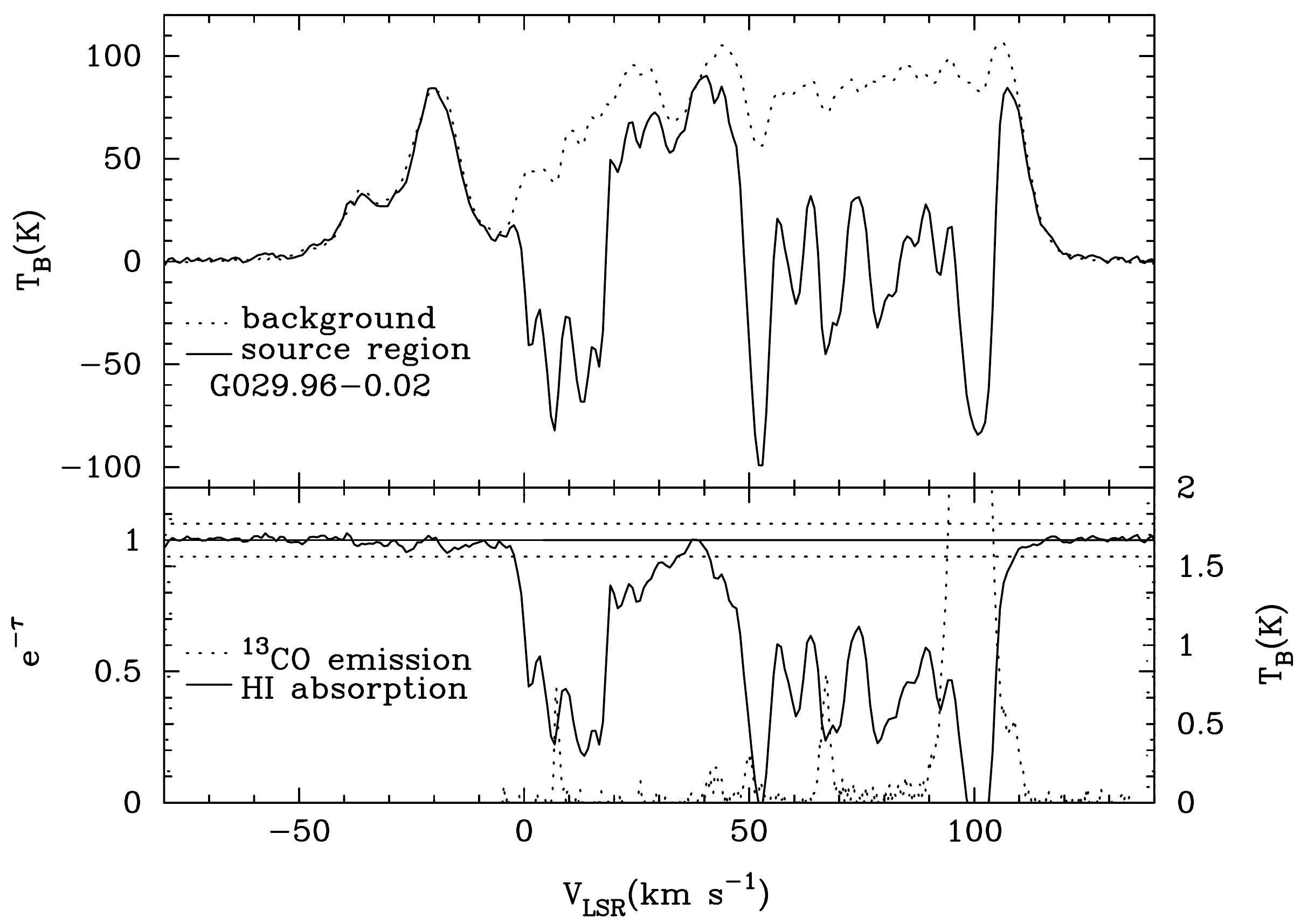}&
   ~~\includegraphics[width = 0.33\textwidth]{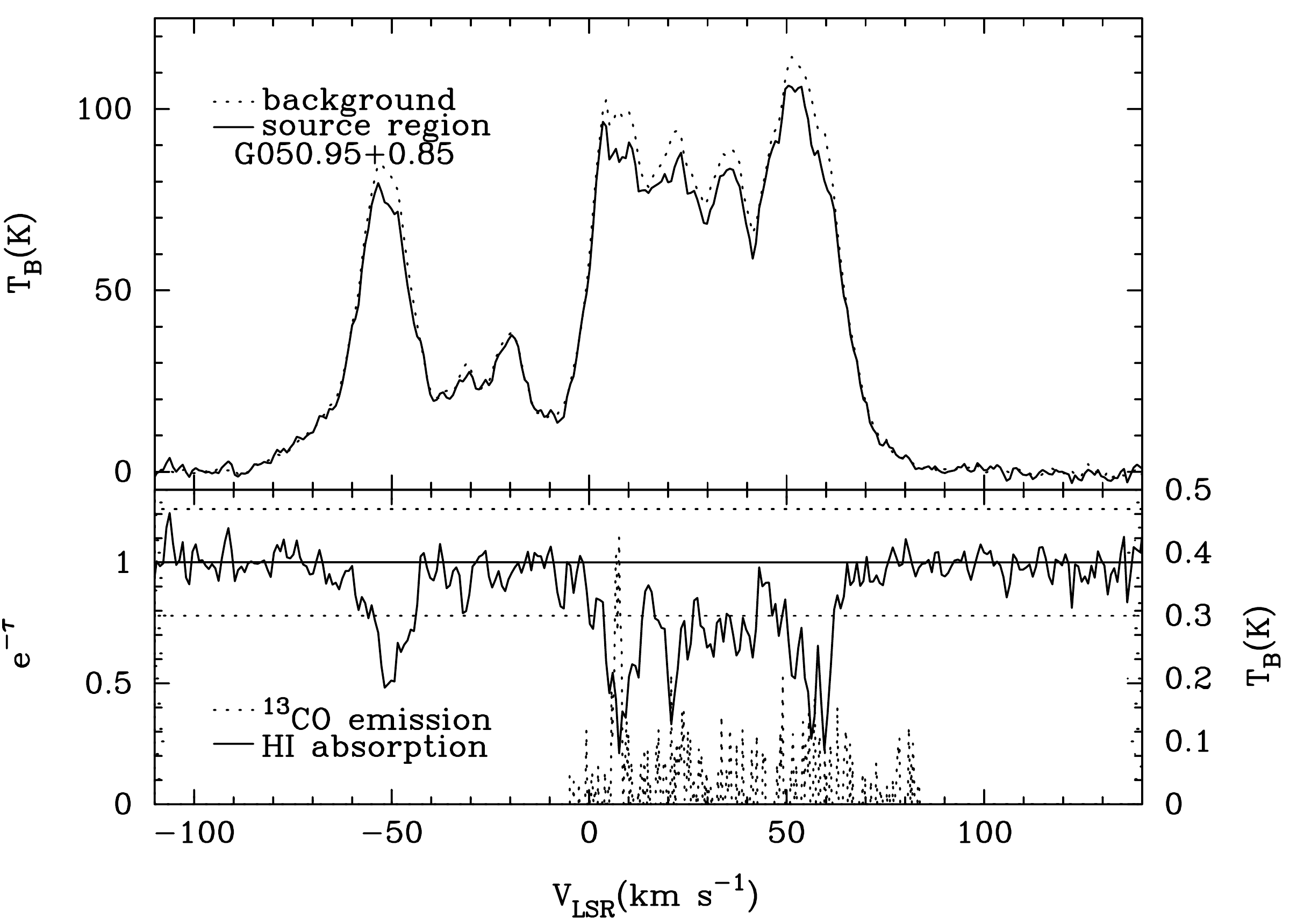}\\  
    \includegraphics[width = 0.30\textwidth]{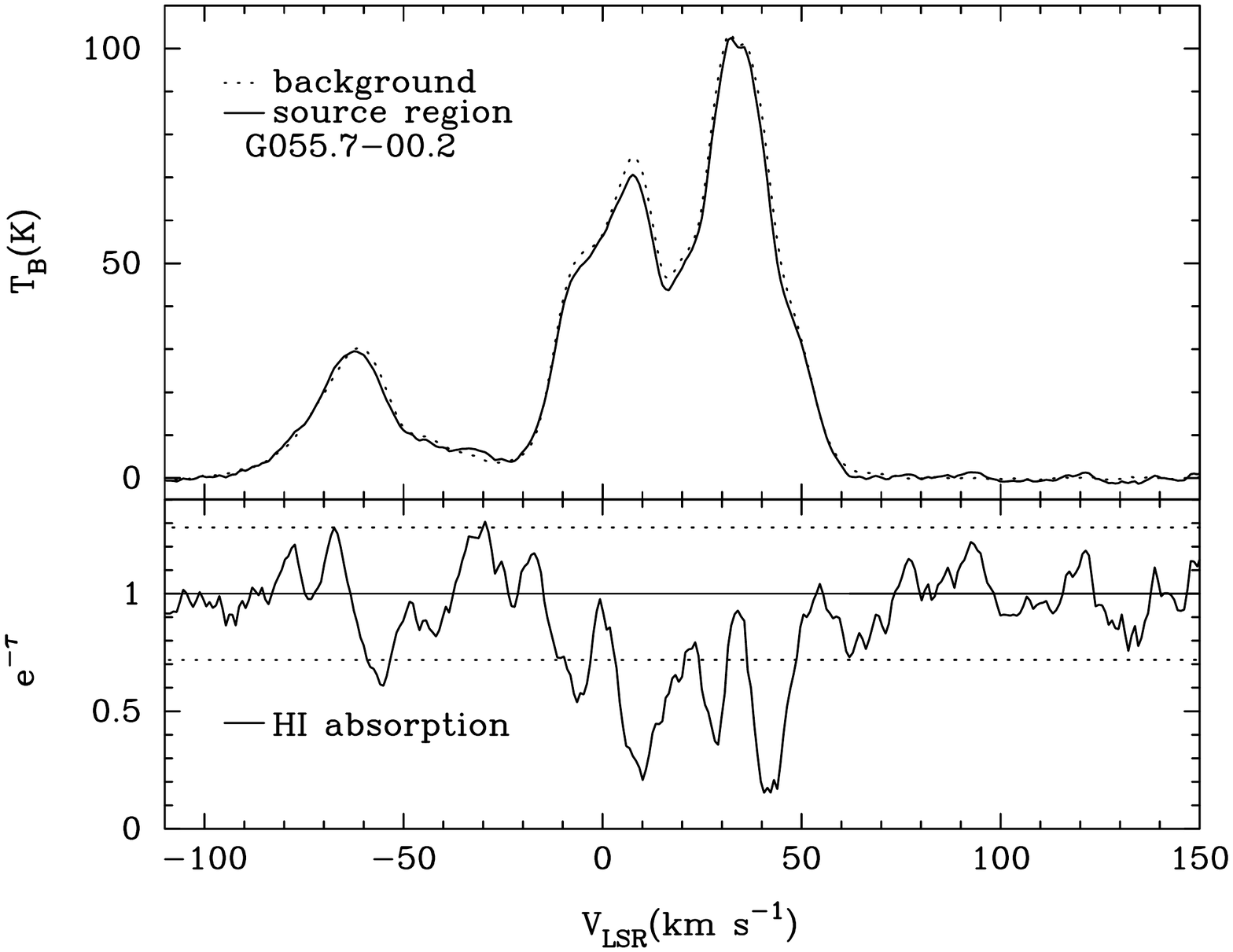}~~~&
     \includegraphics[width = 0.30\textwidth]{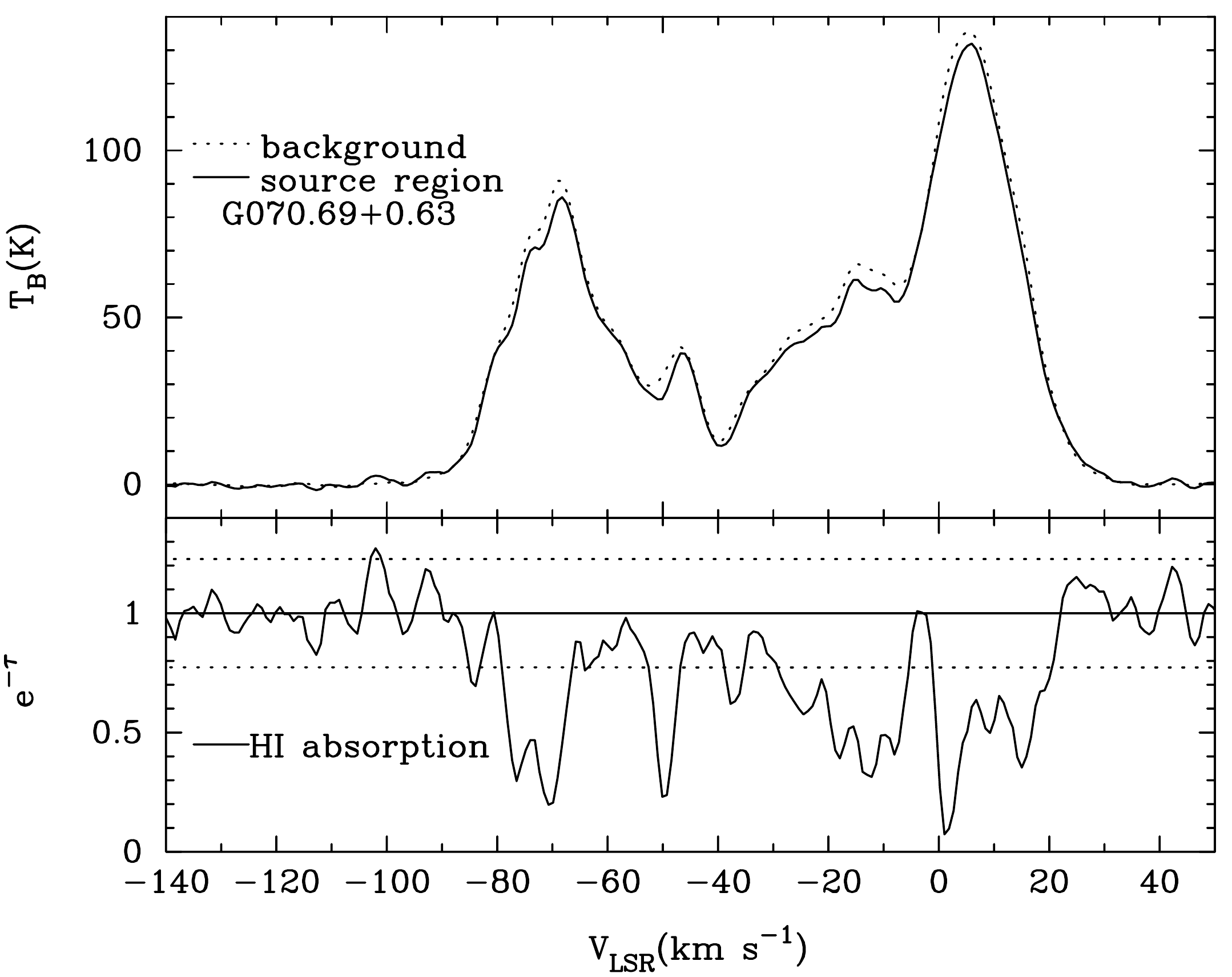}~~~\\
     \includegraphics[width = 0.30\textwidth]{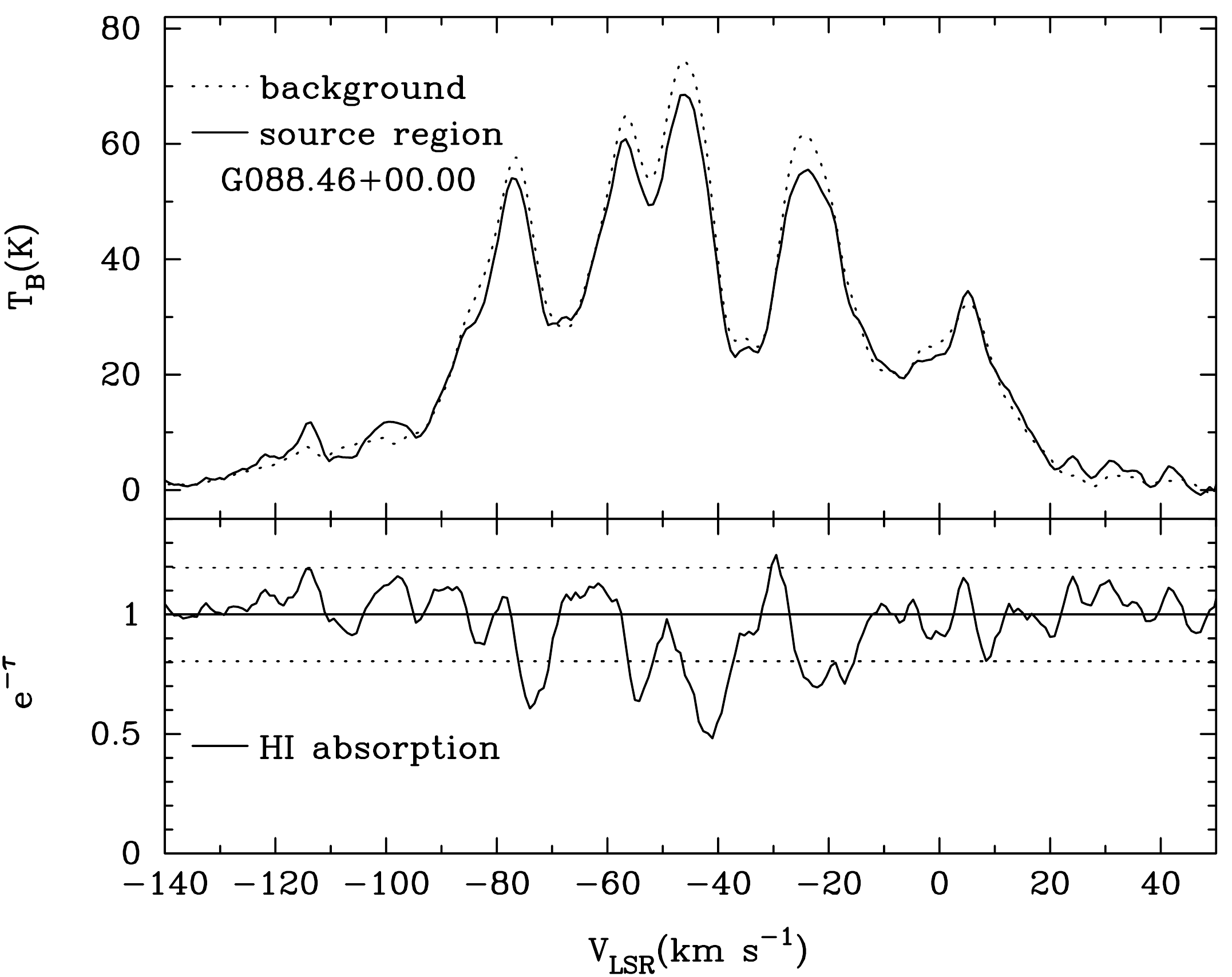}~~~&
\\    
   \end{tabular}
     \caption{Nearby background sources are also used to measure the PN kinematic distances, 
     the dotted horizontal lines in the lower panel of the background sources show the 3$\sigma$ noise level.}
 \label{fig21}
\end{figure*}

\end{document}